\documentclass[a4,12pt]{elsarticle}
\usepackage{graphics}
\usepackage{bm}
\usepackage{subfigure}
\usepackage{bm}
\usepackage{amssymb}
\usepackage{amsmath}
\usepackage{pxfonts}
\usepackage{latexsym}
\usepackage{setspace}
\usepackage{accents}
\usepackage{color} % for highlighting some text
\biboptions{square,comma,sort&compress}

\journal{Journal of Non-Newtonian Fluid Mechanics}

\newcommand\upperconvected[1]{\accentset{\nabla}{#1}}

\begin{document}

\begin{frontmatter}



%\title{}


%\author{}

%\address{}

\title{Viscoelastic fluid flow in a 2D channel bounded above by a deformable finite thickness elastic wall}

%\vspace{-24pt}

\author[chem]{Debadi Chakraborty}\fnref{fn1}
\author[chem]{J. Ravi Prakash\corref{cor1}}
\ead{ravi.jagadeeshan@eng.monash.edu.au}
\cortext[cor1]{Corresponding author. Tel +61 3 9905 3274; Fax +61 3 9905 5686.}
\fntext[fn1]{Present address: Department of Mathematics and Statistics, The 
University of Melbourne, Victoria 3010, Australia}
\address[chem]{Department of Chemical Engineering, Monash University, 
Melbourne, VIC~3800, Australia}

\begin{abstract}
% Text of abstract
{The steady flow of three viscoelastic fluids (Oldroyd-B, FENE-P, and 
Owens model for blood) in a two-dimensional channel, partly bound by a 
deformable, finite thickness neo-Hookean solid, is computed. The limiting 
Weissenberg number beyond which computations fail to converge is found to 
increase with increasing dimensionless solid elasticity parameter $\Gamma$, following the trend Owens $>$ FENE-P $>$ Oldroyd-B. The highly shear thinning nature of Owens model leads to the elastic solid always collapsing into the channel, for the wide range of values of $\Gamma$ considered here. In the case of the FENE-P and Oldroyd-B models, however, the fluid-solid interface can be either within the channel, or bulge outwards, depending on the value of $\Gamma$. This behaviour differs considerably from predictions of earlier models that treat the deformable solid as a zero-thickness membrane, in which case the membrane always lies within the channel. The capacity of the solid wall to support both pressure and shear stress, in contrast to the zero-thickness membrane that only responds to pressure, is responsible for the observed difference. Comparison of the stress and velocity fields in the channel for the three viscoelastic fluids, with the predictions for a Newtonian fluid, reveals that shear thinning rather than elasticity is the key source of the observed  differences in behaviour.}
\end{abstract}


\end{frontmatter}

\section{\label{sec:Introduction}Introduction}
Numerical simulation of blood transportation through the human
cardiovascular system is an intense area of
research~\citep{Grotberg2004,Taylor2004,Galdi2008}. Blood, which is 
rheologically complex, interacts with blood vessels walls both chemically and 
mechanically to give rise to an intricate fluid-structure interaction. From a fluid 
mechanics point of view, blood flow in large diameter blood vessels is commonly 
referred to as the macrocirculation, while flow in small vessels, such as arterioles,
venules, and capillaries is referred to as the microcirculation. 
The Navier-Stokes equations are a good
model for blood flow in the medium to large arteries, since Reynolds
numbers are high, tube diameters are large, and blood can be
considered to be an incompressible viscous Newtonian fluid. Most
studies in the literature on fluid-structure interaction in the
context of blood flow have so far been focused on the
macrocirculation. In small vessels, however, where the shear rate is small,
blood behaves as a non-Newtonian fluid due to its particulate nature. This 
necessitates its shear thinning, viscoelastic and thixotropic
nature to be taken into account~\citep{sequeira07,cristini05,baskurt03,owens06}. 
Another important aspect of the microcirculation is that the vessel wall thickness 
to diameter ratio is very high~\citep{kalita08}. To the best of our knowledge, so far 
there have been no studies of fluid-structure interaction associated with the flow 
of viscoelastic fluids in vessels with finite thickness walls. The aim of this work is 
to examine the flow of a variety of viscoelastic fluid models interacting with a
finite-thickness elastic vessel wall.

The study of fluid flow in collapsible channels and tubes has been motivated by 
the complex and nonlinear dynamics revealed by laboratory experiments~
\citep{bertram86,bertram90,Grotberg2004}. The earliest and simplest theoretical 
models of collapsible-tube flow were lumped-parameter ~\citep{katz69} and one 
dimensional models~\citep{shapiro77,jensen90}, followed by two-dimensional
models where part of one wall is replaced by a tensioned membrane in
a two-dimensional rigid parallel sided channel. The membrane model
assumes that the bending stiffness and extensibility of the wall in
the flow direction can be ignored, and that the movement of the
elastic wall is only in the direction normal to the wall. More
recently, this basic model has been improved by using a plane
strained elastic beam model for the collapsible wall with a
Bernoulli-Euler beam, a Timoshenko beam and a 2D solid
model~\citep{cai03,luo07, Liu2009}. Wall stiffness was found to play 
a major role in attaining a steady state solution, and for very
small wall stiffness, the results of the beam model compared 
favourably with those of the membrane model. Work is ongoing on
extending these models to describe 3D compliant
tubes~\citep{JensenHeil2003a,Xiebookchap03,XiePhD,Liu2009,hazel03,
Marzo2005}. In all these studies, however, the fluid has always been
treated as Newtonian. In this work, we examine the flow in
a 2D collapsible channel of three different viscoelastic fluid models, namely, the 
Oldroyd-B, the FENE-P and the Owens model for blood~\citep{owens06}, by 
considering the deformable wall to be a finite-thickness incompressible neo-Hookean solid. The channel dimensions are chosen to be compatible with the 
microcirculation.

The steady flow of Oldroyd-B, FENE-P and the Owens model fluids in a two-dimensional collapsible channel has been studied recently by~\citet{debadi10}. In contrast to the present work, however, the collapsible wall was modelled as a zero thickness membrane under constant tension~\citep{luo95}.  It was shown that the predictions for the different viscoelastic fluids differ significantly from each other, with the key factor being the extent of shear thinning predicted by the individual models. In particular,
it was shown that viscoelastic fluids behave identically to
Newtonian fluids, provided that the viscosity of the two fluids at
the location of the maximum shear rate in the channel is the same. Subsequently, 
\citet{debadi11} examined the influence of the degree of shear thinning of the
viscoelastic fluid by systematically varying the finite
extensibility parameter $b_{\textbf{M}}$, in the FENE-P model, which 
controls the extent of shear thinning experienced by the fluid, and
is consequently a convenient parameter for examining the influence
of shear thinning. They found that the pressure drop, the molecular conformation 
tensor fields, and the stresses in the flow domain are significantly affected by the 
extent of shear thinning of the FENE-P fluid. Importantly, in both these studies, it 
was found that the significant differences that arise amongst the different 
viscoelastic fluids in the predicted value of the tangential shear stress on the 
membrane surface, has no influence on the shape of the deformable membrane,
because of the boundary condition adopted in the work. Essentially it was 
assumed that the shape of the membrane is governed only by
the normal stresses acting on it.

In order to use a more realistic model for the collapsible wall, \citet{debadi2012} 
replaced the zero thickness membrane with a finite thickness neo-Hookean solid, which can account for the effect of shear stress on the shape of the wall. The fluid was, however, assumed to be Newtonian. The model formulation followed the seminal work of \citet{carvalho97}, who examined roll cover deformation in roll coating flows, with the rubber roll cover modelled as either an incompressible neo-Hookean or a Mooney-Rivlin solid. Computational predictions of the deformation of the collapsible wall and pressure drop in the channel were found to be in good agreement with experimental measurements carried out in a polydimethylsiloxane microfluidic device, composed of a single microchannel with a thin flexible layer along one side of the channel. 

The present work extends the model of \citet{debadi2012} by replacing the Newtonian fluid with Oldroyd-B, FENE-P and the Owens model viscoelastic fluids. The constitutive equations of all the viscoelastic fluids are written in conformation tensor form \citep{pasquali04}, and the fully  coupled, steady state, fluid and solid equations are solved using the DEVSS-TG/SUPG (discrete elastic viscous stress split-traceless gradient) finite element method. Such a general formulation enables us to study the complex fluid-structure interaction that arises both from the capacity of the deformable wall to support shear and normal stresses, and from the shear thinning and viscoelastic nature of the fluid. 

The plan of the paper is as follows. The problem formulation, with
details of the governing equations for the viscoelastic fluids and
incompressible neo-Hookean solid, the boundary conditions and the
relevant dimensionless variables are presented in Section~\ref{sec:formulation}. The results of viscoelastic and Newtonian fluid computations are compared in Section~\ref{sec:RD} . In particular, the dependence of the shape of the fluid-solid
interface, and of the pressure, stress, conformation tensor and velocity fields on the different parameters, is examined. Finally, concluding remarks are drawn in Section~\ref{sec:conclusion}.

\section{\label{sec:formulation}Problem formulation}
\begin{figure}[t]
\begin{center}
\includegraphics[width=0.90\textwidth]{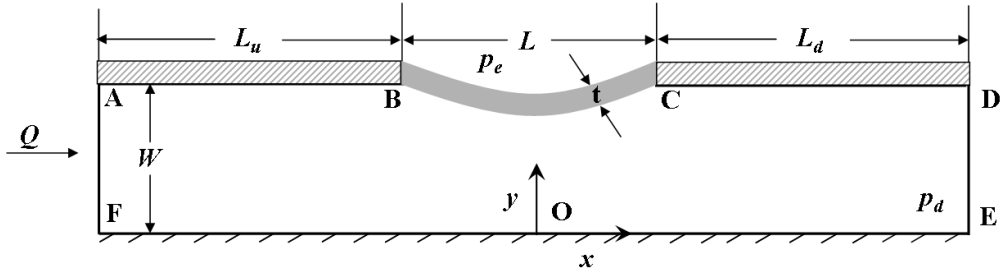}
\caption{Geometry of the domain. $L_u = 7W$, $L = 5W$, and $L_d = 7W$. The wall thickness $t$ is varied between $0.1 W$ and $0.4 W$.}
\label{fig1}
\end{center}
\end{figure}
The geometry of the flow is that of a 2D channel, with one of the
walls containing an elastic segment as illustrated in
Fig.~\ref{fig1}. In units of channel width $W$, the dimensions of
the channel are $L_u = 7W$, $L = 5W$, and $L_d = 7W$. In most of the simulations, the solid wall has a thickness $t = 0.4W$.

\subsection{\label{sec:} Governing Equations}
We have nondimensionalized the various physical quantities, by
scaling lengths and displacements with $W$, velocities with
$GW/\eta_0$ and pressure and stresses with $G$. Here $G$ is the
 shear modulus of the solid, $\eta_{0} (= \eta_\text{s} +
\eta_{\text{p},0})$ is the zero shear rate solution viscosity,
$\eta_{\text{s}}$ is the solvent viscosity and $\eta_{\text{p},0}$
is the contribution of the micro-structural elements to the zero
shear rate viscosity. (For a Newtonian fluid, $\eta_{0}$ is just the
constant Newtonian viscosity). Non-dimensionalization of the
governing equations and boundary conditions yields the following
dimensionless numbers:
\begin{equation}
Re = \frac{\rho W U_0}{\eta_{0}} \, ; \quad \beta  =
\frac{\eta_\text{s}}{\eta_0} \, ; \quad Wi = \frac{\lambda_0 U_0}{W}
\, ; \quad  \Gamma =  \frac{\eta_0 U_{0}}{G W} \, ; \quad P  =
\frac{p}{G} \label{dimensionlessnumbers}
\end{equation}
where $Re$ is the Reynolds number, $\beta$ is the viscosity ratio,
$Wi$ is the \emph{inlet} Weissenberg number, $\Gamma$ is the
dimensionless solid elasticity parameter, $P$ is the dimensionless
pressure, $\rho$ is the density of the liquid, $U_0$ is the average
inlet velocity,  $\lambda_{0}$ is the constant characteristic
relaxation time of the microstructure. It is also convenient to
define a \textit{local} Weissenberg number $\widetilde{Wi} =
\lambda_0 \, \dot\gamma$, which measures the non-dimensional shear
rate at any location in the flow.

Upon introduction of these dimensionless variables, we can recast
governing equations in the following
dimensionless form:
\begin{equation}
\bm{\nabla}\cdot\textbf{v}=0  \quad \text{(Mass
balance)}\label{continuity}
\end{equation}
\begin{equation}
\frac{Re}{\Gamma}\,\textbf{v}\cdot\bm{\nabla}\textbf{v}=\bm{\nabla}\cdot\textbf{T}
\quad \text{(Momentum balance)}\label{momentum}
\end{equation}
\begin{equation}
{\textbf{v}} \cdot \bm{\nabla}\textbf{M} -
\bm{\nabla}{\textbf{v}^{T}} \cdot \textbf{M} - \textbf{M} \cdot
\bm{\nabla}{\textbf{v}} = - \frac{\Gamma}{Wi}\left\lbrace f
(\text{tr} \, \textbf{M}) \, \textbf{M}  - \textbf{I} \right\rbrace
\quad \text{(Conformation tensor )}\label{conformation}
\end{equation}
\begin{equation}
\textbf{T} = -P\textbf{I}+\bm{\tau^{\text{s}}}+\bm{\tau^{\text{p}}}
\quad \text{(Cauchy stress tensor)}\label{Cauchystress}
\end{equation}
\begin{equation}
\bm{\tau^{\text{s}}}=\beta(\bm{\nabla}\textbf{v}+\bm{\nabla}\textbf{v}^T)
\quad \text{(Viscous stress tensor)}\label{viscousstress}
\end{equation}
\begin{equation}
\bm{\tau^{\text{p}}}=(1 - \beta)\,\frac{\Gamma}{Wi}\left\lbrace f
(\text{tr}  \, \textbf{M}) \, \textbf{M} - \textbf{I}\right\rbrace
\quad \text{(Elastic stress tensor)}\label{elasticstress}
\end{equation}
\begin{equation}
\bm{\nabla}_{\textbf{X}}\cdot\textbf{S}=0 \quad \text{(Equation of
motion for solid)}\label{dcauchy-stress}
\end{equation}
\begin{equation}
\textbf{S}=\textbf{F}^{-1}\cdot\bm{\sigma} \quad \text{(First
Piola-Kirchhoff stress tensor )}
\end{equation}
\begin{equation}
\bm{\sigma}=-\pi \textbf{I}+\textbf{B} \quad \text{(Cauchy stress
tensor for a neo-Hookean material)} \label{dNH}
\end{equation}
In these equations, $\textbf{v}$ is the velocity, $\bm{\nabla}$
denotes the gradient, $\textbf{I}$ is the
identity tensor, {$\pi$ is the pressure}, and
$\textbf{B}$ is the left Cauchy-Green tensor, expressed as
$\textbf{B} = \textbf{F}\cdot\textbf{F}^{\textbf{T}}$. The
deformation gradient tensor \textbf{F} relates the undeformed state
[$\textbf{X}$ = $(X, Y, Z)$] to the deformed state [$\textbf{x}$ =
$(x, y, z)$] and is expressed as:
\begin{equation}
\textbf{F}=\frac{\partial \textbf{x} }{\partial \textbf{X} }
\end{equation}

The form of $ f (\text{tr}   \, \textbf{M})$ is model specific. For
the Oldroyd-B model, $ f (\text{tr}   \, \textbf{M}) = 1$, while for
the FENE-P model~\citep{pasquali04},
 \begin{equation} \label{funcf}
f(\text{tr}  \,  \textbf{M}) = \frac{b_{\textbf{M}} -
1}{b_{\textbf{M}} -  \dfrac{\text{tr}  \,  \textbf{M}}{3}}
\end{equation}
where, $b_{\textbf{M}}$ is the {finite extensibility} parameter,
defined as the ratio of maximum length squared of the
micro-structural element to its average length squared at
equilibrium.

For the Owens model, $ f (\text{tr}   \, \textbf{M}) = 1$ also holds, however, the 
\emph{constant} relaxation time $\lambda_{0}$ in Eq.~(\ref{conformation}) is replaced by a \emph{function} $\lambda$, which represents the relaxation time of
the elastic stress due to blood cell aggregates. Note that this replacement is not carried out in ~Eq.~(\ref{elasticstress}), where the relaxation time remains constant and equal to $\lambda_{0}$. The function $\lambda$ depends
on the average size of the blood cell aggregates, $n$, which is
controlled by the competition of spontaneous aggregation and
flow-induced disaggregation. \citet{iolov11} have recently developed a finite 
element method for solving the Owens model in its complete generality. Since our 
focus here is on developing a fluid-structure interaction model that accounts for a 
viscoelastic fluid model and a finite thickness elastic wall, we assume for 
simplicity that the dynamics of $n$ are fast with respect to other changes of the flow, i.e., $n = n_\text{st}(\dot \gamma)$, which is its equilibrium value based on the local shear rate $\dot{\gamma} = \sqrt{2 \, \textbf{D}:\textbf{D} }$,
where $\textbf{D} = \frac{1}{2}(\bm{\nabla}\textbf{v}+\bm{\nabla}\textbf{v}^T)$
is the rate of strain tensor. This choice preserves the viscoelastic and shear thinning character of blood but does not capture its thixotropic
behaviour~\citep{owens06}. This simplification makes it unnecessary
to solve an additional equation for the variation of $n$ in the flow
domain. Under this assumption, the relaxation time $\lambda$ is
\begin{equation}
\lambda  = \left(\frac{\lambda_\text{H}}{\eta_{\text{p},\infty}} \right) \eta_\text{p}
(\dot{\gamma})
\label{lambda}
\end{equation}
where, $\lambda_\text{H}$ is the relaxation time of individual blood cell
aggregates, {$\eta_{\text{p},\infty}$ is the infinite shear-rate viscosity}, and $\eta_\text{p}(\dot{\gamma})$ is their contribution to blood viscosity given by the Cross model,
\begin{equation}
\eta_\text{p}(\dot{\gamma}) = \eta_{\text{p},0}
\left(\frac{1+\theta_1 \dot{\gamma}^{m}}{1+\theta_2
\dot{\gamma}^{m}}\right) \label{etap}
\end{equation}
where $m$ is a power law index, and the ratio of parameters
$\theta_1$ and $\theta_2$ satisfies the expression,
$\theta_1/\theta_2 =
\eta_{\text{p},\infty}/\eta_{\text{p},0}$~\citep{fang06}. More details on the current 
implementation of Owens model are given in Ref.~\cite{debadiPhD}. The values
of all model parameters used here are reported in section~\ref{sec:MC}. 

In free boundary flow problems, one of the major difficulties lies
in the fact that the location of the free boundaries is unknown
\emph{a priori} and their solution is a part of the total solution.
Here we use a boundary fitted finite element based elliptic mesh
generation method \citep{chris92, desantos91, benjamin94} which
involves solving the following elliptic differential equation for the mapping:
\begin{equation}
\bm{\nabla}\cdot \tilde{\textbf{D}}\cdot\bm{\nabla}{\bm{\xi}} = 0
\label{ellipt}
\end{equation}
where $\bm{\xi}$ is a vector of positions in the computational
domain and the dyadic, $\tilde{\textbf{D}}$, is a function of
$\bm{\xi}$, analogous to a diffusion coefficient, which controls the
spacing of the coordinate lines \citep{benjamin94}.\\

As mentioned earlier, the formulation of the fluid-structure
interaction problem posed here follows the procedure introduced
previously by~\citet{carvalho97} in their examination of roll cover
deformation in roll coating flows. However, it turns out that the
weighted residual form of Eq.~(\ref{dcauchy-stress}) used in their
finite element formulation is incorrect. While the error does not
lead to significant discrepancies for small deformations, it is
serious for large deformations. The correct form of the
weighted-residual equation is presented in \citet{debadi2012}.

\subsection{\label{sec:boundary} Boundary conditions and discretization}

We prescribe the following boundary conditions:
\begin{enumerate}
\item No slip boundary conditions (\textbf{v} = \textbf{0}) are applied on the rigid 
walls.
\item Zero displacements are prescribed at the left side and right side of
the solid.
\item At the upstream boundary, a {fully developed} dimensional velocity profile is 
specified in the form, $v_y = 0$ and $v_x =U_0 {f(y/W)}$ where $U_0$ is the average inlet velocity. In dimensionless form this can be represented as $v_x
=\Gamma {f(y/W)}$ where, $v_x$ is now non-dimensional. Since, for all the Wi considered here, the upstream velocity profiles for the Oldroyd-B and FENE-P fluids do not differ significantly from that of a Newtonian fluid, a Newtonian velocity profile is used. However, for the Owens model fluid, which is strongly shear thinning in nature, a different procedure is used. Since the viscosity of the Owens model fluid obeys power law scaling with shear rate at relatively low values of shear rate in simple shear flow (see section~\ref{sec:Fluids}), we assume that the velocity profile at the upstream boundary is identical to the analytically computed velocity profile for a power-law fluid (with power law index $m$), flowing in a 2D channel. We find in our simulations (since the entrance length has been assumed to be sufficiently long), that the fully developed velocity profile of the Owens model fluid at the inlet to the collapsible channel is relatively unchanged from the analytical velocity profile imposed at the upstream boundary.
\item At the downstream boundary, the
fully developed flow boundary condition is imposed,
$\textbf{n}\cdot\bm{\nabla}\textbf{v = 0}$ where $\textbf{n}$ is the
unit normal to the outlet.
\item {The conformation tensor equation for all the viscoelastic models 
is hyperbolic in nature and therefore boundary conditions are needed only on inflow  boundaries.} At the upstream inflow, the conformation tensor does not change  along the streamlines because the flow is fully developed~\citep{pasquali02, xie04}. Thus,
\begin{equation}
\textbf{v}\cdot\bm{\nabla}\textbf{M} = \textbf{0}
\end{equation}
\item A force balance and a no-penetration condition are prescribed at the 
interface between the liquid and solid domain.
\begin{equation}
\textbf{n}.\textbf{T} = \textbf{n}. \bm{\sigma}\quad \text{and}
\quad {\textbf{v}_{solid} = \textbf{v}_{fluid}} \label{interface}
\end{equation}
where $\textbf{n}$ is the unit normal to the deformed solid surface.
\item A force balance is prescribed at the top surface.
\begin{equation}
\textbf{n}.\bm{\sigma} = -P_{e} \, \textbf{n}\label{dmembrane}
\end{equation}
where $P_{e}$ is the dimensionless external pressure.

\item The non-dimensional pressure of the fluid at the downstream boundary, 
$P_d$, is set equal to zero.
\end{enumerate}

Equations~(\ref{continuity}), (\ref{momentum}), (\ref{conformation})
and (\ref{ellipt}) are converted into a set of algebraic equations by the DEVSS-TG finite element method \citep{guenette95,pasquali02}, which introduces the traceless interpolated velocity gradient $\textbf{L}$~\citep{pasquali02}
\begin{equation}
\textbf{L}-\bm{\nabla}\textbf{v} + \frac{1}{\text{tr} \, \textbf{I}}
\, (\bm{\nabla}\cdot\textbf{v}) \, \textbf{I}
 =  0 \label{velgrad}
\end{equation}
In the transport equations the rate of strain tensor $\textbf{D}$ is calculated from the interpolated velocity gradient $\textbf{L}$.

The weighted residual form of Eqs.~(\ref{continuity}), (\ref{momentum}), (\ref{conformation}), (\ref{dcauchy-stress}), (\ref{ellipt}) and (\ref{velgrad}), yields a large set of coupled non-linear algebraic equations, which is solved by Newton's method with analytical Jacobian, frontal solver, and first order arclength continuation in parameters~\citep{pasquali02,zevallos05,bajaj07,debadi10}. {Note that while Eqs.~(\ref{continuity}), (\ref{momentum}),  (\ref{conformation}), (\ref{ellipt}) and (\ref{velgrad}) are implemented in the fluid domain, Eq.~(\ref{dcauchy-stress}) is solved in the solid domain.}

\section{\label{sec:RD} Results and discussions}
A thorough validation of the finite-element code, in the context of a Newtonian fluid, has been carried out by~\citet{debadi2012}, who have compared results obtained with the present formulation with several earlier results obtained in different contexts.

It is appropriate to briefly discuss the fluid models used in the present work 
before presenting the results of our simulations, since the differences in behaviour amongst them is essentially due to differences in their rheology.

\subsection{\label{sec:Fluids} Fluid models and choice of parameters}
Each of the three fluids examined here has distinct qualitative features: (i) The 
Oldroyd-B fluid is {viscoelastic}, but does not shear thin. Furthermore, 
its uniaxial extensional viscosity is unbounded. (ii) The FENE-P fluid is 
{viscoelastic}, shear thins, and has a bounded  uniaxial extensional 
viscosity. (iii) The Owens model fluid is {viscoelastic} and shear thins, 
but has an unbounded  uniaxial extensional viscosity.  Additionally, a notable 
feature of the Owens model, which belongs to the class of White-Metzner fluids, 
is that the dependence of viscosity $\eta_\text{p}$ on shear rate $\dot \gamma$ 
can be prescribed arbitrarily through the choice of parameters in the Cross model 
(see Eq.~(\ref{etap})). In particular, the viscosity can be prescribed independently 
of the relaxation time. In contrast, for the FENE-P model, the dependence of $\eta_\text{p}$ on the shear rate $\dot \gamma$ is completely determined by the choice of the parameters, $\eta_{\text{p},0}$, $\lambda_0$, and the finite extensibility parameter $b_\mathbf{M}$. Unlike in the case of the Owens model, no further control can be exerted on the shape of the viscosity function.

\begin{figure}[t]
\begin{center}
\includegraphics[width=0.90\textwidth]{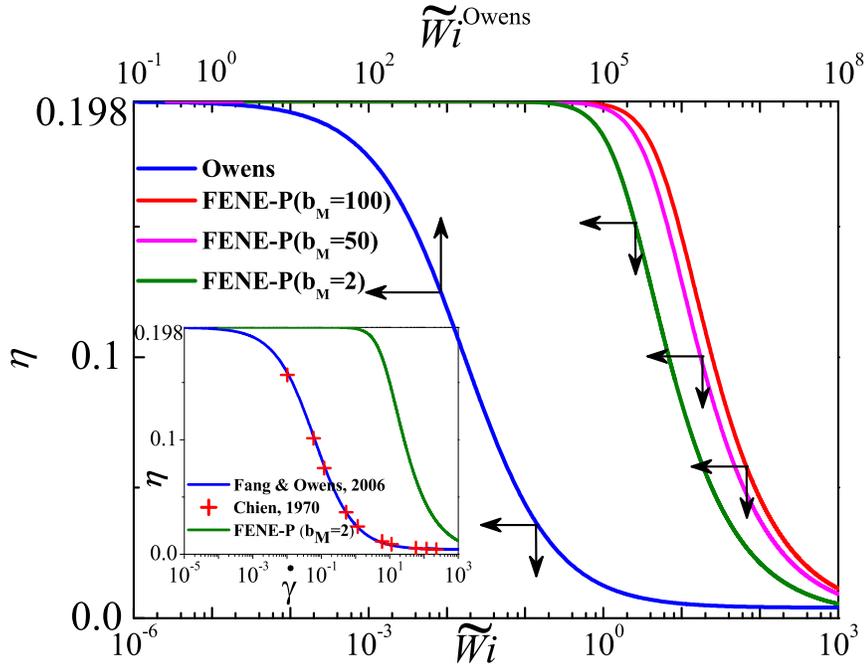}
\vskip-30pt
\caption{\small  \label{figflurheo} The contribution of the microstructure to the total viscosity, $\eta$, for the Owens model and FENE-P fluids in steady shear flow as a function of Weissenberg number $\widetilde{Wi}$. The inset shows the shear rate dependence of viscosity in the Owens model, fitted  to the experimental results for blood reported by~\citet{chien70}, and the predictions of the FENE-P model for $b_\mathbf{M}=2$ and $\lambda_0 = 0.263$.}
\end{center}
\end{figure}
The difference in the prediction of viscosity by the FENE-P and
Owens models, as a function of the Weissenberg number
$\widetilde{Wi} = \lambda_0 \dot{\gamma}$, in steady shear flow, is
displayed in figure~\ref{figflurheo}. For the Owens model, we set
$\eta_{\text{p},0} = 0.197$ Pa s, $\eta_{\text{p},\infty} = 0.003$
Pa s, $\eta_\text{s}=0.001$ Pa s, $\theta_2 = 8$, and $m = 0.75$.
These parameter values were chosen by~\citet{fang06} to fit
experimental data for the steady-state viscosity of blood, as
reported by~\citet{chien70}. The fitted curve and experimental data
are reproduced in the inset to figure~\ref{figflurheo}. Additionally,
\citet{fang06} suggest $\lambda_\text{H}= 0.004$ s, which leads to
$\lambda_0 = 0.263$. In order to compare the two fluid models, we
assume that the FENE-P model has the same value of
$\eta_{\text{p},0}$ and $\lambda_0$, and that $\eta_\text{s}$ is the
same. This assumption is based on the expectation that any choice of
viscoelastic model would have to be compatible with known
experimental information on the rheology of the fluid, which would,
at the least, include a knowledge of the zero shear rate viscosity
and the relaxation time. Note that for the FENE-P model, while there
is no necessity to prescribe $\lambda_0$ when the shear rate
dependence of viscosity is expressed in terms of $Wi$, it is
necessary when represented in terms of  $\dot{\gamma}$. As is well
known, the FENE-P model predicts increasing shear thinning with
decreasing values of the finite extensibility parameter
$b_\mathbf{M}$. The entire family of curves for the FENE-P model
shown in figure~\ref{figflurheo}, with values of $b_\mathbf{M}$ ranging
from 100 to 2, does not shear thin as rapidly as the Owens model. In
particular, it is clear from the inset that for the parameters
recommended by \citet{fang06}, the FENE-P model is unable to capture
the rapidity with which blood shear thins, even for
$b_\mathbf{M}=2$. In all cases, in line with expectation, shear
thinning first occurs for the FENE-P fluid when $\widetilde{Wi}
\sim \textit{O}(1)$. For viscoelastic fluids, the onset of shear
thinning at $\widetilde{Wi} \sim \textit{O}(1)$ implies that for shear rates $\dot \gamma > 1/\lambda_{0}$, the fluid no longer responds in a Newtonian manner, and that the longest time scale for micro-structural relaxation is comparable to $\lambda_{0}$. Since the Owens model shear thins at $\widetilde{Wi} \ll 1$, this suggests that the characteristic time scale for micro-structural rearrangement is much larger than $\lambda_{0}$. As mentioned earlier, $\lambda_{0}$ corresponds to the relaxation time for an aggregate of blood cells, which according to~\citet{owens06}, are typically of a size that represents the greatest proportion of erythrocytes. By defining a Weissenberg number
$\widetilde{Wi}^{\text{Owens}}$ for which the Owens model fluid
shear thins when it is of $\textit{O}(1)$ (see the upper horizontal
axis of figure~\ref{figflurheo}), we can estimate that the appropriate
relaxation time is of order $10^{4}$, which must correspond to much
larger structures than a typical blood cell aggregate. We do not
explore this aspect further here, rather, for the purposes of the
present paper, we assume that the FENE-P and Owens models are
distinct constitutive models, which have the same zero shear rate
material properties, but shear thin significantly differently. As
will be discussed in greater detail in the sections below, the
difference between the models leads to significant differences in
their behaviour.

For all the computational results reported here, we set
$\eta_{\text{p},0}$, $\eta_{\text{p},\infty}$ $\eta_\text{s}$,
$\theta_2 $, and $m$ at the values recommended by  \citet{fang06}.
However, we vary $\lambda_0$ (by varying $\lambda_\text{H}$) in
order to control the inlet Weissenberg number. For the FENE-{P}
fluid, we set $b_\mathbf{M}=100$, which is a value commonly used in
simulations. As we are interested in small blood vessels, we choose
the width $W$ of the channel to be 100 $\mu$m and $U_0 = 0.01$ m/s,
inline with the data reported in~\citet{Robertson2008a}. The value
of $Re$ in small blood vessels is well below 1. We have not seen any
significant difference in the profile shape of the collapsible wall
for values of $Re$ in the range of $0$-$1$, so we set $Re = 0$ by
setting $\rho = 0$.

\citet{deng98} and \citet{intengan1999} have reported the values of Young's modulus ($E$) for the human artery to be in the range $200$-$4000$ kPa. Zhang and co-workers~\citep{zhang04,zhang05,zhang06a,zhang06b} have reported
the Young's modulus of the porcine artery to be in the range
$110$-$140$ kPa, while using two different values for the external
pressure ($8$ kPa and $9.3$ kPa) in their experimental measurements
of the Young's modulus. In order to adequately represent the
microcirculation, we choose a wide range of values for the external
pressure $p_e$ from $1.2$ to $16$ kPa and $G$ (which is equal to $E/3$) in the 
range $30$ to $400$ kPa. Since $P_e=p_e/G$, we keep $P_e$ fixed at a constant value of $0.04$ even though both $p_e$ and $G$ are varied. On the
other hand,  we vary $\Gamma$ in the range $4.95\times10^{-5}$ to
$6.6\times10^{-4}$. For most of the simulations reported here, a fixed value of $0.4W$ is chosen for the thickness of the solid wall ($t$), as the artery wall thickness to vessel diameter ratio is  typically very high in small blood vessels~\citep{kalita08}. However, in section~\ref{sec:thick}, a few results of simulations with varying wall thickness are discussed to elucidate the effect of wall thickness.

\subsection{\label{sec:MC} Mesh convergence and the high Weissenberg 
number problem}

\begin{figure}
\begin{center}
\includegraphics[width=0.90\textwidth]{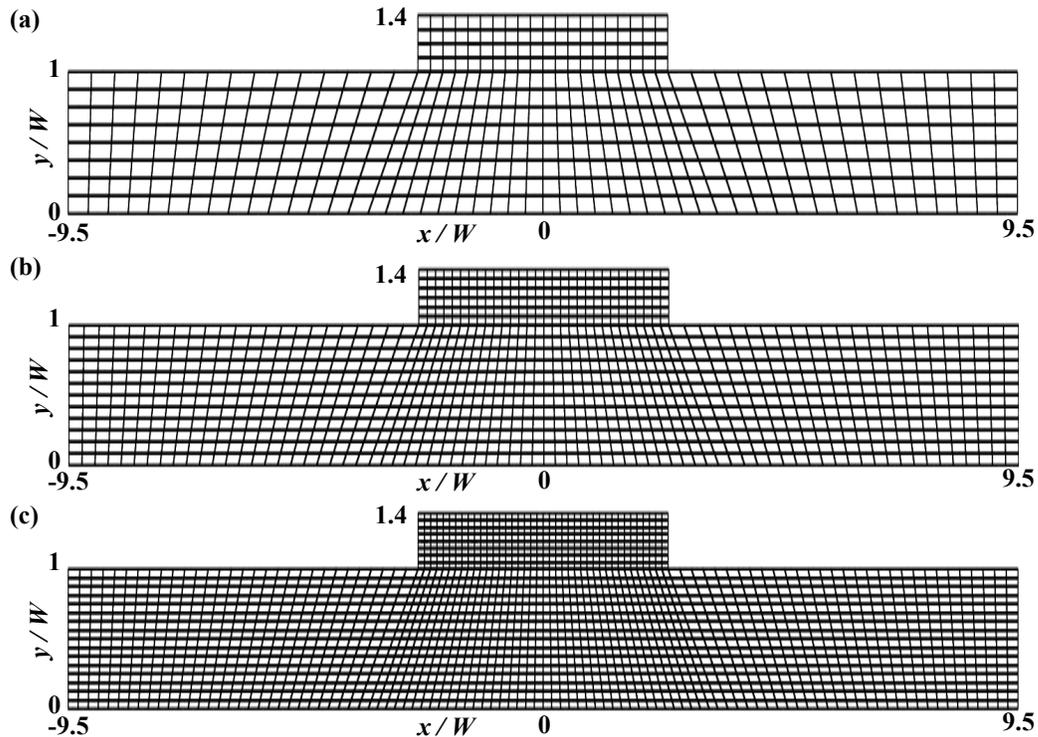}
\caption{\small Meshes considered in the current study. (a) M1, (b) M2, and (c) M3, for $t = 0.4W$.}
\label{figmesh}
\end{center}
\end{figure}

\citet{debadi10} have established that the flow in a collapsible
channel with a zero-thickness membrane suffers from the high
Weissenberg number problem and have shown that there is a limiting
Weissenberg number for each of the fluid models beyond which
computations fail. Furthermore, this limiting $Wi$ value has been shown
to increase with mesh refinement. Here we study mesh convergence
over a range of parameters for the current geometry, using three
different meshes M1-M3 for $t = 0.4W$, as illustrated in
figure~\ref{figmesh}, with the mesh details given in
Table~\ref{table1}.

\begin{table}[t]
\begin{center}
\begin{tabular}{cccc}
%\hline
 Mesh &Number of elements & Number of nodes &
Degrees of freedom \\
& & & \\
M1 &400 &1705 &10972\\
M2 &900 &3757 &24072\\
M3 &1600 &6609 &42252\\
% \hline
 \end{tabular}
   \caption{\small Meshes considered  in the current study.}\label{table1}
  \end{center}
%  \vskip30pt
  \end{table}

\begin{figure}
\begin{center}
\begin{tabular}{cc}
\resizebox{7.5cm}{!} {\includegraphics*[angle=0]{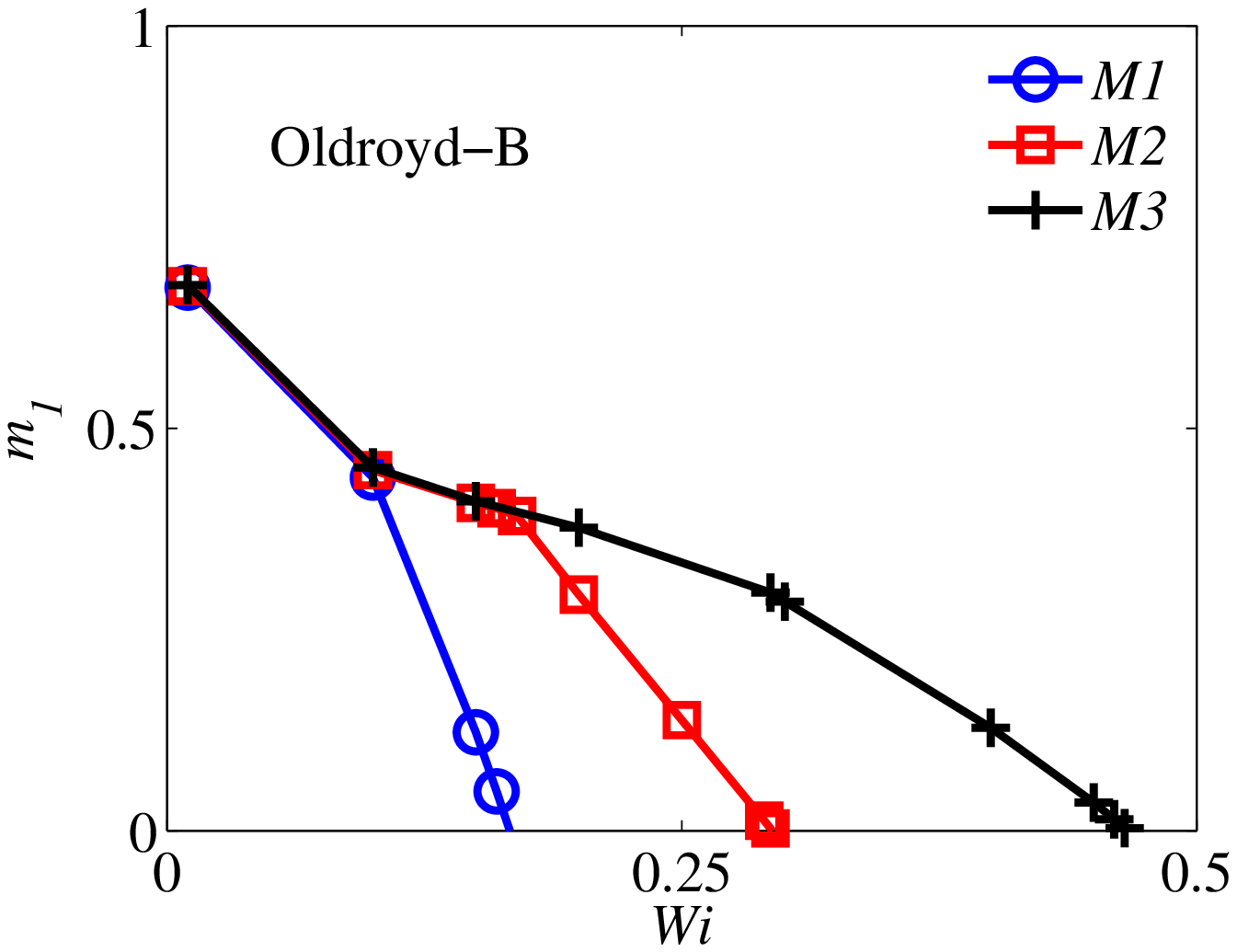}} &
\resizebox{7.5cm}{!} {\includegraphics*[angle=0]{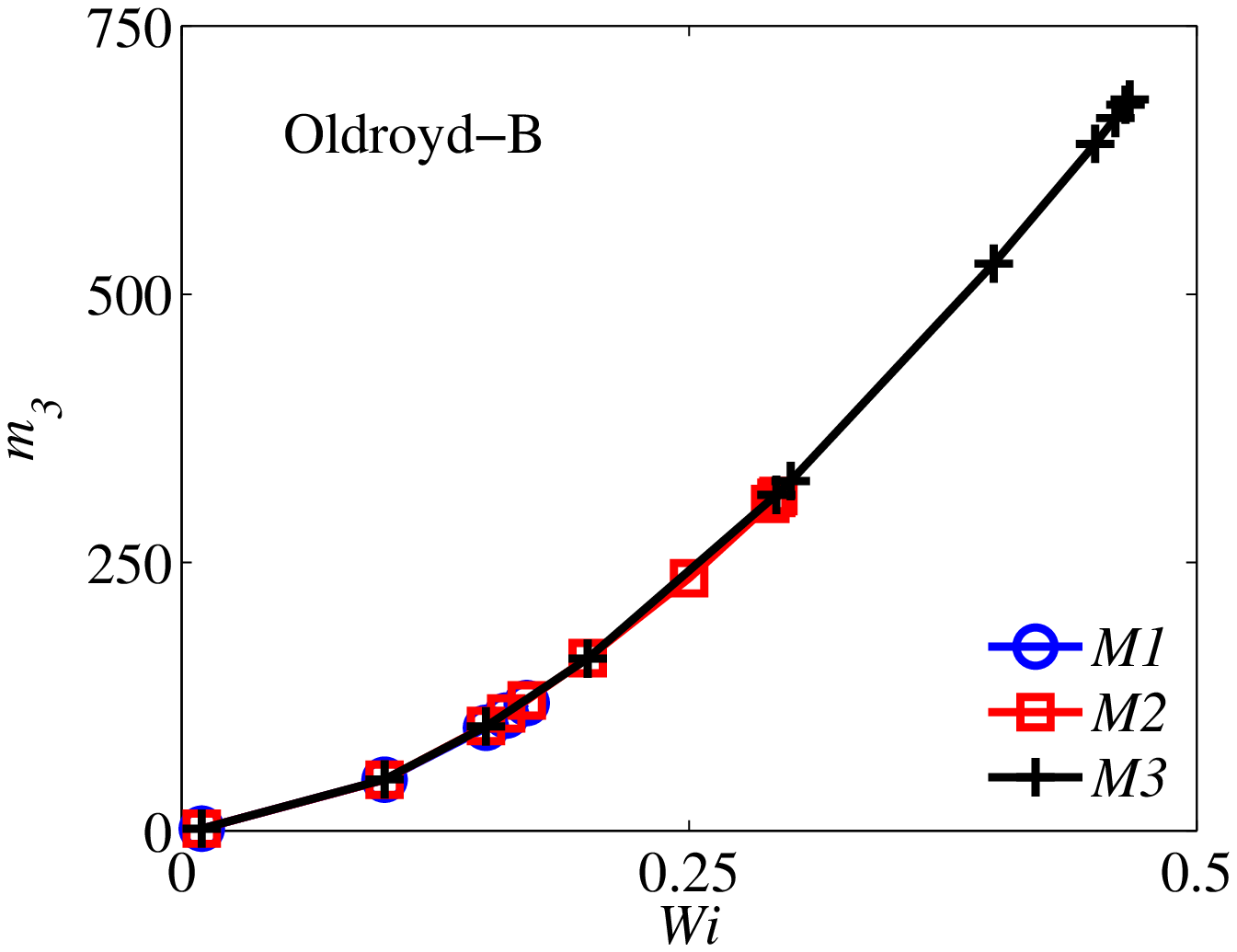}}\\
(a) & (d)  \\
\resizebox{7.5cm}{!} {\includegraphics*[angle=0]{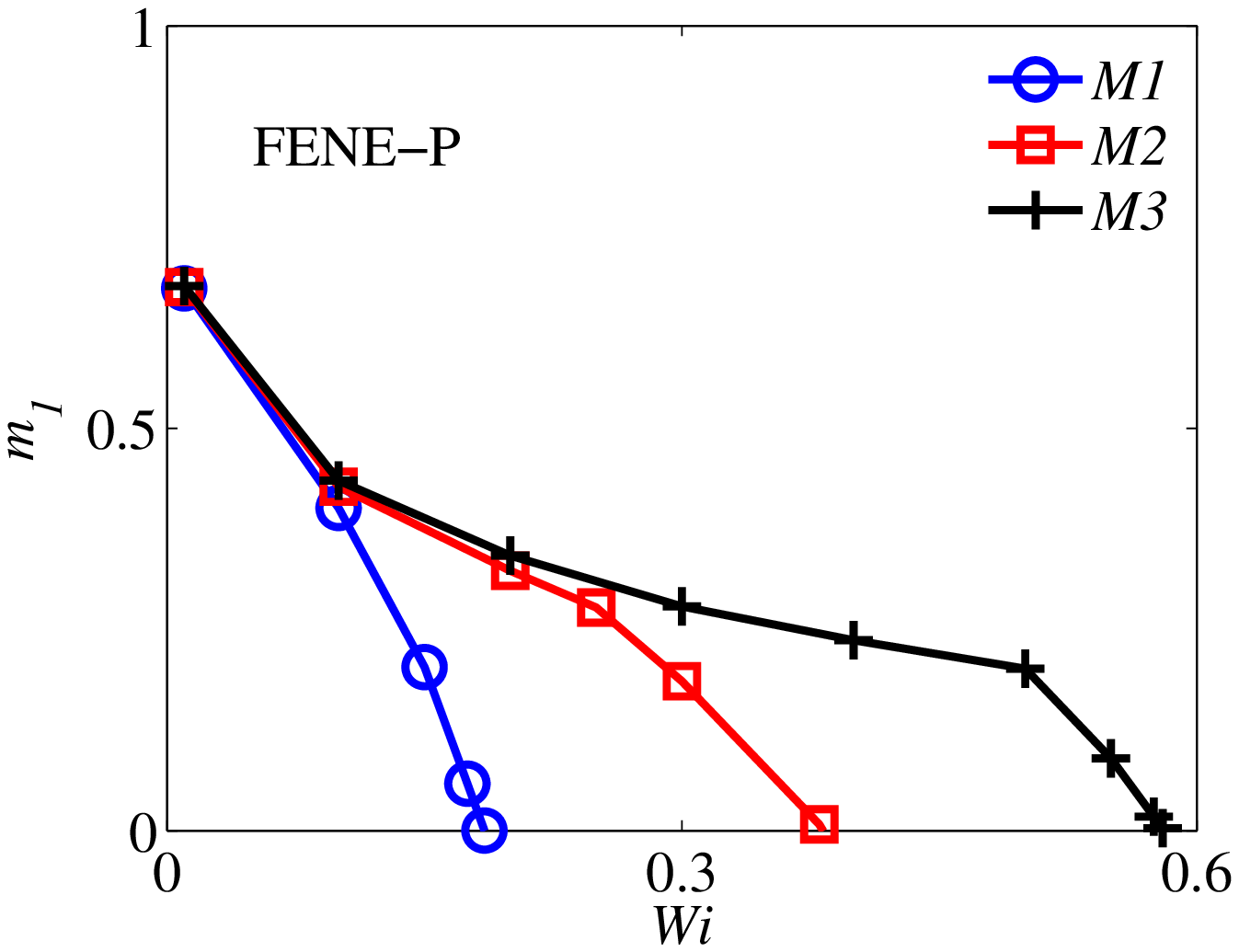}} &
\resizebox{7.5cm}{!} {\includegraphics*[angle=0]{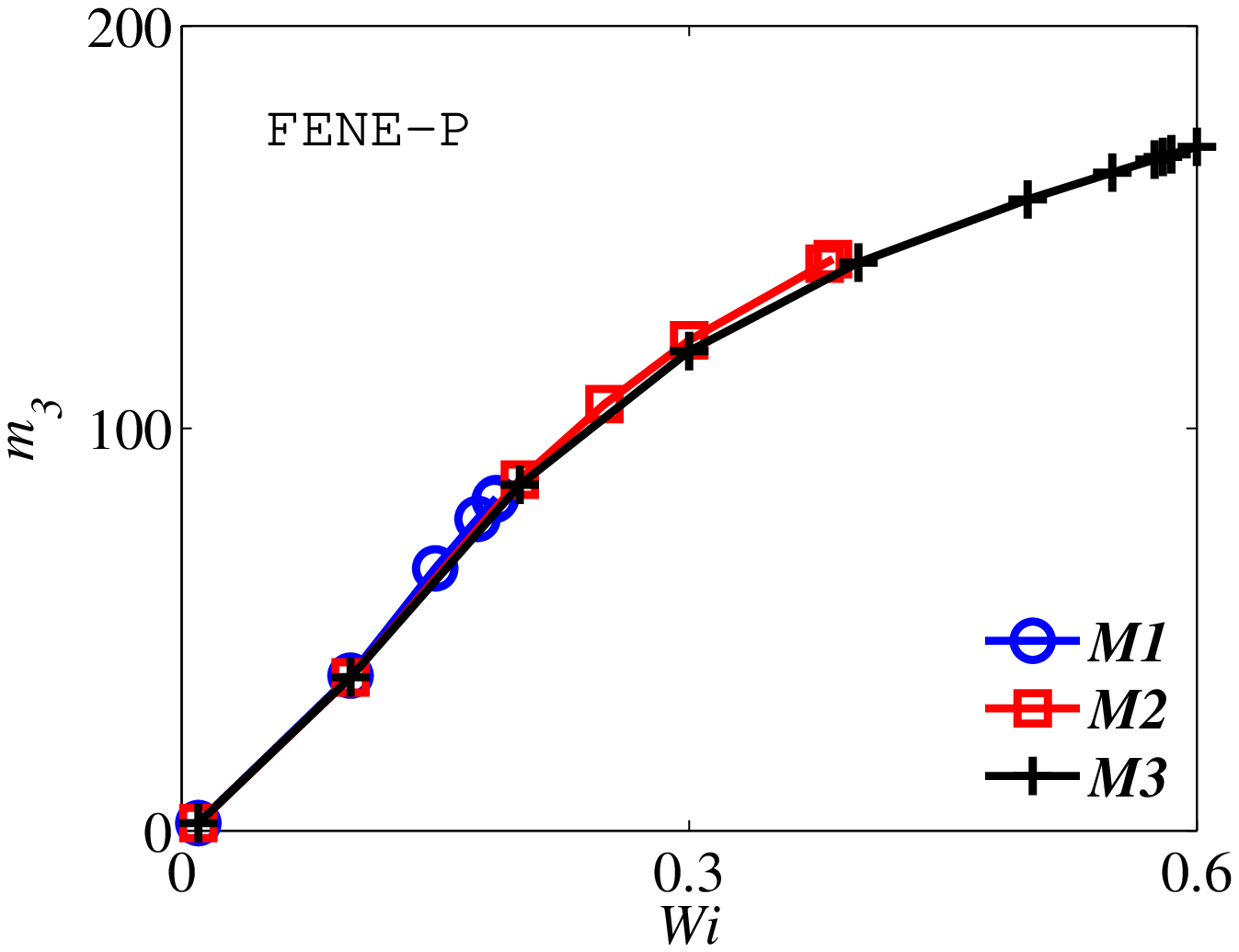}}\\
(b) & (e)  \\
\resizebox{7.5cm}{!} {\includegraphics*[angle=0]{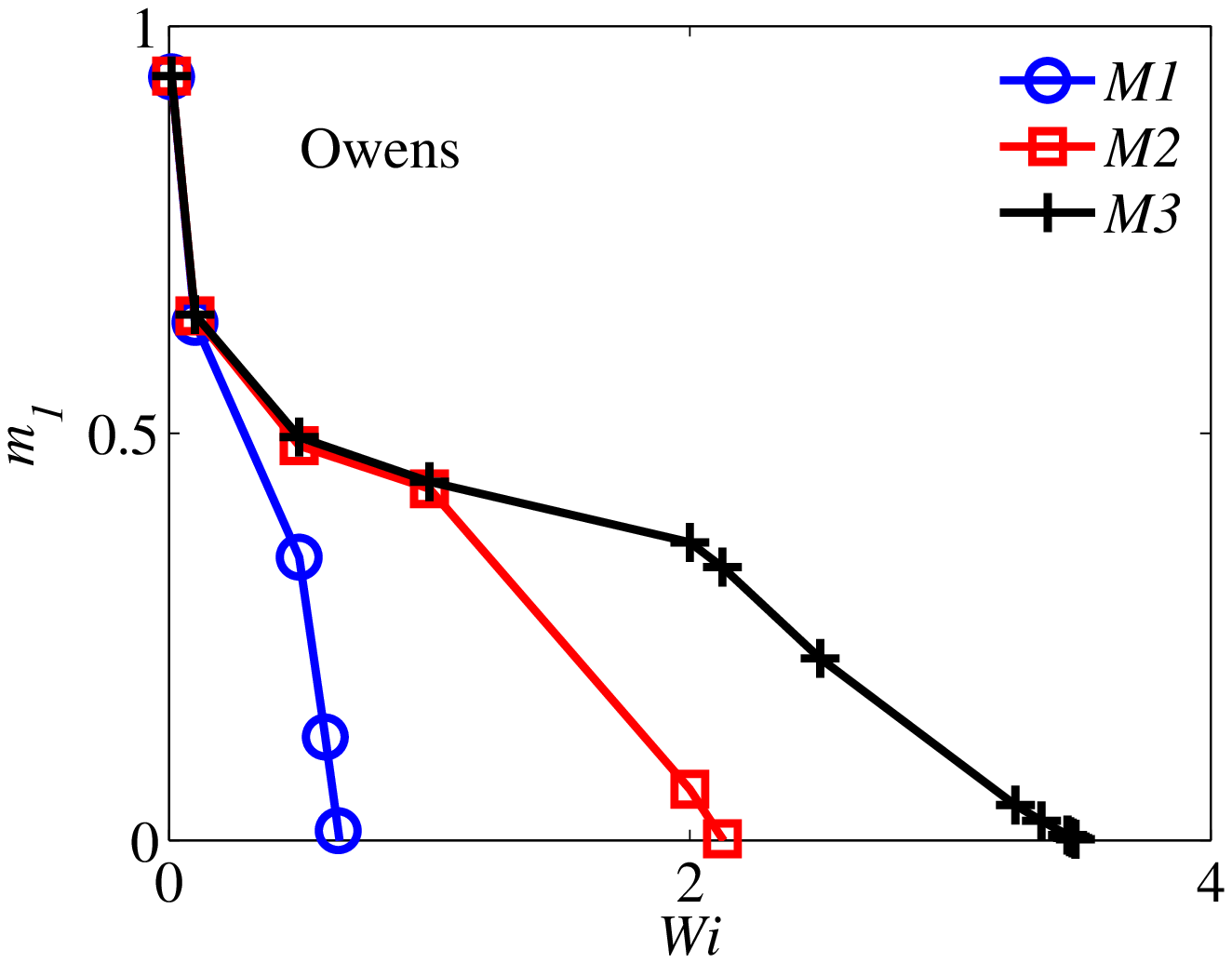}} &
\resizebox{7.5cm}{!} {\includegraphics*[angle=0]{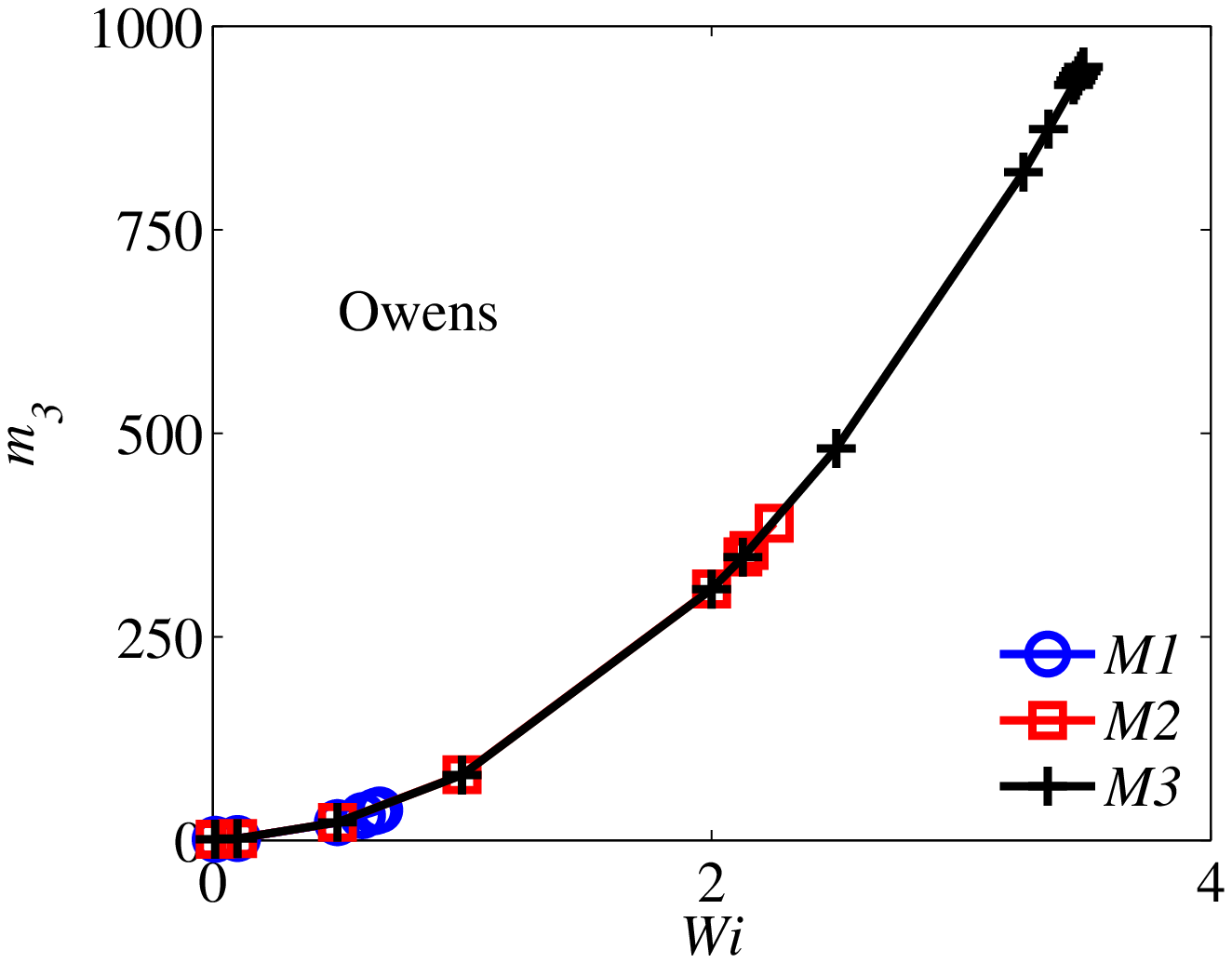}}\\
(c) & (f)  \\
\end{tabular}
\end{center}
%\begin{spacing}{1.5}
\caption{\small \label{figeigen}Minimum value of the smallest eigenvalue
($m_1$) and maximum value of the largest eigenvalue ($m_3$) in the
entire flow domain, for the Oldroyd-B ((a) and (d)), FENE-P ((b) and
(e)), and Owens model ((c) and (f)), as a function of $Wi$ at
$\Gamma$ = $4.95\times10^{-5}$ and $P_e$ = $0.04$ for $t = 0.4W$.}
%\end{spacing}
\end{figure}

In viscoelastic flow, mesh convergence is generally studied by
examining the values of the invariants of the conformation dyadic,
$\textbf{M}$. The eigenvalues ${m_i}$ of the conformation dyadic
represent the square stretch ratios along the principal directions
of stretching $\bm{m_i}$ for an ensemble of molecules~\citep{pasquali02,pasquali04}. It has been well established that the breakdown of viscoelastic computations is typically due to the smallest eigenvalue becoming negative in some regions of the flow domain~\citep{zanden88,singh93,pasquali02,
bajaj07, debadi10}.

Figure~\ref{figeigen} shows the maximum eigenvalue $m_3$ and minimum
eigenvalue $m_1$ of the conformation tensor as a function of $Wi$
for the Oldroyd-B, FENE-P and Owens fluids at $\Gamma$ =
$4.95\times10^{-5}$ and $P_e$ = $0.04$ for $t = 0.4W$.
Figures~\ref{figeigen}~(a)-(c) clearly exhibit the breakdown of
viscoelastic computations at a particular value of Weissenberg
number on each mesh, since the minimum value of $m_1$ becomes
negative. This limiting $Wi$ increases with increase in mesh
refinement.

An increase in $Wi$ leads to a higher maximum $m_3$ and lower
minimum $m_1$ across the flow domain. While the breakdown value of
$Wi$ at each mesh can be anticipated from the sudden change of the
slope of the curves in the minimum $m_1$ plots
(figure~\ref{figeigen}~(a)-(c)), the curves of the maximum $m_3$ on
various meshes overlap with each other (figures~\ref{figeigen}~(d)-(f)).
The limiting value of $Wi$ on the M2 mesh for the Oldroyd-B, FENE-P
and Owens fluids is respectively $0.29$, $0.38$ and $2.13$, while
the mesh converged value of $Wi$ for these models is $0.17$, $0.20$
and $1.0$ respectively. In all our analysis, we have ensured that
mesh converged values of $Wi$ are used for any particular mesh.
\begin{figure}
\centering \subfigure[] {
\label{figMxx:sub:a}
\includegraphics[height=6.5cm]{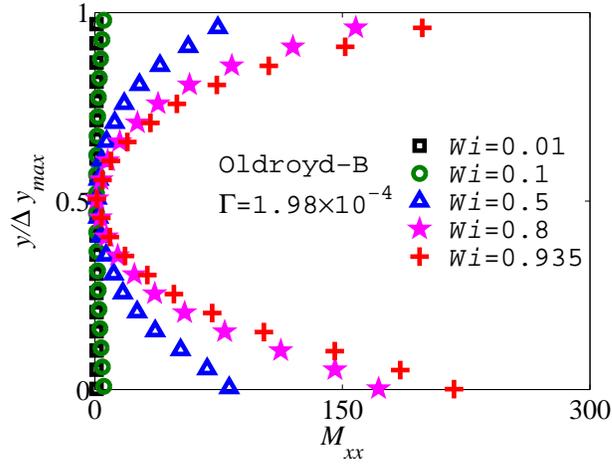}
}
 \subfigure[] {
\label{figMxx:sub:b}
\includegraphics[height=6.5cm]{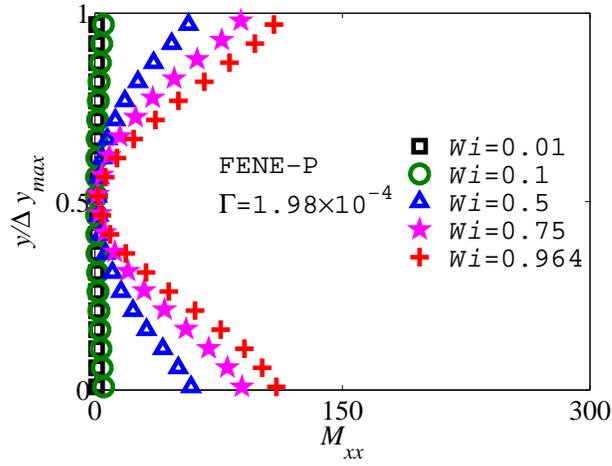}
}
\subfigure[] {
\label{figMxx:sub:c}
\includegraphics[height=6.5cm]{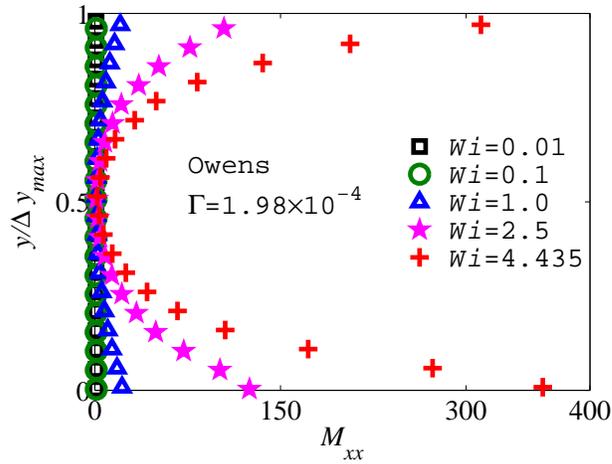}
}
%\begin{spacing}{1.5}
\caption{\small  \label{figMxx}   Profile of $M_{xx}$ across the
narrowest channel gap for the Oldroyd-B, FENE-P and Owens models,
for a range of Weissenberg numbers, at $\Gamma = 1.98\times10^{-4}$, {$P_e = 0.04$ and $t = 0.4W$. For the FENE-P fluid, we set $b_\mathbf{M}=100$.} The distance from the bottom channel is scaled by the narrowest gap width $\Delta y_{max}$ (see figure~\ref{figmaxdef:sub:a} for a definition) of the particular model.}
%\end{spacing}
\end{figure}

\begin{figure}
\begin{center}
\includegraphics[width=0.80\textwidth]{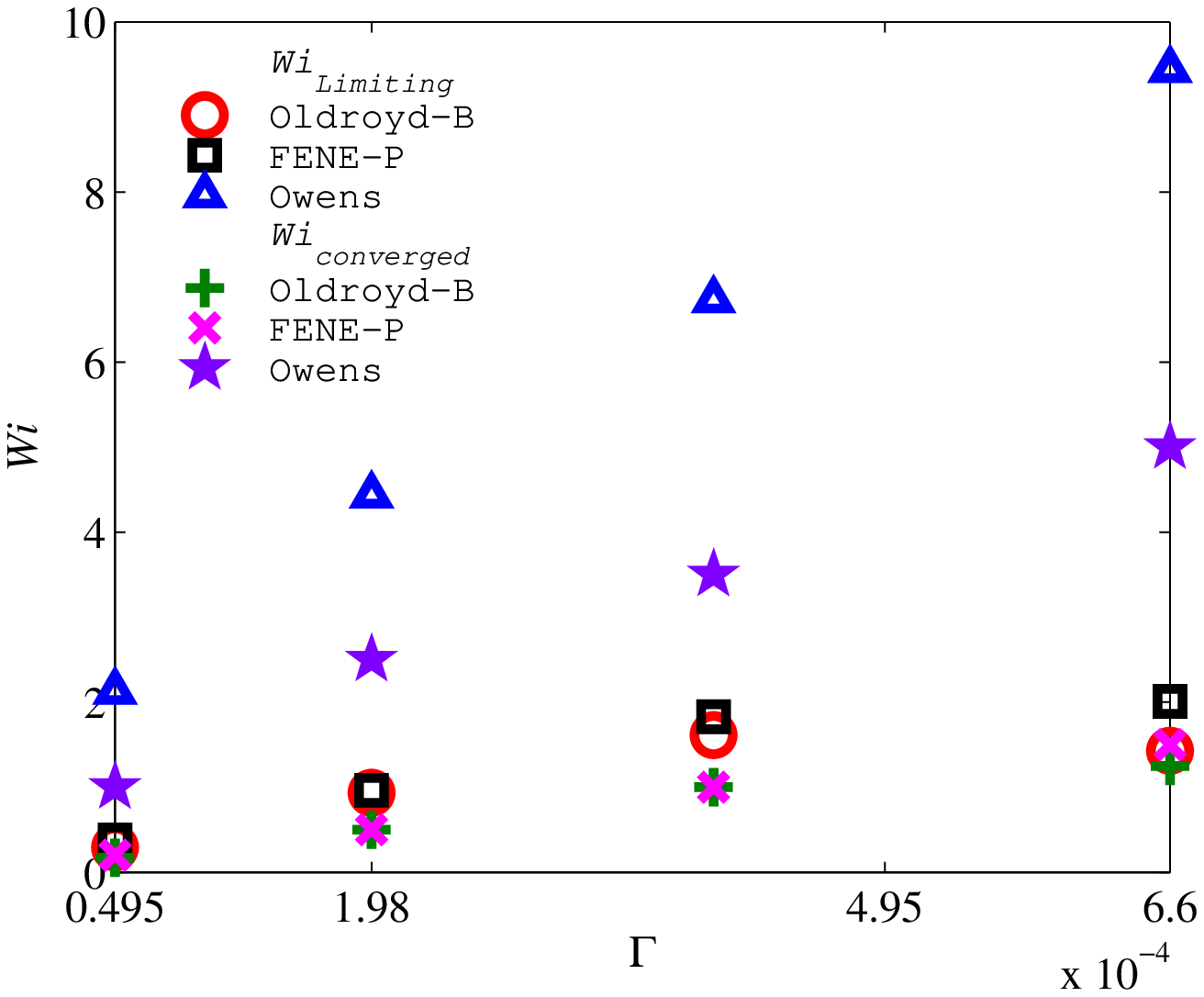}
\caption{\label{figmaxWi} Maximum mesh converged value of $Wi$ and the limiting $Wi$, for the three fluid models, for computations carried out with the M2 mesh, at $P_e = 0.04$ and $t = 0.4W$, as a function of $\Gamma$. }
\end{center}
\end{figure}

While figure~\ref{figeigen} displays the maximum and minimum
eigenvalues in the entire flow field, figure~\ref{figMxx} displays the mean streamwise molecular stretch $M_{xx}$  across the channel at the point where the gap between the flexible and rigid walls is a minimum, for a fixed value of $\Gamma$, and increasing values of $Wi$. With increasing Weissenberg number, $M_{xx}$ grows nearly symmetrically from a relatively low value in the middle of the gap, to a significantly larger value near the bottom (rigid) and top (flexible) walls. Note that in the Oldroyd-B and Owens models, $M_{xx}$ is unbounded due to the infinite extensibility of the Hookean spring in the Hookean dumbbell
model which underlies these fluid models. Conversely, the existence of
a upper bound to the mean stretchability of the spring in the FENE-P
model restricts the maximum value for $M_{xx}$, which for
$b_{\textbf{M}}= 100$ is 300. The profiles of $M_{xx}$
for the different fluids in figure~\ref{figMxx} clearly reflect this micro-mechanical
aspect of the models, and confirm that as in other benchmark problems for
non-Newtonian flow, numerical computations in a 2-D collapsible channel
also fail due the development of large stresses and stress gradients
in certain regions of the flow field, which are related to large changes
in the conformations of the molecules.

In their earlier study with a zero-thickness membrane, \citet{debadi10}
have shown that the extent of collapse of the membrane also has a significant effect on the limiting Weissenberg number. As the gap in the channel becomes narrower with decreasing tension in the membrane, the fluid is 'squeezed' leading
to a greater deformation of the molecules, with a concomitant numerical breakdown at smaller values of $Wi$. One of the parameters that controls the deformability of the wall in the current work is $\Gamma$ (the other being wall thickness $t$, whose influence on the limiting $Wi$ is discussed subsequently in section~\ref{sec:thick}). Figure~\ref{figmaxWi} displays the limiting and the mesh converged values of the Weissenberg number for the M2 mesh, as a function of $\Gamma$. The limiting Weissenberg number follows the trend Owens $>$ FENE-P $>$ Oldroyd-B. For the Oldroyd-B and FENE-P models, the limiting and mesh converged Weissenberg numbers appear to increase with an increase in $\Gamma$ for values of $\Gamma \lesssim 4.0 \times 10^{-4}$, before levelling off at higher values of $\Gamma$. On the other hand, for the Owens model fluid, $Wi_{Limiting}$ and $Wi_{Converged}$ increase monotonically with $\Gamma$. The Owens model fluid also exhibits the biggest difference between the converged and limiting values of $Wi$. As will be apparent when we discuss the shape of the fluid solid interface in figure~\ref{figintshape}, an increase in $\Gamma$ leads to an increase in the magnitude of the narrowest channel gap, and consequently an increase in the limiting value of $Wi$.

\subsection{\label{sec:VF} Interface shape and velocity contours}

\begin{figure}
\begin{center}
\begin{tabular}{cc}
\resizebox{8.0cm}{!} {\includegraphics*[width=8.0cm]{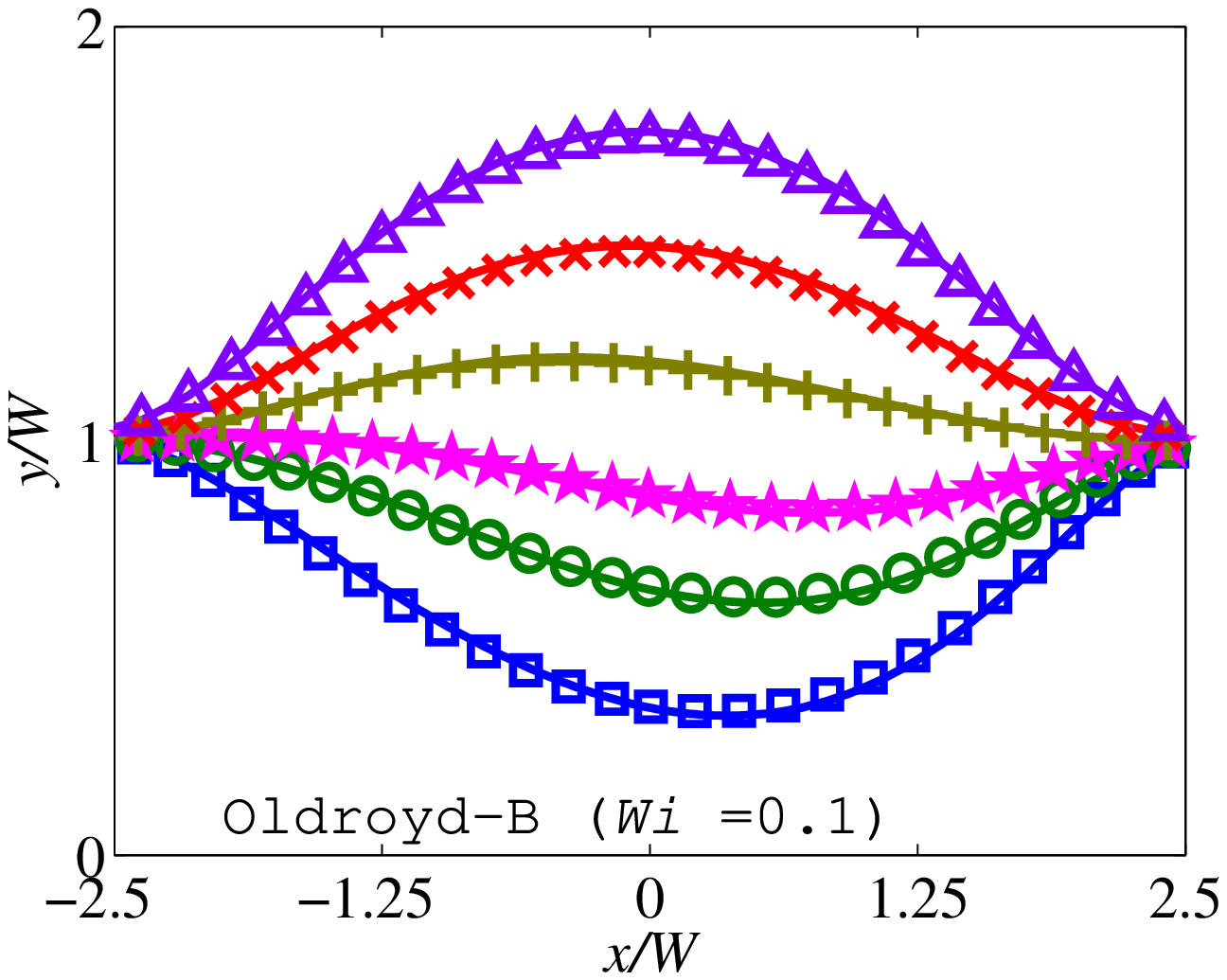}}
&
\resizebox{8.0cm}{!} {\includegraphics*[width=8.0cm]{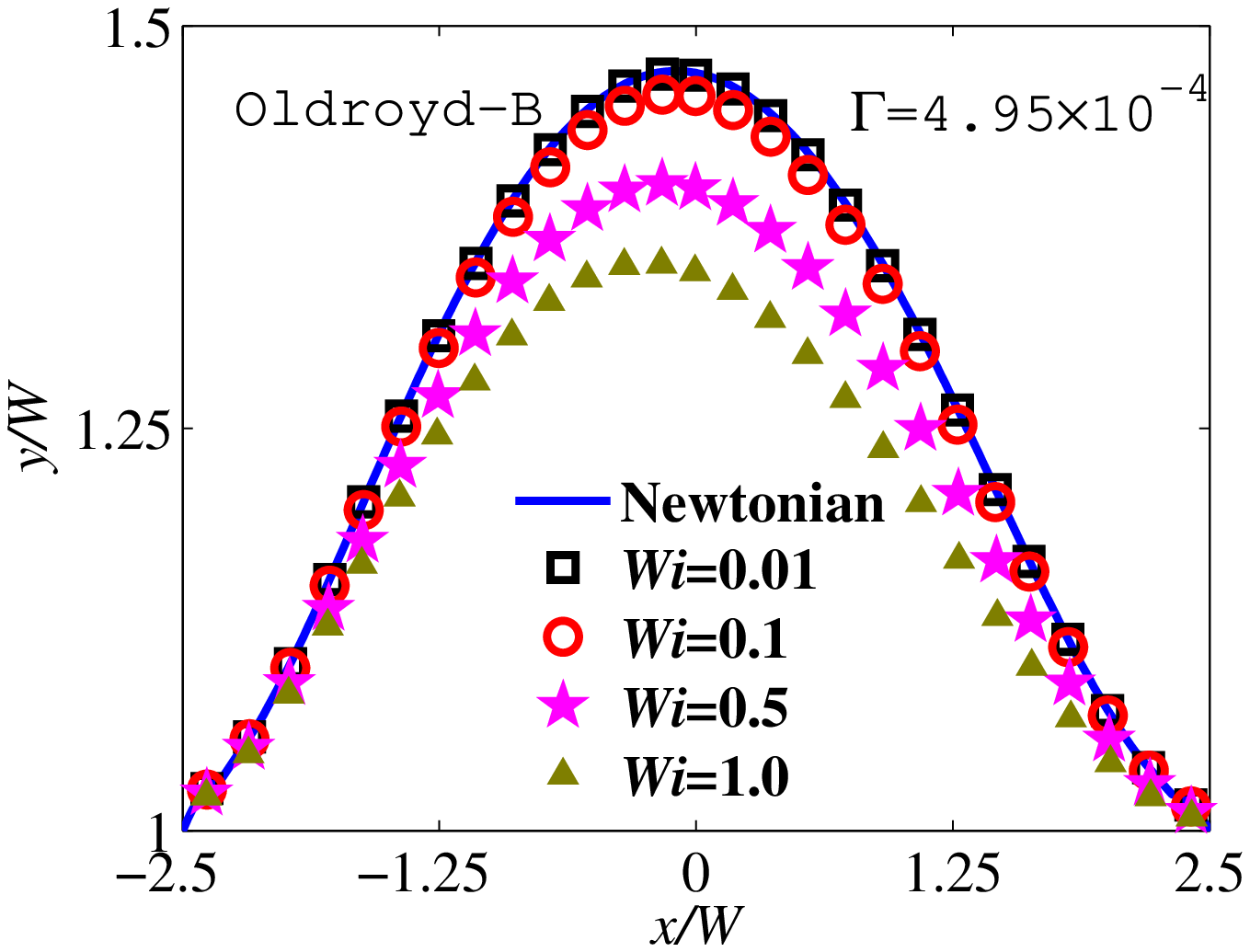}}\\
(a) & (d)  \\
\resizebox{8.0cm}{!} {\includegraphics*[width=8.0cm]{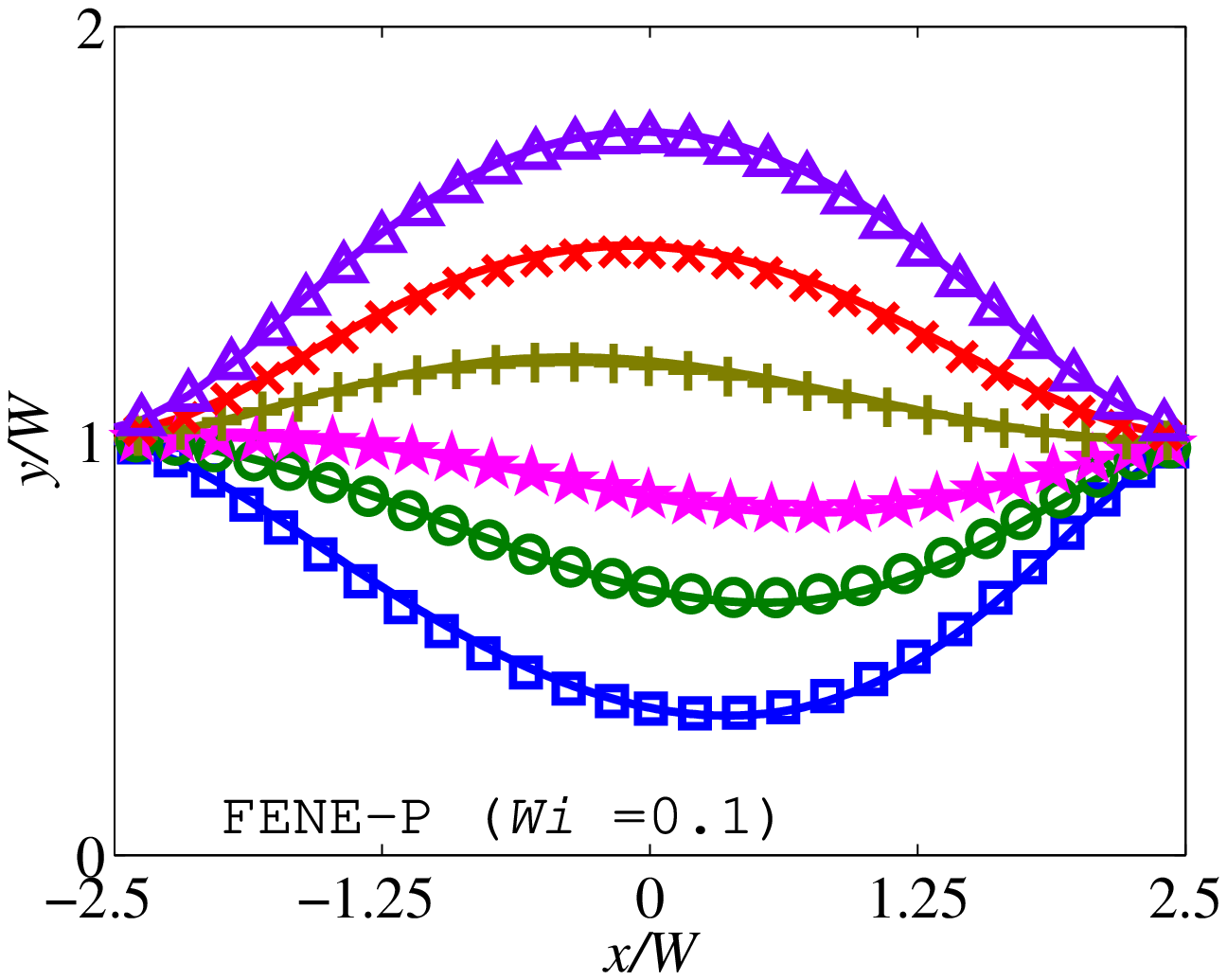}}
&
\resizebox{8.0cm}{!} {\includegraphics*[width=8.0cm]{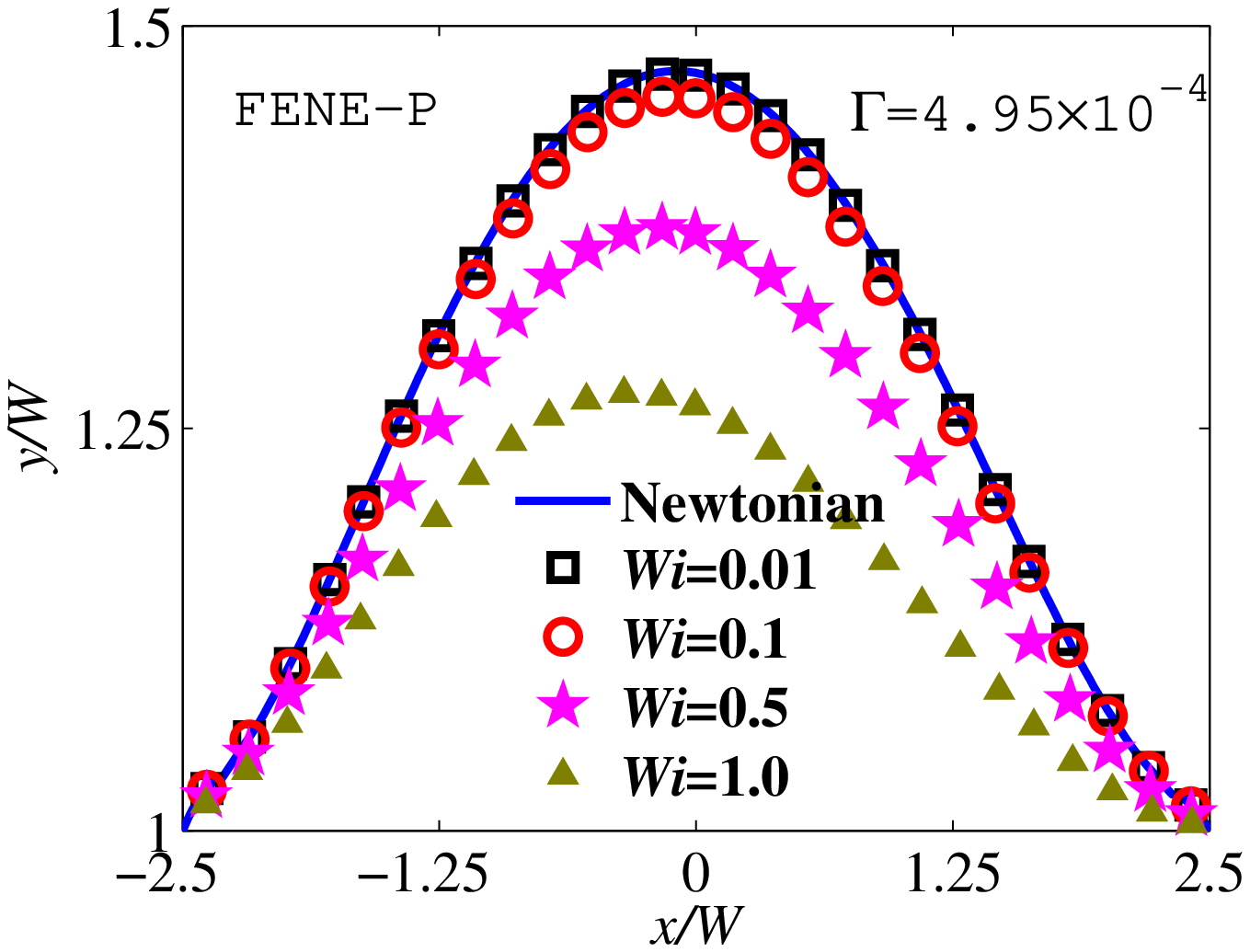}}\\
(b) & (e)  \\
\resizebox{8.0cm}{!} {\includegraphics*[width=8.0cm]{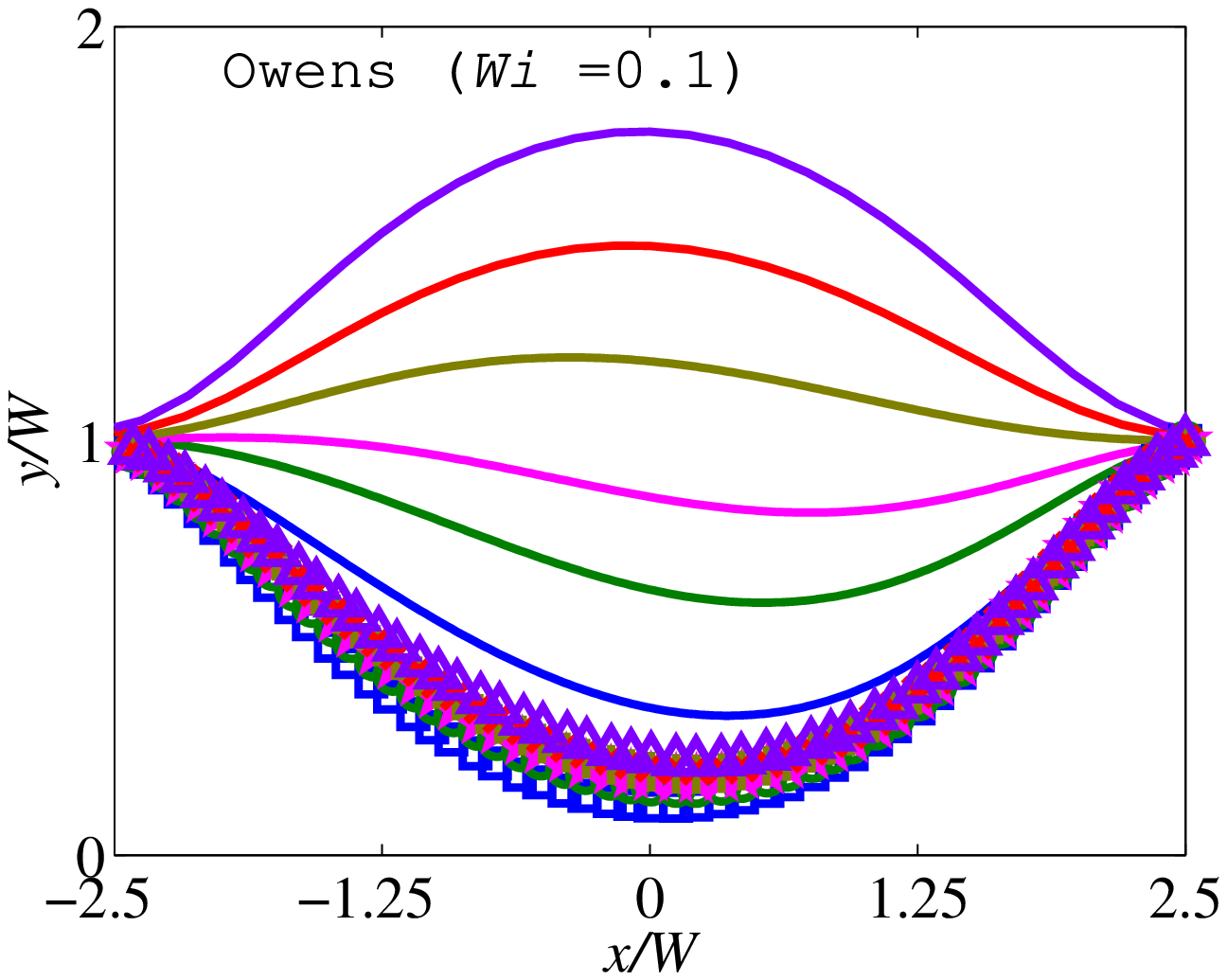}}
&
\resizebox{8.0cm}{!} {\includegraphics*[width=8.0cm]{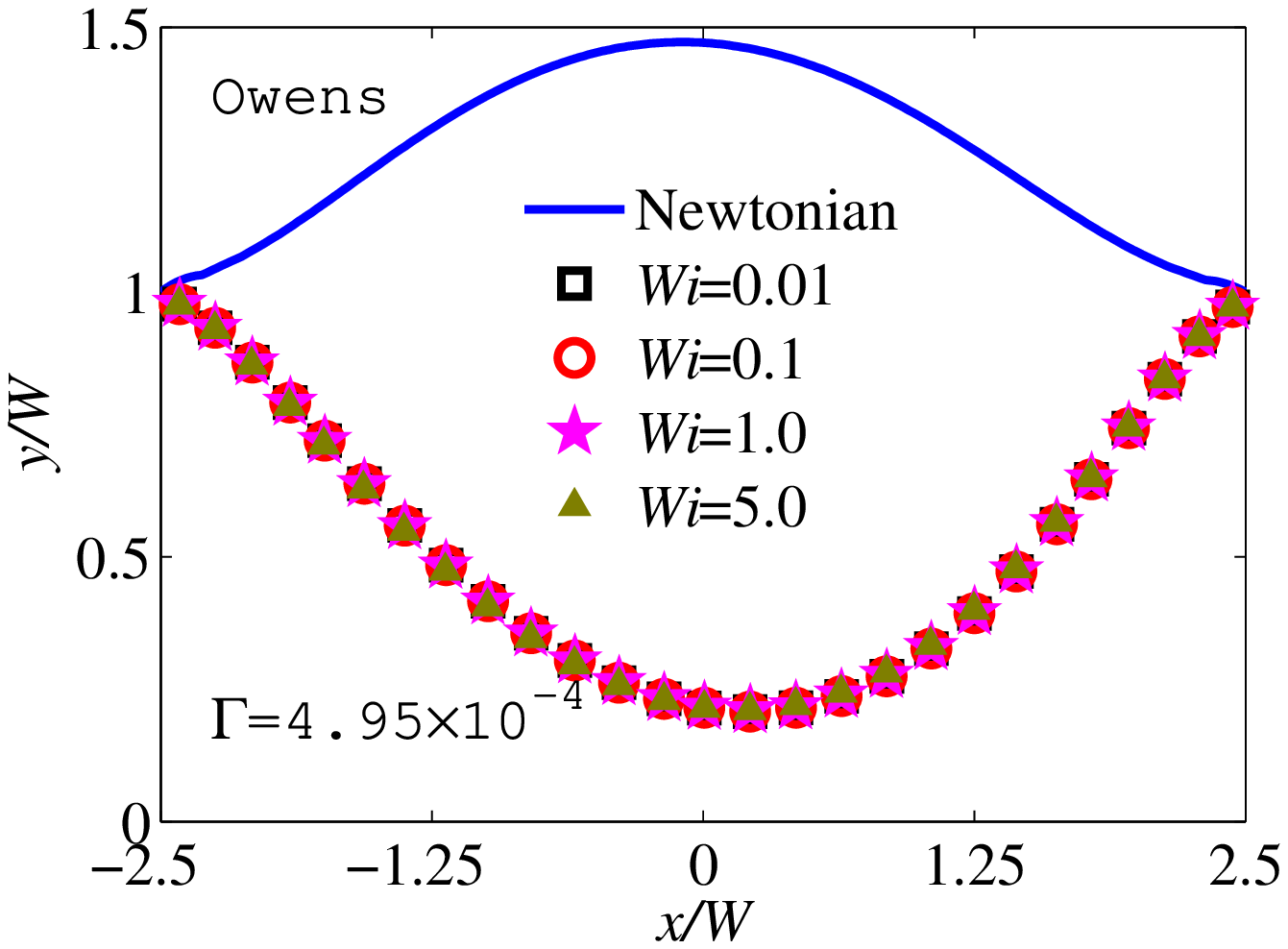}}\\
(c) & (f)  \\
\end{tabular}
\end{center}
%\begin{spacing}{1.5}
\caption{\small \label{figintshape} The shape of the fluid-solid interface
in a 2-D collapsible channel for the Oldroyd-B ((a) and (d)), FENE-P
((b) and (e)) and Owens models ((c) and (f)), compared with the
profile for a Newtonian fluid. Note that $Wi$ is $0.1$ in (a)--(c)
and $\Gamma$ is $4.95\times10^{-4}$ in (d)--(f). In (a)--(c)
different symbols represent different values of $\Gamma$
({\large${\color{blue}\square}$}: $4.95\times10^{-5}$,
{\Large${\color[rgb]{0.0,0.5,0.0}\circ}$}: $1.98\times10^{-4}$,
{\Large${\color{magenta}\star}$}: $3.0\times10^{-4}$,
{\large${\color[rgb]{0.5,0.5,0}\text{+}}$}: $3.96\times10^{-4}$,
{\large${\color{red}\text{x}}$}: $4.95\times10^{-4}$ and
{\large${\color[rgb]{0.5,0,1}\triangle}$}: $6.6\times10^{-4}$).
Lines with the same colour as the symbols represent the
predictions of a Newtonian fluid for identical values of $\Gamma$.}
%\end{spacing}
\end{figure}

Figure~\ref{figintshape} explores the deformation of the finite-thickness solid wall, while interacting with the different fluids, at a fixed value of $t=0.4W$. While figures~\ref{figintshape}~(a)-(c) investigate the shape of the fluid-solid interface for different values of $\Gamma$ at $Wi = 0.1$, figures~\ref{figintshape}~(d)-(f) examine the dependence of the interface profile on $Wi$ for $\Gamma = 4.95\times10^{-4}$. The extraordinary variation in the shape of the elastic solid with varying elasticity parameter $\Gamma$ is immediately apparent from figures~\ref{figintshape}~(a)-(c). In particular, there is a stark contrast in the response of the solid to the flow of different viscoelastic fluids. Except in the case of the Owens model, increasing $\Gamma$ leads to a movement of the
deformable solid from being within the channel (concave downwards)
to bulging out of the channel (convex upwards) due to action of the
forces exerted by the flowing fluid. At the relatively low value of
$Wi =0.1$ there is no discernible difference between the Newtonian,
Oldroyd-B and FENE-P fluids. For the Owens model on the other hand,
while the elastic solid remains concave downwards for the entire range of
$\Gamma$ values, there is a decrease in the minimum channel gap with increasing $\Gamma$. This behaviour is related to the significant
difference in the force field generated in the Owens model fluid due
to flow, as discussed shortly.

\begin{figure}
\centering \subfigure[] {
    \label{figmaxdef:sub:a}
    \includegraphics[width=9.0cm]{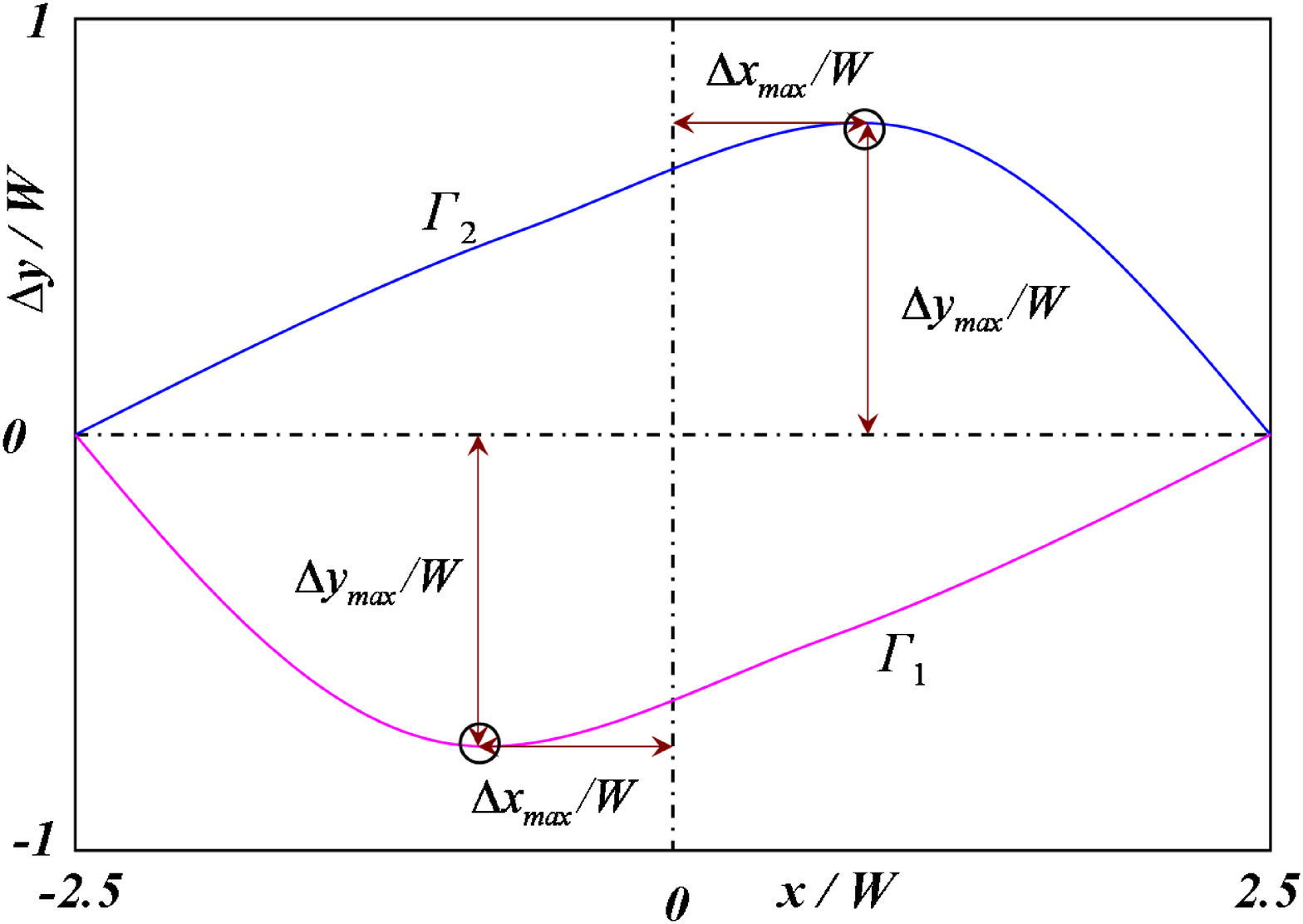}
} \hspace{1cm} \subfigure[] {
    \label{figmaxdef:sub:b}
    \includegraphics[width=9.0cm]{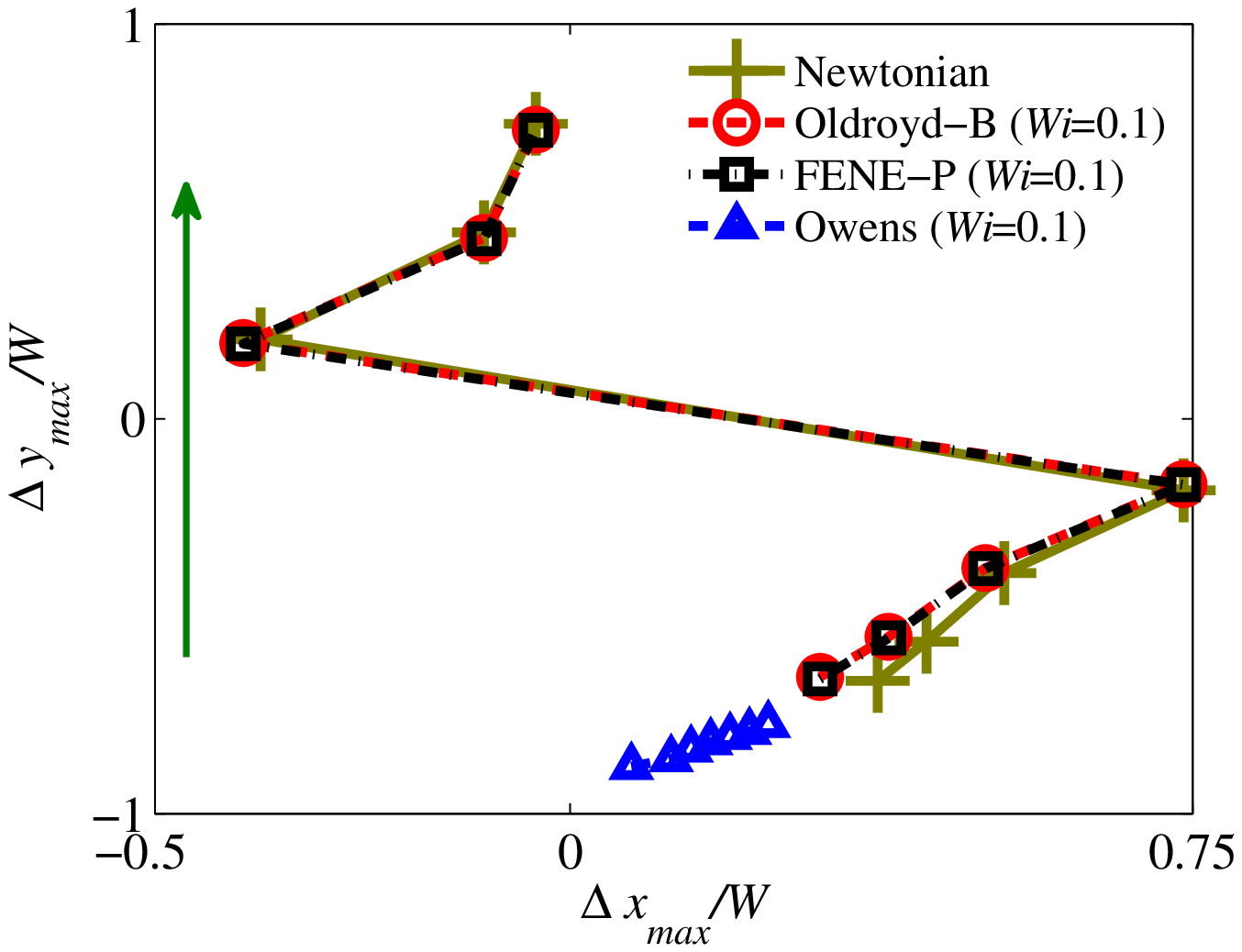}
}\hspace{1cm} \subfigure[] {
    \label{figmaxdef:sub:c}
    \includegraphics[width=9.0cm]{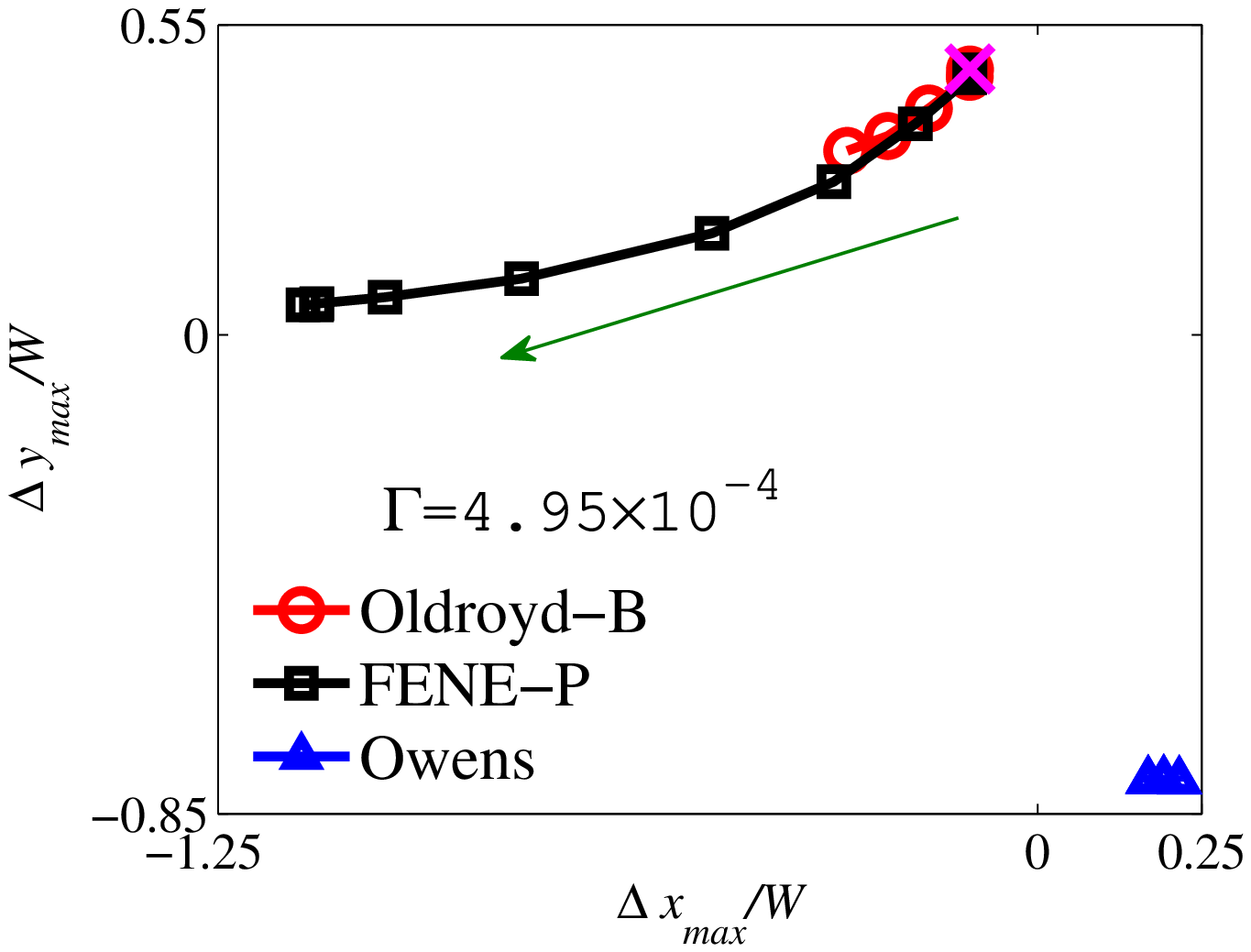}
}
\begin{spacing}{0.5}
\caption{\small  \label{figmaxdef} (a) Schematic diagram defining the position of maximum deformation ($\Delta x_{max}, \Delta y_{max}$). (b) Dependence of ($\Delta x_{max}, \Delta y_{max}) $ on $\Gamma$ at a fixed value of $Wi =0.1$, and (c) on  $Wi$ at a fixed value of $\Gamma = 4.95\times10^{-4}$. In (b) and (c) the arrows indicate the direction of increasing $\Gamma$ and $Wi$, respectively. The range of $Wi$ for the Oldroyd-B, FENE-P and Owens models are 0.01-1.508, 0.01-2.372 and 0.01-7.9, respectively.}
\end{spacing}
\end{figure}

It is appropriate to note here that in our earlier investigation of
viscoelastic flow in a 2D channel with a zero-thickness
membrane~\citep{debadi10}, the fluid-solid interface was always observed to be
concave downwards for all the viscoelastic fluids, at all values of membrane tension. Indeed, in contrast to the situation for a finite thickness solid, with decreasing tension, the zero-thickness membrane moves further into the channel, with a concomitant decrease in the narrowest channel gap.

At a fixed value of elasticity parameter $\Gamma$, while figure~\ref{figintshape}~(f) indicates that $Wi$ has no effect on the shape of the deformable solid in the case of the Owens model (which remains concave downwards), it has a noticeably different effect for the Oldroyd-B and FENE-P fluids. Both fluids cause the elastic solid to bulge outwards. However, the extent  of this bulge decreases more rapidly for the FENE-P fluid with increasing $Wi$. In the case of the Owens model, at these values of $Wi$, shear thinning is nearly complete, and there is
consequently no change discernible in the membrane shape. On the other hand,
the onset of shear thinning for the FENE-P model is responsible for the observed
variation in the predicted membrane shape from that for an Oldroyd-B fluid.

Figure~\ref{figintshape} indicates that the deformation of
the solid wall occurs along both axial and vertical directions for
all the fluid models, with the extent of movement depending on the values of
$\Gamma$ and $Wi$. By defining the position of maximum deformation as the 
point on the elastic solid furthest in the vertical direction from the horizontal surface, this dependence can be examined more systematically.
The precise location of the position of maximum deformation
is given by the co-ordinate pair ($\Delta x_{max}$, $\Delta y_{max}$),
as shown schematically in figure~\ref{figmaxdef}~(a), which measures the 
maximum displacement from the centre of the elastic solid when it is horizontal.
Figures~\ref{figmaxdef}~(b) and \ref{figmaxdef}~(c) track the position of
maximum deformation for varying $\Gamma$ and $Wi$, and correspond to
the set of figures~\ref{figintshape}~(a) to (c) and \ref{figintshape}~(d) to (f),
respectively. The movement of the elastic solid from being concave
downwards to convex upwards in the case of varying $\Gamma$, and the
downward movement with increasing $Wi$ are clearly captured in this
description. The relative immobility of the solid in the case of a
flowing Owens model fluid is also clearly revealed.

\begin{figure}
\begin{center}
\begin{tabular}{cc}
\resizebox{8.0cm}{!} {\includegraphics*[width=10cm]{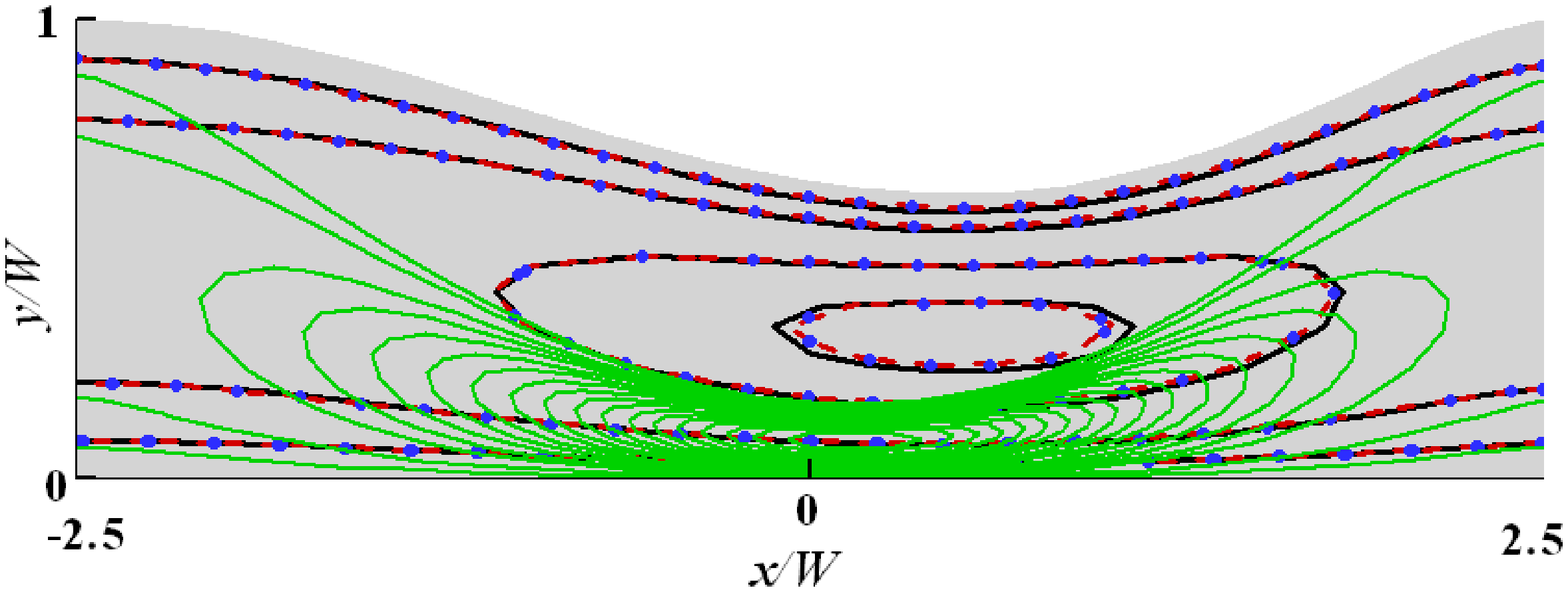}} &
\resizebox{8.0cm}{!} {\includegraphics*[width=10cm]{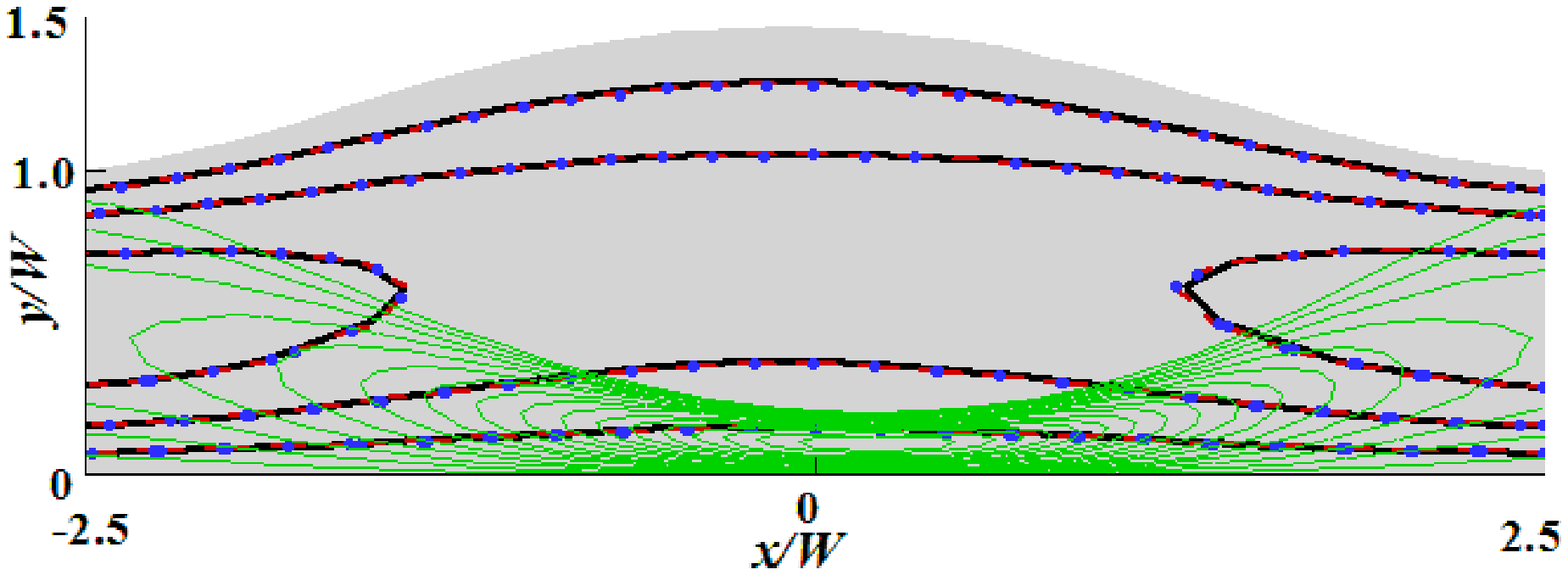}}\\
(a) & (c)  \\
\resizebox{8.0cm}{!} {\includegraphics*[width=10cm]{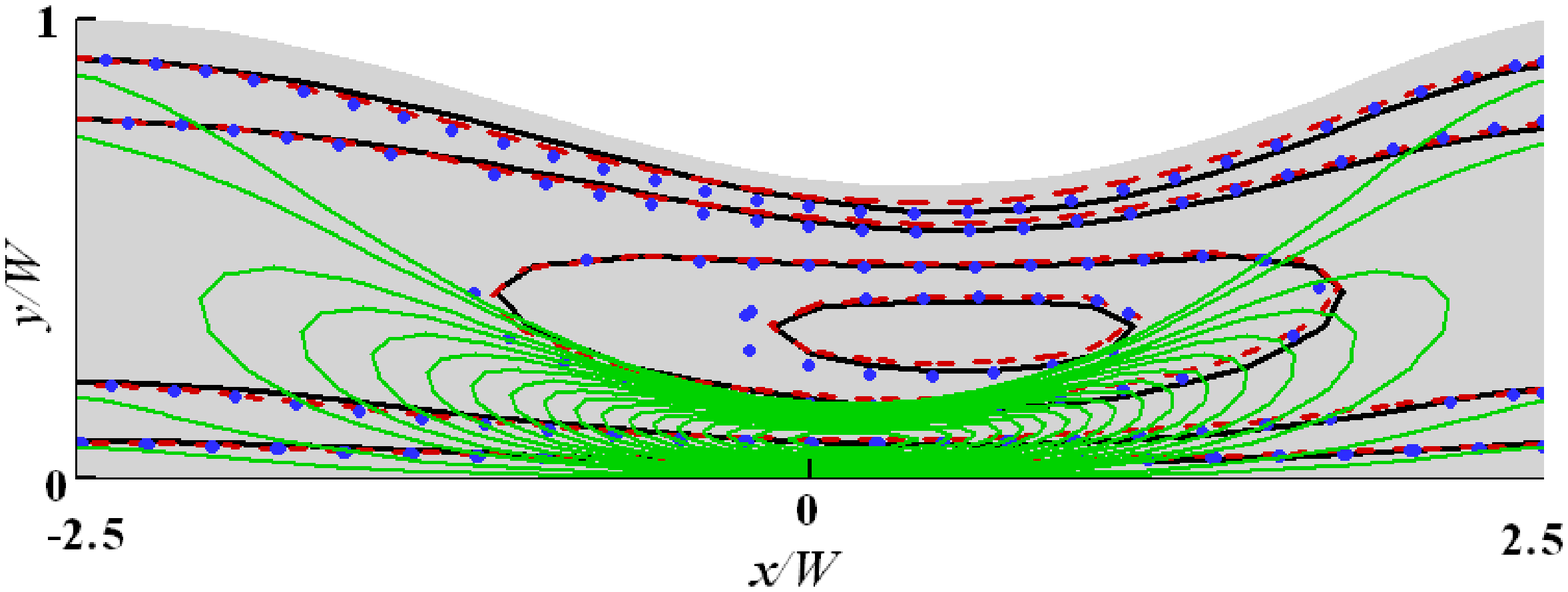}} &
\resizebox{8.0cm}{!} {\includegraphics*[width=10cm]{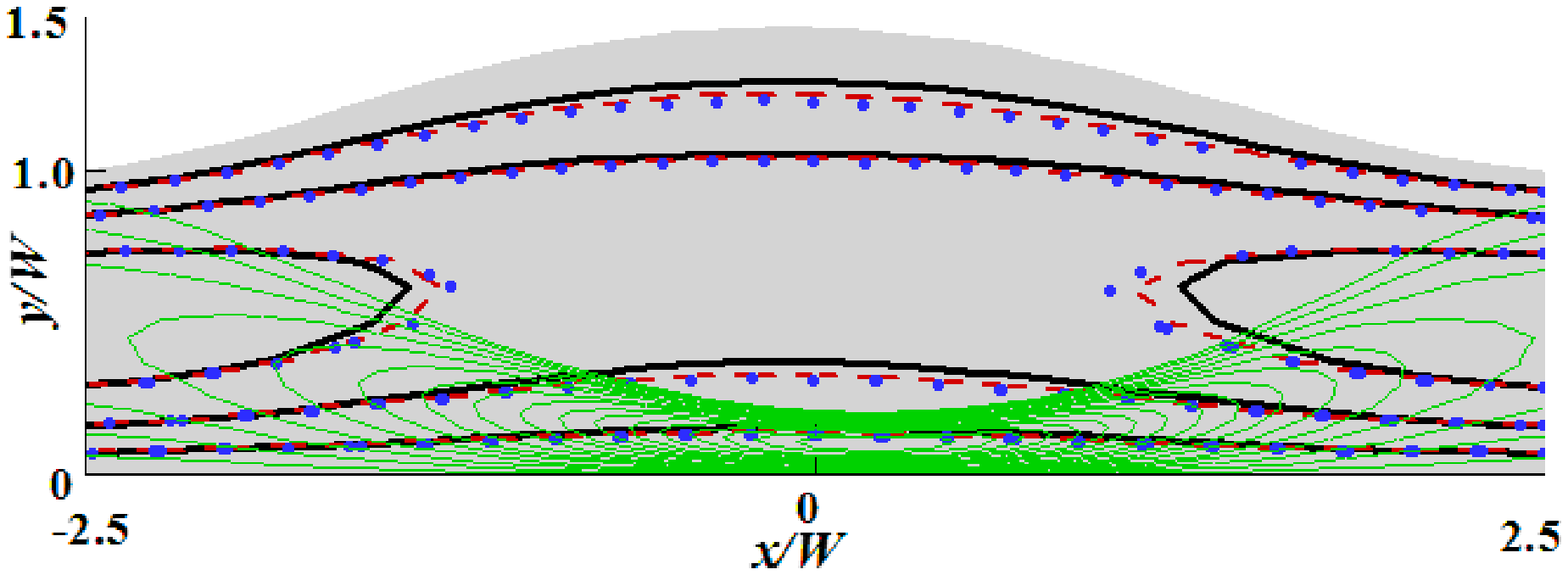}}\\
(b) & (d)  \\
\end{tabular}
\end{center}
%\begin{spacing}{1.5}
\caption{\small \label{figvelcon} {Contours of axial velocity ($v_x$) in 
the flow domain, for Newtonian (black), Oldroyd-B (red), FENE-P (blue) and 
Owens (green) fluids at $P_e = 0.04$, $t = 0.4W$, for two different values of 
Weissenberg number $Wi =0.1$ ((a) and (c)) and $Wi =0.5$ ((b) and (d)). Note 
that $\Gamma =1.98\times10^{-4}$ in (a)-(b) and $\Gamma = 4.95\times10^{-4}$ in (c)-(d).} The upper boundary in these figures reflects the shape of the interface at the corresponding parameter values.}
%\end{spacing}
\end{figure}

Figure~\ref{figvelcon} compares the velocity contours predicted by a
Newtonian fluid with those of an Oldroyd-B, FENE-P and Owens' fluid
at $P_e$ = $0.04$ and $t = 0.4W$, for two different values of $\Gamma$ and $Wi$. The qualitative and nearly quantitative similarity of the velocity contours for the Oldroyd-B and FENE-P models with those for a Newtonian fluid at $Wi=0.1$ and $Wi=0.5$ suggests that fluid rheology does not have a significant influence on the velocity field at these values of $Wi$. On the other hand, the dominant influence appears to be the value of the elasticity parameter $\Gamma$, which determines the shape of the fluid-solid interface. While the velocity contours for all the fluids are qualitatively similar to each other when the elastic solid lies within the channel (figures~\ref{figvelcon}~(a) and (b) for the Oldroyd-B and FENE-P models, and figures~\ref{figvelcon}~(a) to (d) for the Owens model), there is a qualitative change when the elastic solid lies outside the channel. The latter situation occurs only for the Oldroyd-B and FENE-P models for $\Gamma = 4.95\times10^{-4}$, as displayed in figures~\ref{figvelcon}~(c) and (d).

\begin{figure}
\begin{center}
\begin{tabular}{cc}
\resizebox{8.0cm}{!} {\includegraphics*[width=8.0cm]{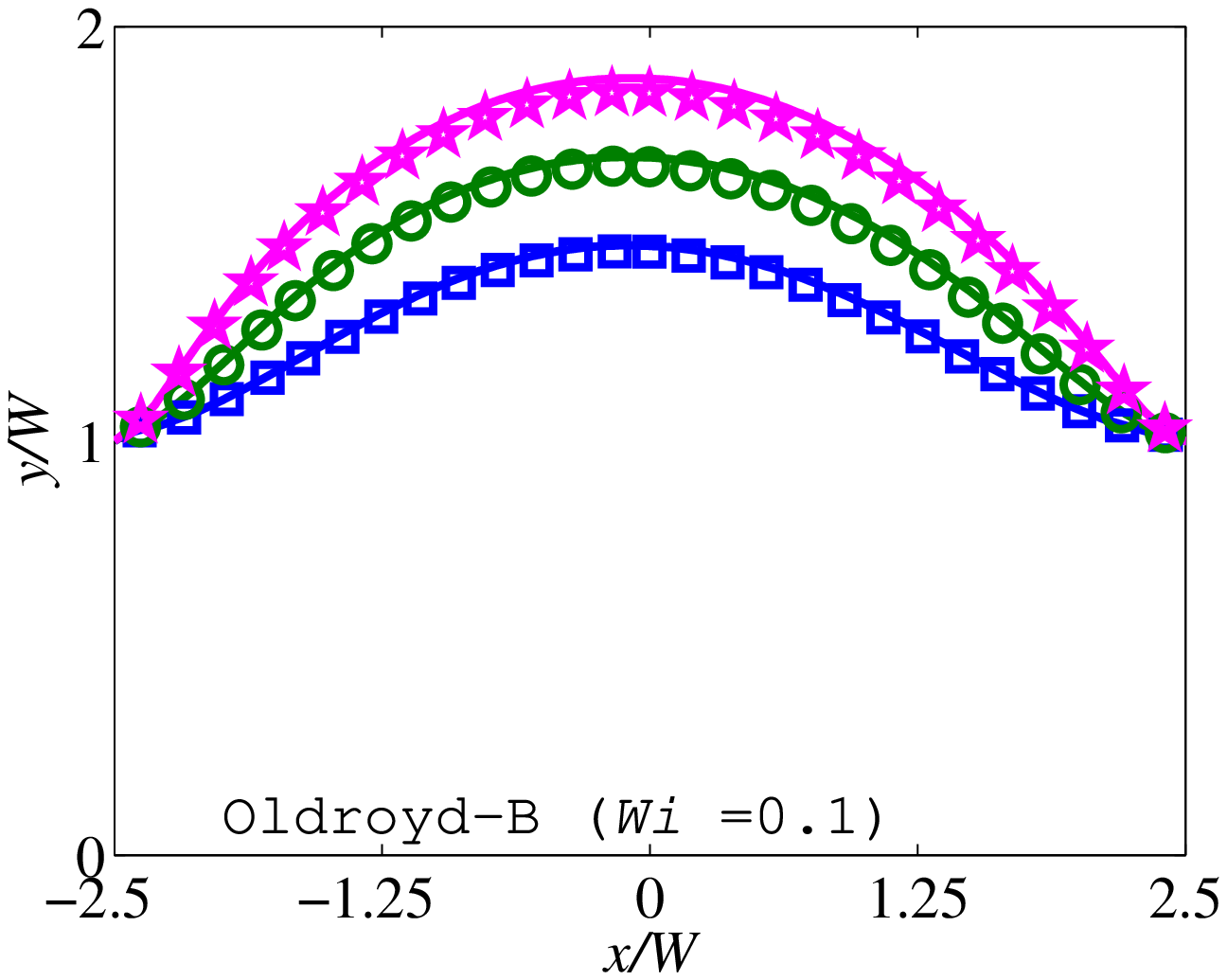}}
&
\resizebox{8.0cm}{!} {\includegraphics*[width=8.0cm]{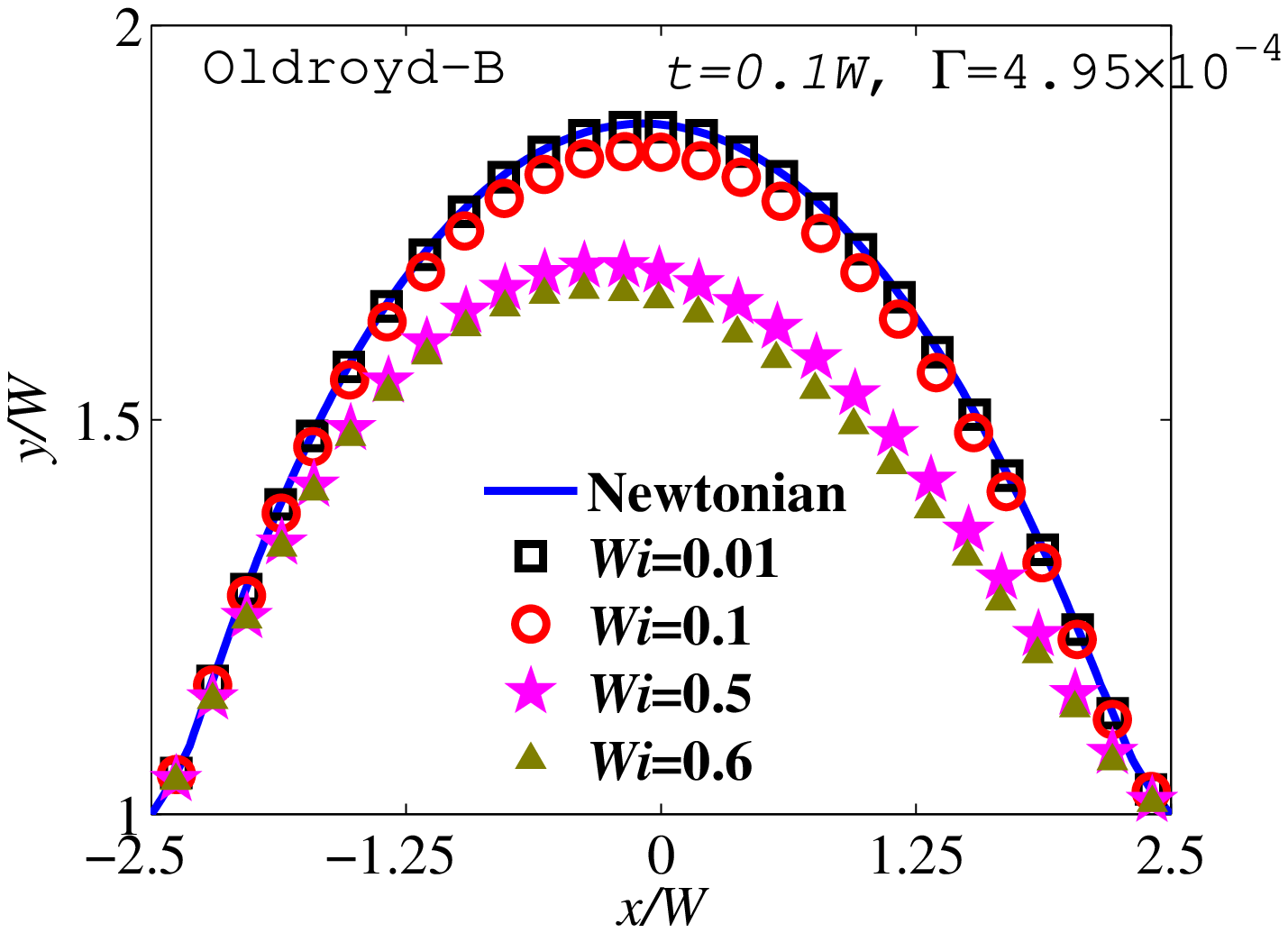}}\\
(a) & (d)  \\
\resizebox{8.0cm}{!} {\includegraphics*[width=8.0cm]{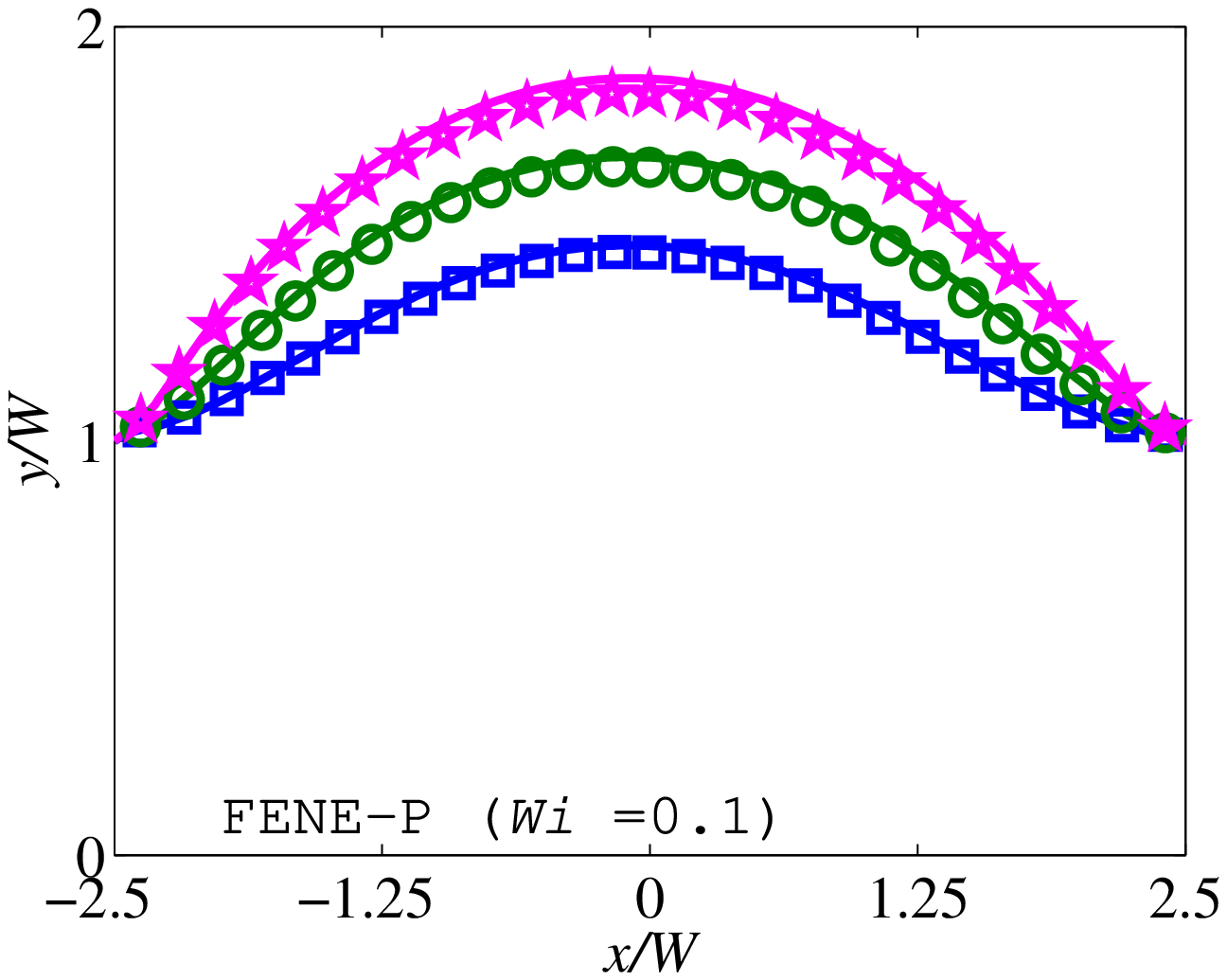}}
&
\resizebox{8.0cm}{!} {\includegraphics*[width=8.0cm]{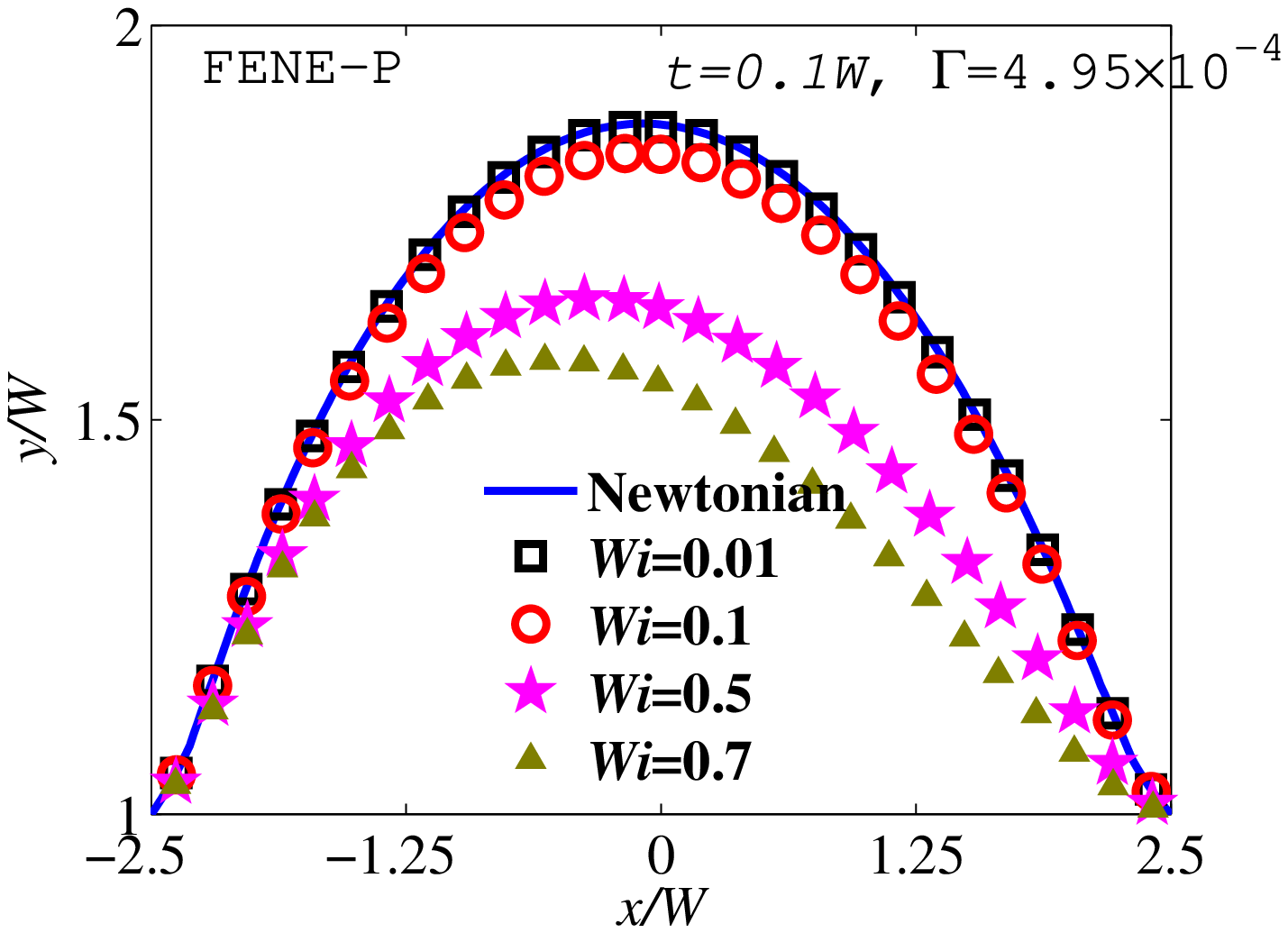}}\\
(b) & (e)  \\
\resizebox{8.0cm}{!} {\includegraphics*[width=8.0cm]{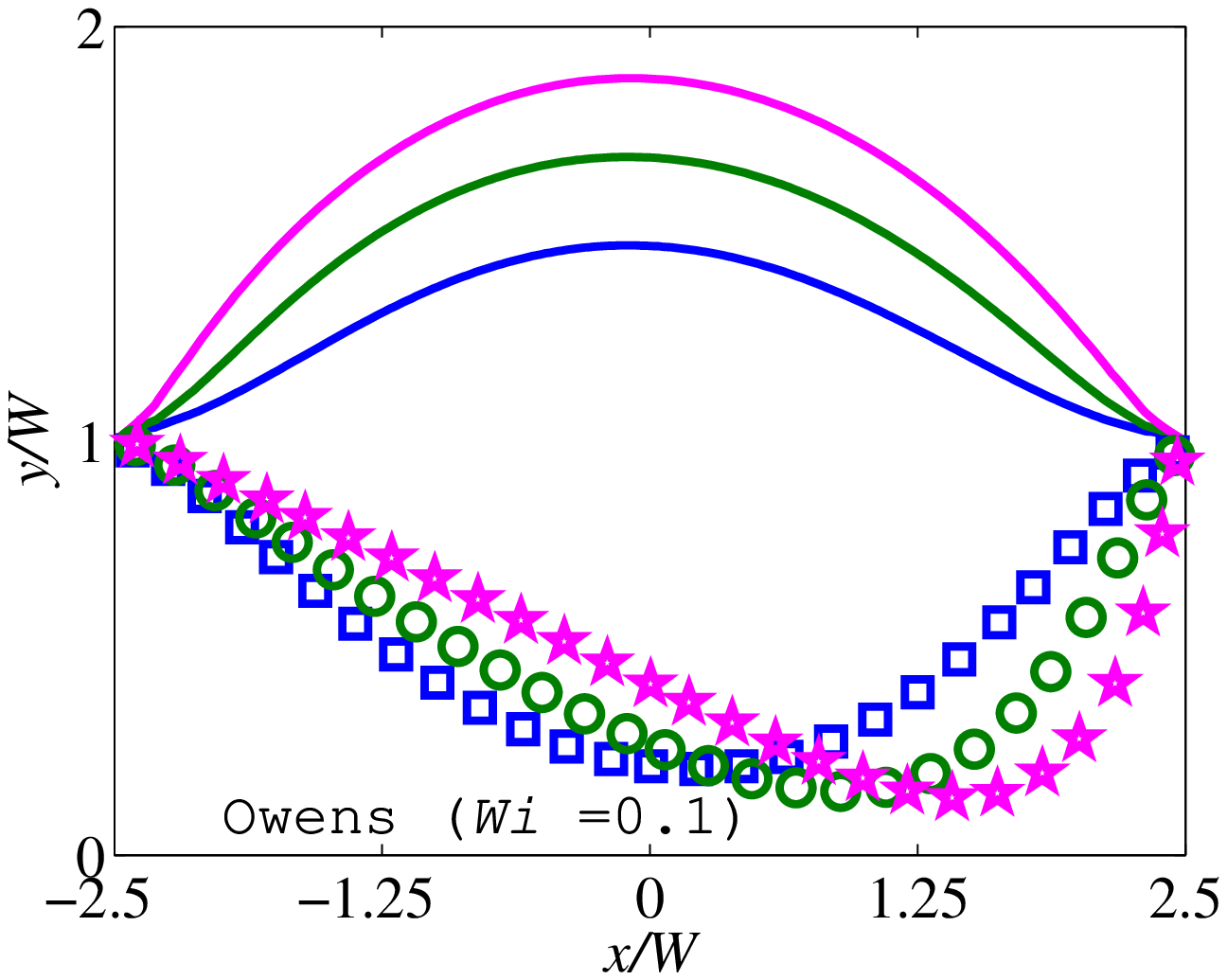}}
&
\resizebox{8.0cm}{!} {\includegraphics*[width=8.0cm]{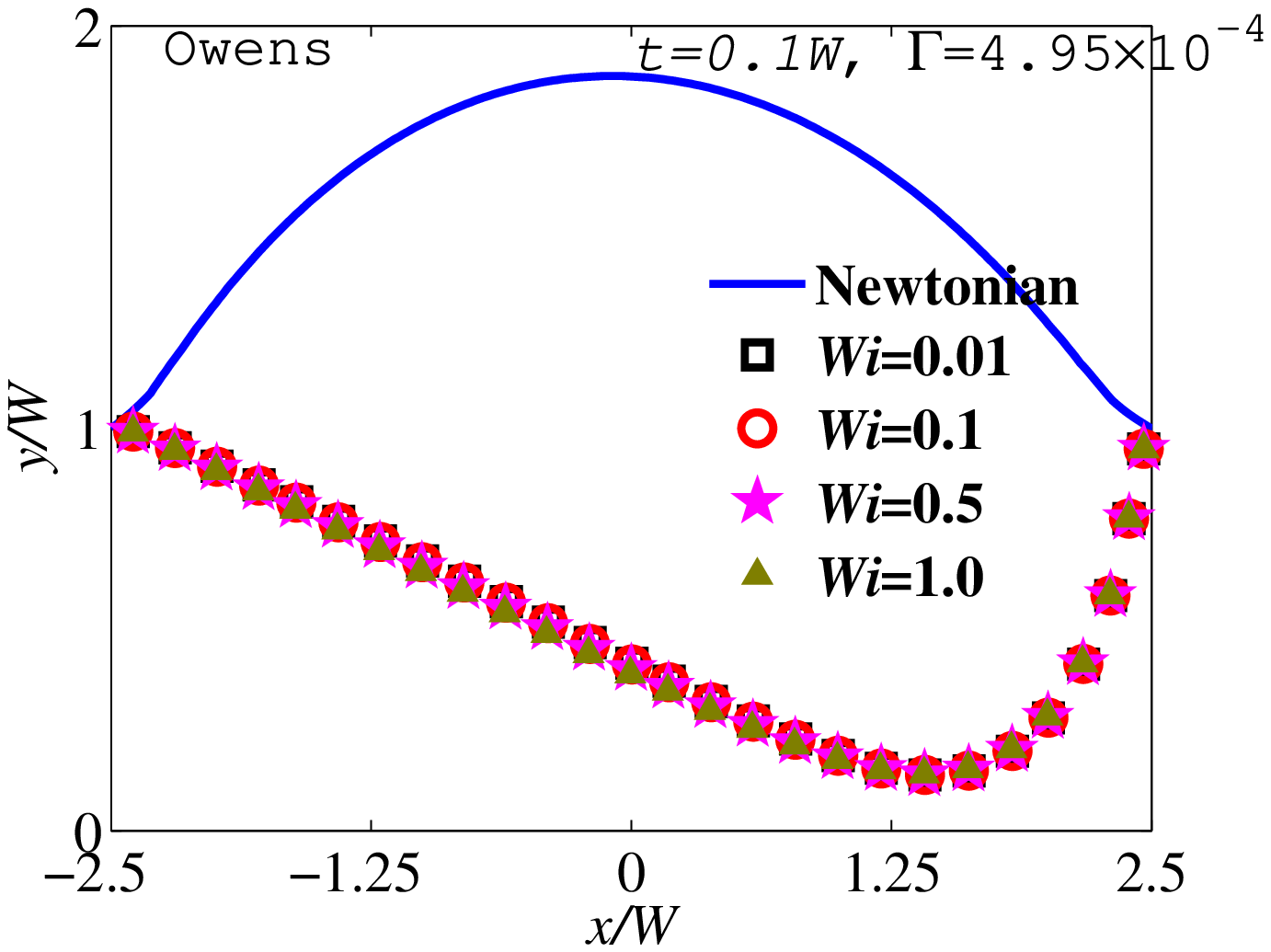}}\\
(c) & (f)  \\
\end{tabular}
\end{center}
%\begin{spacing}{1.5}
\caption{\small \label{figintshape2} The shape of the fluid-solid interface for the Oldroyd-B ((a) and (d)), FENE-P ((b) and (e)) and the Owens model fluids ((c) and (f)), compared with the profile for a Newtonian fluid, at different values of wall thickness $t$ and $Wi$, at a fixed value of $\Gamma = 4.95\times10^{-4}$. The value of $Wi$ is fixed at $0.1$ in (a)--(c), while $t$  is fixed at $0.1 W$ in (d)--(f). In (a)--(c) different symbols represent different values of $t$ ({\Large ${\color{magenta}\star}$}: $0.1 W$, {\Large${\color[rgb]{0.0,0.5,0.0}\circ}$}: $0.2 W$, {\large${\color{blue}\square}$}: $0.4 W$). Lines with the same colour as the symbols represent the predictions of a Newtonian fluid for identical values of $t$.}
%\end{spacing}
\end{figure}

\subsection{\label{sec:thick} Influence of wall thickness on interface shape and limiting Weissenberg number}

The wall thickness $t$ does not appear directly in the governing equations, or in the boundary conditions, rather it determines the size of the solid domain. We can anticipate, however, that its influence on the interface shape and limiting Weissenberg number will be similar to that of the elasticity parameter $\Gamma$. In very simple terms, increasing $\Gamma$ for a given state of stress at the fluid-solid interface due to the flow of viscoelatic fluid in the channel, leads to a larger strain in the solid, since (with all other parameter values fixed) an increase in $\Gamma$ implies a decrease in the shear modulus of the solid, $G$. Similarly, for a given state of stress at the fluid-solid interface, decreasing the wall thickness $t$ would lead to a larger strain in the solid, since there is ``less'' solid material over which to distribute the resultant stress in the solid. 

These arguments are borne out in figures~\ref{figintshape2}, which explores the deformation of the finite-thickness solid wall, while interacting with the different fluids, at a fixed value of $\Gamma = 4.95\times10^{-4}$. While figures~\ref{figintshape2}~(a)-(c) investigate the shape of the fluid-solid interface for different values of $t$ at $Wi = 0.1$, figures~\ref{figintshape2}~(d)-(f) examine the dependence of the interface profile on $Wi$ for $t=0.1W$. 

Figures~\ref{figintshape2}~(a) and (b) indicate that for $\Gamma = 4.95\times10^{-4}$, at all values of $t$ examined here, the solid wall bulges outward, and there is no discernible difference between the Oldroyd-B and FENE-P models, and a Newtonian fluid. For all these fluids, increasing $t$ leads to a decrease in the extent of deformation of the solid wall. The profiles are similar to those observed previously in figures~\ref{figintshape}~(a) and (b) for decreasing values of $\Gamma$. In the case of the Owens model fluid, the fluid-solid interface lies within the channel for all the examined values of wall thickness. However, there is a distinct change observed in the shape of the interface, with the profile becoming more symmetric as the value of $t$ increases. 

The change in interface shape with increasing $Wi$, at fixed values of $t$ and $\Gamma$, displayed in figures~\ref{figintshape2}~(d)--(f) is very similar to that observed previously in figures~\ref{figintshape}~(d)--(f), with the extent of shear thinning playing the dominant role in determining the shape. The essential difference between the two sets of figures appears to be in the loss of symmetry in the interface profile for the smaller value of $t$.

\begin{figure}[t]
\begin{center}
\includegraphics[width=0.80\textwidth]{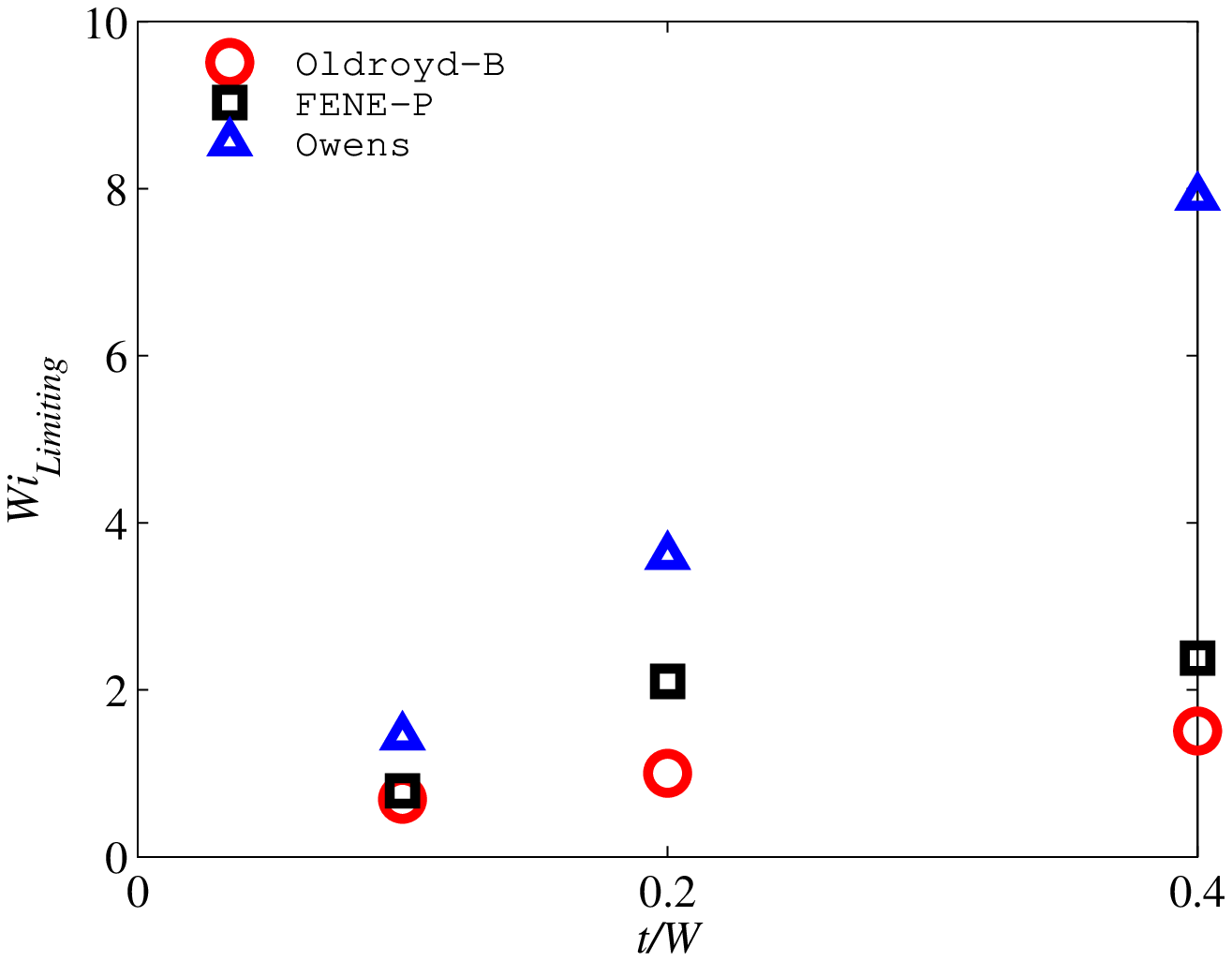}
\caption{\label{figmaxWi_t} Limiting Weissenberg number for the three fluid models, for computations carried out with the M2 mesh, at $P_e = 0.04$ and $\Gamma = 4.95 \times10^{-4}$, as a function of the non-dimensional wall thickness $t/W$. }
\end{center}
\end{figure}

As discussed earlier, both the mesh converged and limiting Weissenberg numbers for the Owens model increase with increasing $\Gamma$ because of an increase in the narrowest channel gap (see figures~\ref{figmaxWi} and  figure~\ref{figintshape}~(c)). This argument is consistent with the behaviour of $Wi_{Limiting}$ displayed in figure~\ref{figmaxWi_t} for the Owens model, for increasing values of $t/W$ (see also figure~\ref{figintshape2}~(c)). In both these situations corresponding to the Owens model, the elastic wall always lies within the channel for all the values of the various parameters considered here. We have seen previously in figure~\ref{figmaxWi}, for both the Oldroyd-B and FENE-P models, that $Wi_{Limiting}$ and $Wi_{Converged}$ are not sensitive to changes in $\Gamma$ for values of $\Gamma$ that correspond to the situation where the elastic wall lies outside the channel. On the other hand, figure~\ref{figmaxWi_t} suggests that even though the elastic wall lies outside the channel for $\Gamma = 4.95 \times10^{-4}$ (see figures~\ref{figintshape2}(a) and (b)), $Wi_{Limiting}$  increases with $t/W$, until $t/W = 0.2$, before levelling off. 

\subsection{\label{sec:PS} Pressure and stresses}

\citet{patankar02} have shown analytically that for any constitutive model of the 
form,
\begin{equation}
a_{1}  \textbf{D} + a_{2} \upperconvected{\textbf{D}} + a_{3} \textbf{T} + a_{4} 
\upperconvected{\textbf{T}} =\bm{0}
\label{pat}
\end{equation}
where, $a_1$, $a_2$, $a_3 $, and $a_4$ are
scalar functions of the invariants of $\textbf{D}$ and $\textbf{T}$,
and $\upperconvected{\textbf{D}}$ and $\upperconvected{\textbf{T}}$
are the upper convected time derivatives of $\textbf{D}$ and
$\textbf{T}$, the \emph{normal component} of extra stress on a rigid body
surface will be zero. \citet{debadi10} have shown numerically that this
is true even in the case of flow past a deformable zero-thickness membrane, for 
all the three viscoelastic fluids considered here. In the present instance as well, we find that the {normal component} of stress on the elastic wall is solely due to pressure.

\begin{figure}
\begin{center}
\begin{tabular}{cc}
\resizebox{8.0cm}{!} {\includegraphics*[width=8.0cm]{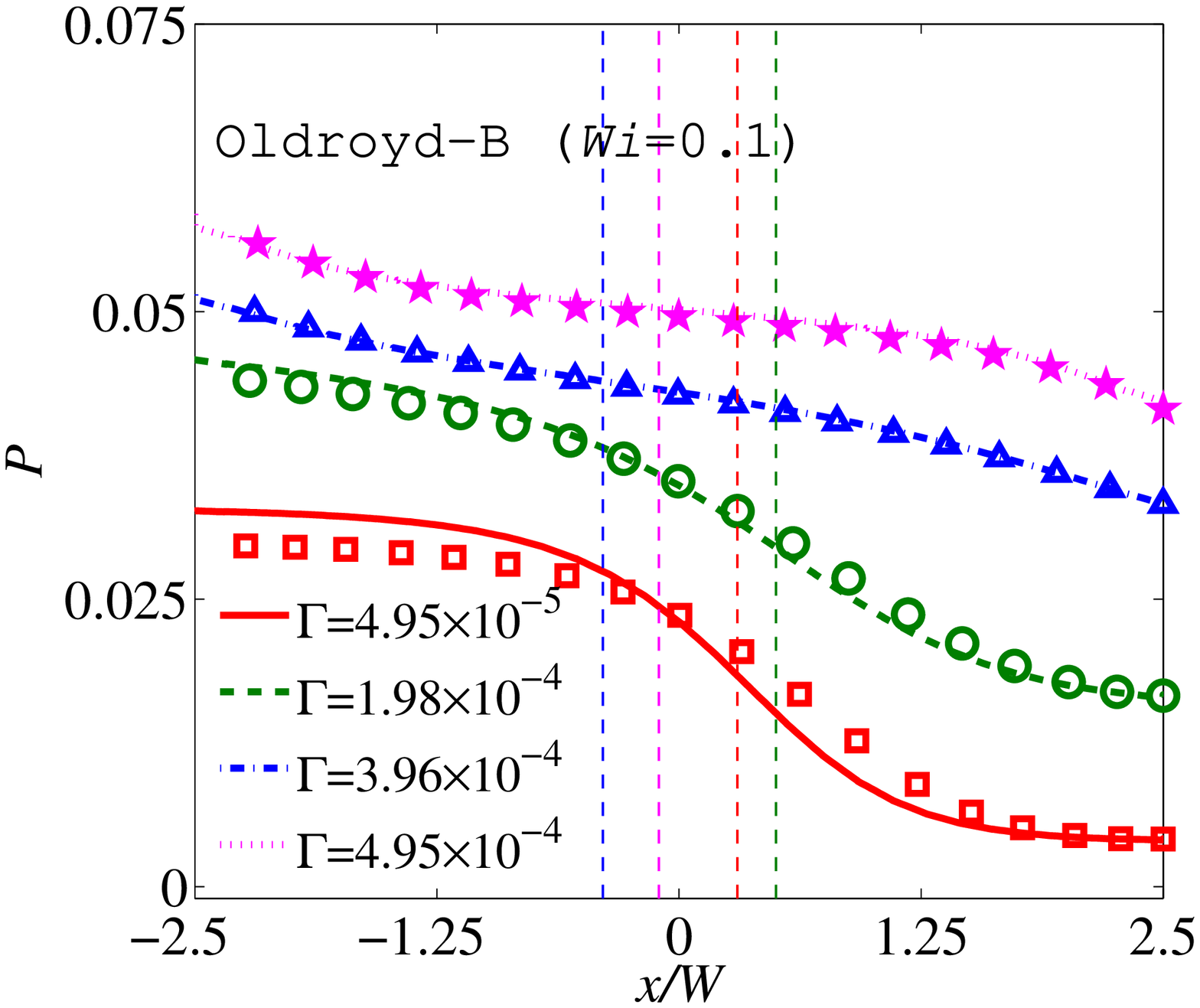}}
&
\resizebox{8.0cm}{!} {\includegraphics*[width=8.0cm]{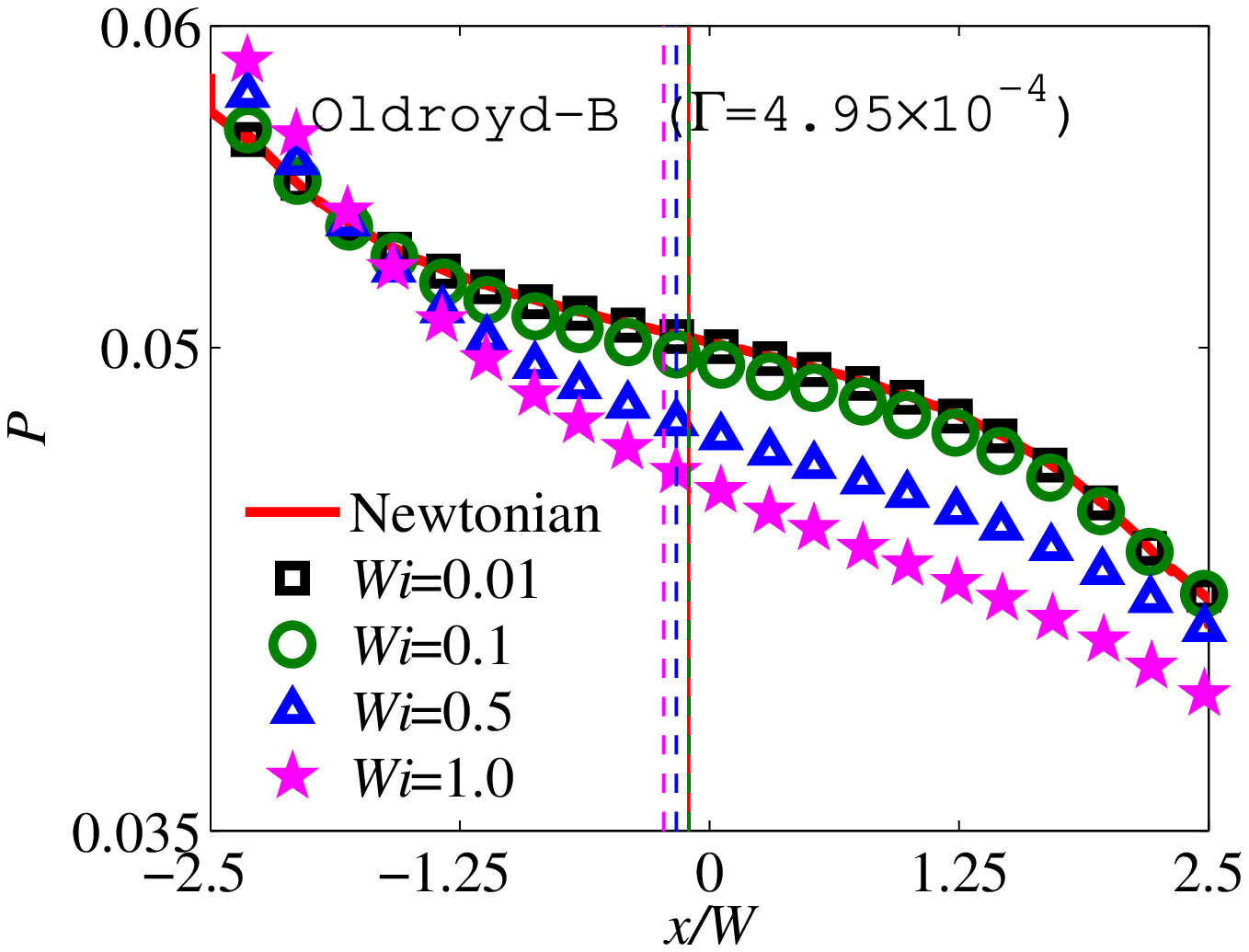}}\\
(a) & (d)  \\
\resizebox{8.0cm}{!} {\includegraphics*[width=8.0cm]{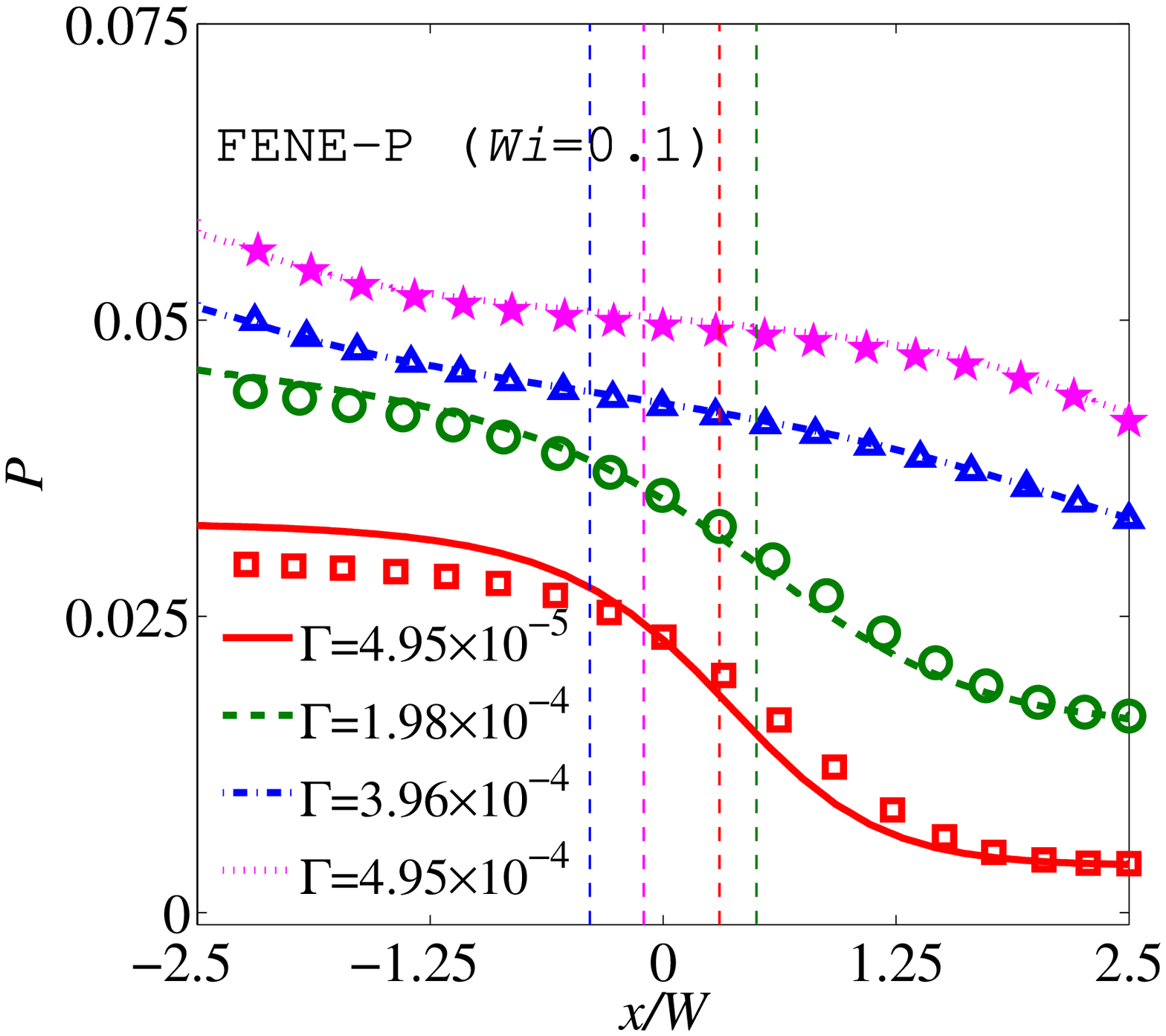}}
&
\resizebox{8.0cm}{!} {\includegraphics*[width=8.0cm]{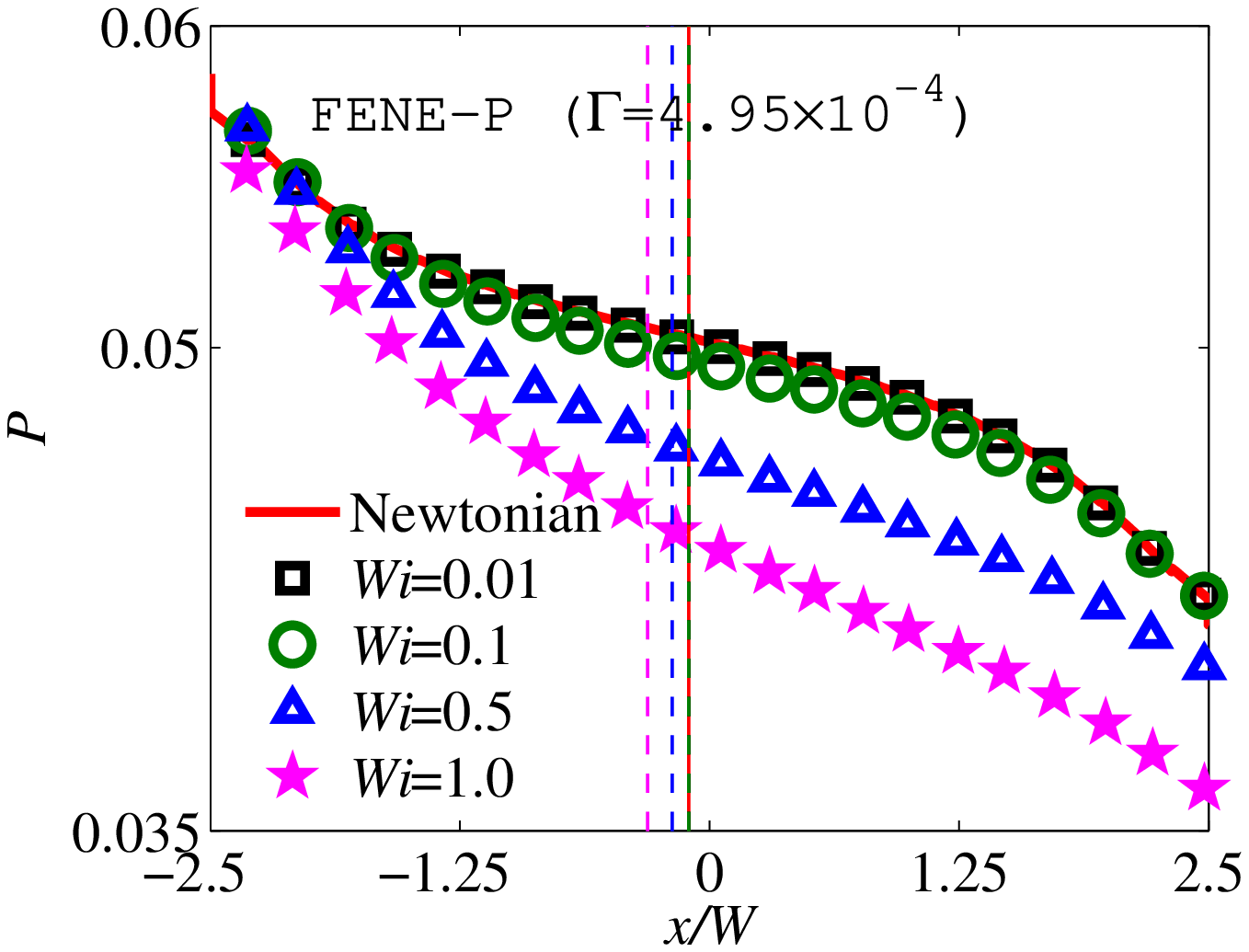}}\\
(b) & (e)  \\
\resizebox{8.0cm}{!} {\includegraphics*[width=8.0cm]{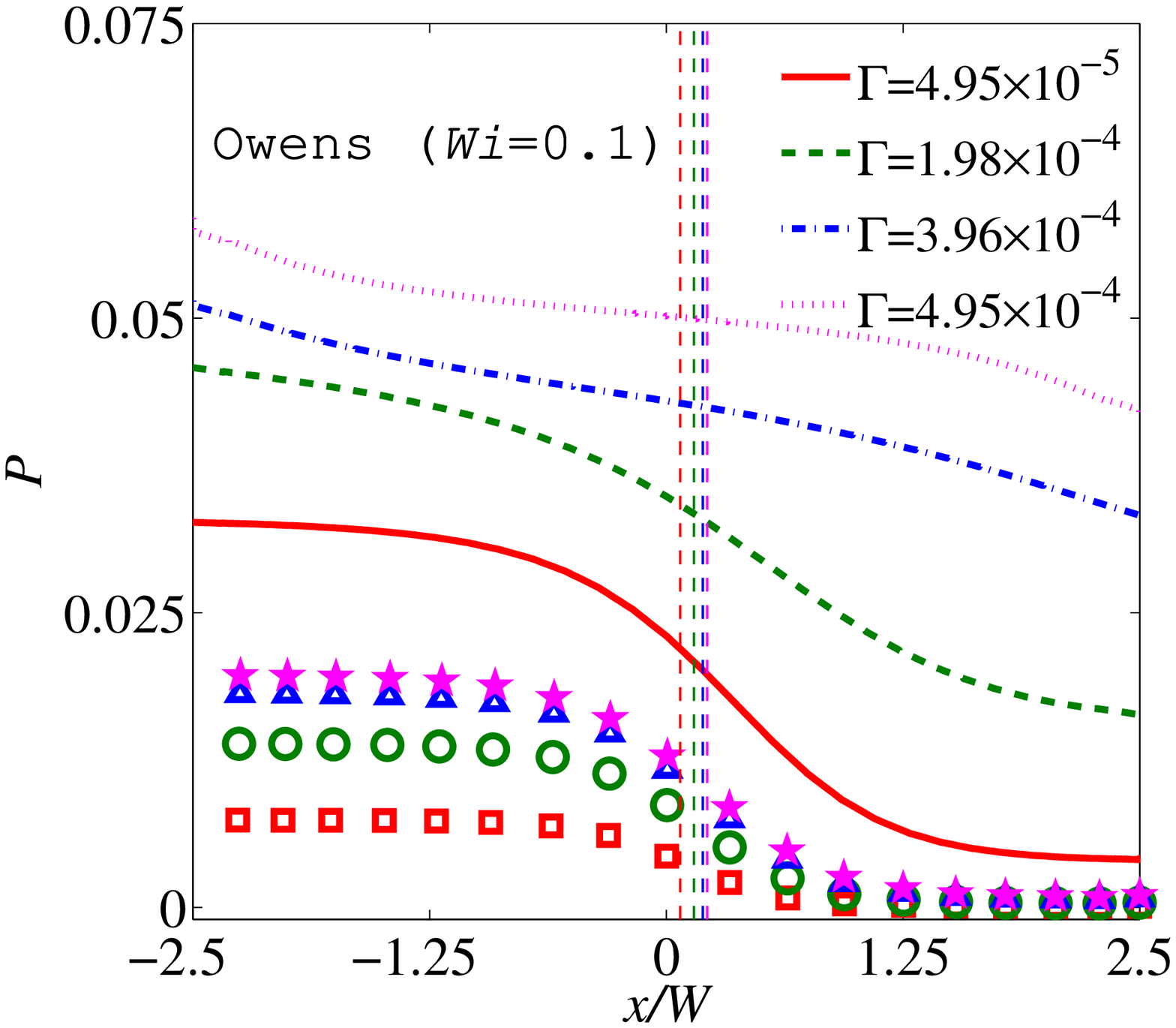}}
&
\resizebox{8.0cm}{!} {\includegraphics*[width=8.0cm]{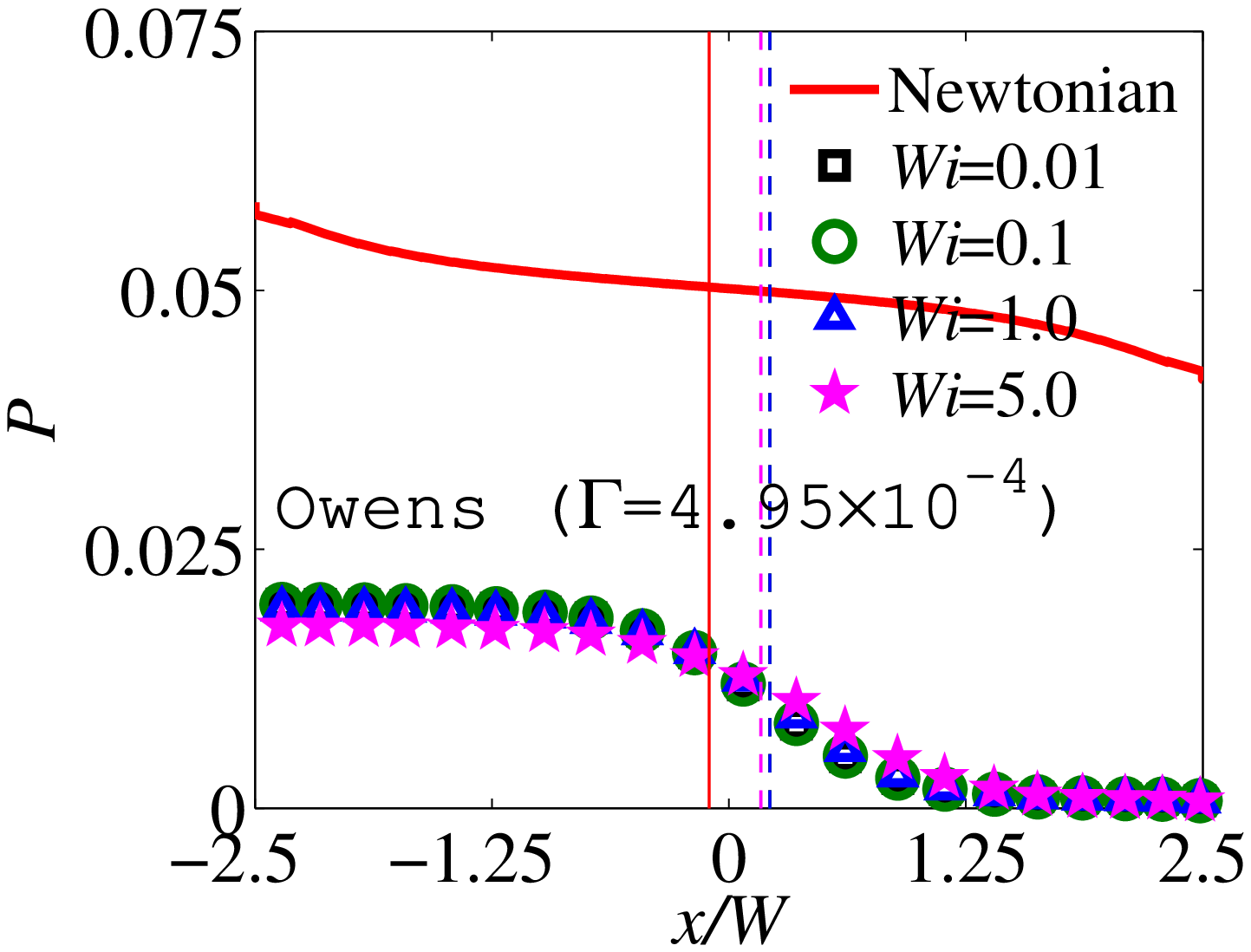}}\\
(c) & (f)  \\
\end{tabular}
\end{center}
\vskip-15pt
%\begin{spacing}{1.5}
\caption{\small  \label{figpresprof} Dependence of the pressure profile
along the flexible membrane on $Wi$ and $\Gamma$, for the Oldroyd-B
((a) and (d)), FENE-P ((b) and (e)) and Owens models ((c) and (f)),
respectively. The lines in (a)--(c) are for a Newtonian fluid. Note
that $\Gamma$ = $1.98\times10^{-4}$ in (d)-(f) and $Wi=0.1$ in
(a)-(c). }
%\end{spacing}
\end{figure}

Figure~\ref{figpresprof} examines the effect of $\Gamma$ and $Wi$
on the non-dimensional pressure $P$ exerted by the different fluids
on the elastic solid. At $Wi = 0.1$, the increase in $P$
with increasing $\Gamma$ for the Oldroyd-B and FENE-P models is nearly
identical to that for a Newtonian fluid, as can be seen from
figures~\ref{figpresprof}~(a) and~(b). Notably, for all these fluids, a distinct 
change occurs in the shape of
the pressure profile for $\Gamma > 1.98 \times 10^{-4}$.
For values of $\Gamma$ less than or equal to
this value, the pressure profile is relatively constant upstream of the position
of maximum deformation, before decreasing relatively rapidly downstream to a
constant value. (The vertical lines in the figure denote the $x$-position of 
maximum
deformation, with the colour coordinated to match the corresponding $\Gamma$
value). On the other hand, for values of $\Gamma > 1.98 \times 10^{-4}$, the
decrease in pressure from the location where the fluid flows under the
deformable solid to the location where it exits, is much more uniform. As can be
seen from figures~\ref{figintshape}~(a) and~(b), the change
in the shape of the pressure profile is correlated with the change in
interface shape that occurs around $\Gamma \sim 3 \times 10^{-4}$, which is
approximately the value at which the elastic solid moves from being
concave downwards within the channel to
bulging outwards from the channel. In the case of the Owens model, even though
the pressure increases with increasing $\Gamma$, the shape of the
pressure profile remains unchanged, since the elastic solid is always concave
downwards in shape (see figures~\ref{figpresprof}~(c) and \ref{figintshape}~(c)). 
Another notable aspect is that the magnitude of pressure at any point along the interface is significantly  lower
for the Owens model compared to that for all the other fluids. This can be 
attributed to
the significant decrease in viscosity that occurs for the Owens model fluid when it
flows under the deformable solid.

Figure~\ref{figpresprof}~(d)-(e) display the effect of $Wi$ on the pressure profile 
for a fixed value of $\Gamma = 4.95 \times 10^{-4}$. At this value of $\Gamma$, as seen earlier in figures~\ref{figintshape}~(d)-(f), for all the values of $Wi$ considered here, the elastic solid bulges outwards from the channel due to interaction with the Oldroyd-B and FENE-P fluids, while it is concave downwards for the flow of an Owens model fluid.  In the former two cases, with increasing $Wi$, there is a clear decrease in the pressure that the fluid exerts on the downstream end of the elastic solid, with the decrease being more substantial for the FENE-P fluid. This correlates with the decrease in the bulge of
the elastic solid seen earlier in figures~\ref{figintshape}~(d) and~(e). For the 
Oldroyd-B fluid, there also appears to be a slight increase in pressure at the upstream end of the elastic solid. In the case of the Owens model fluid, neither the interface shape nor the pressure profile are significantly altered by the variation in $Wi$.

\begin{figure}
\begin{center}
\begin{tabular}{cc}
\resizebox{8.0cm}{!} {\includegraphics*[width=10cm]{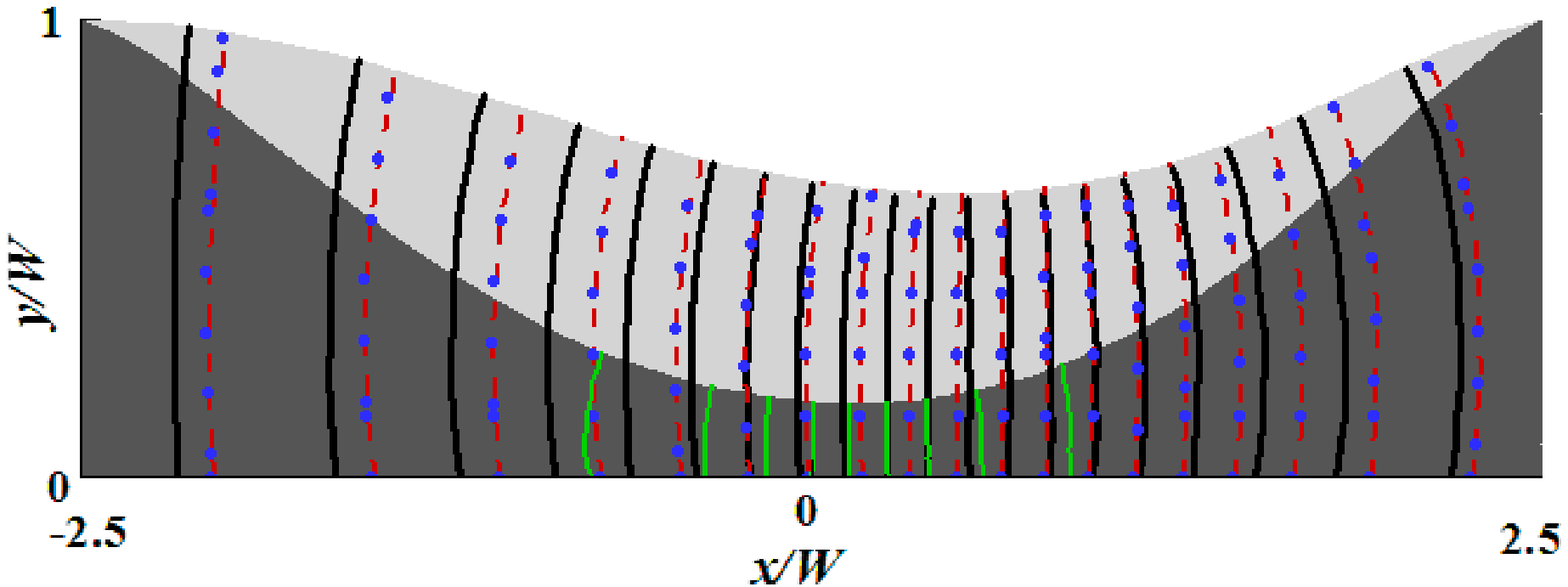}} &
\resizebox{8.0cm}{!} {\includegraphics*[width=10cm]{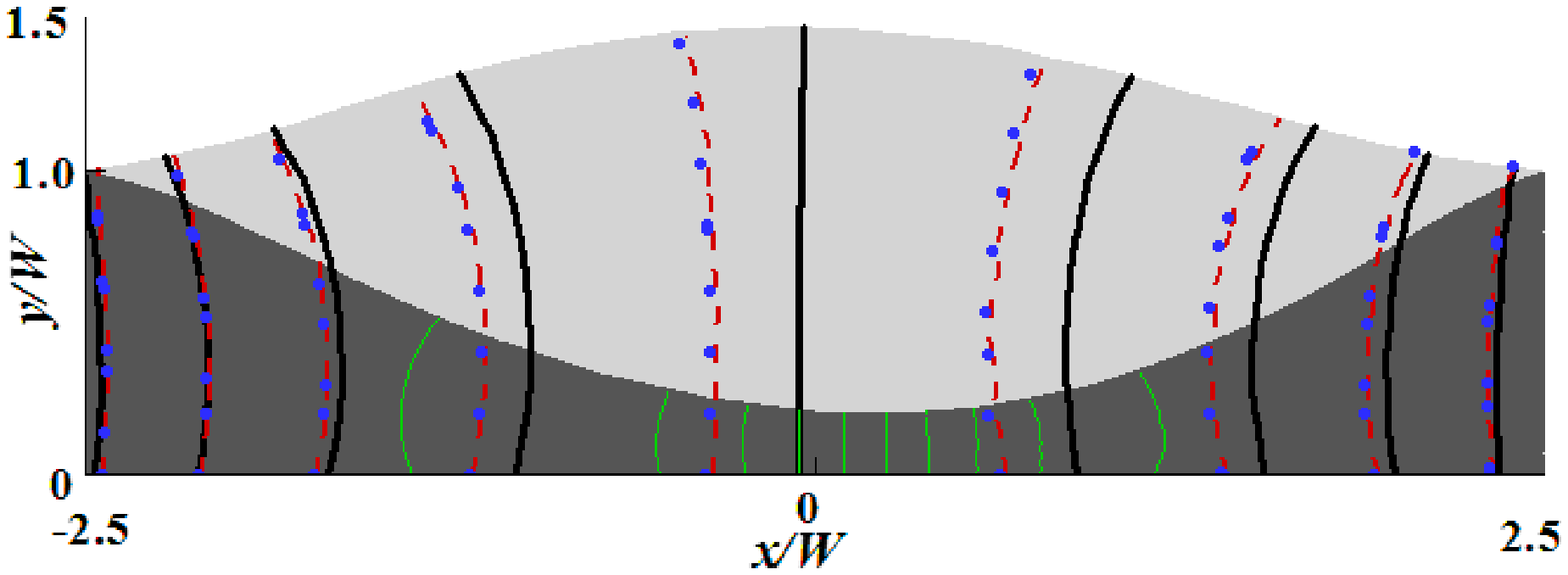}}\\
(a) & (c)  \\
\resizebox{8.0cm}{!} {\includegraphics*[width=10cm]{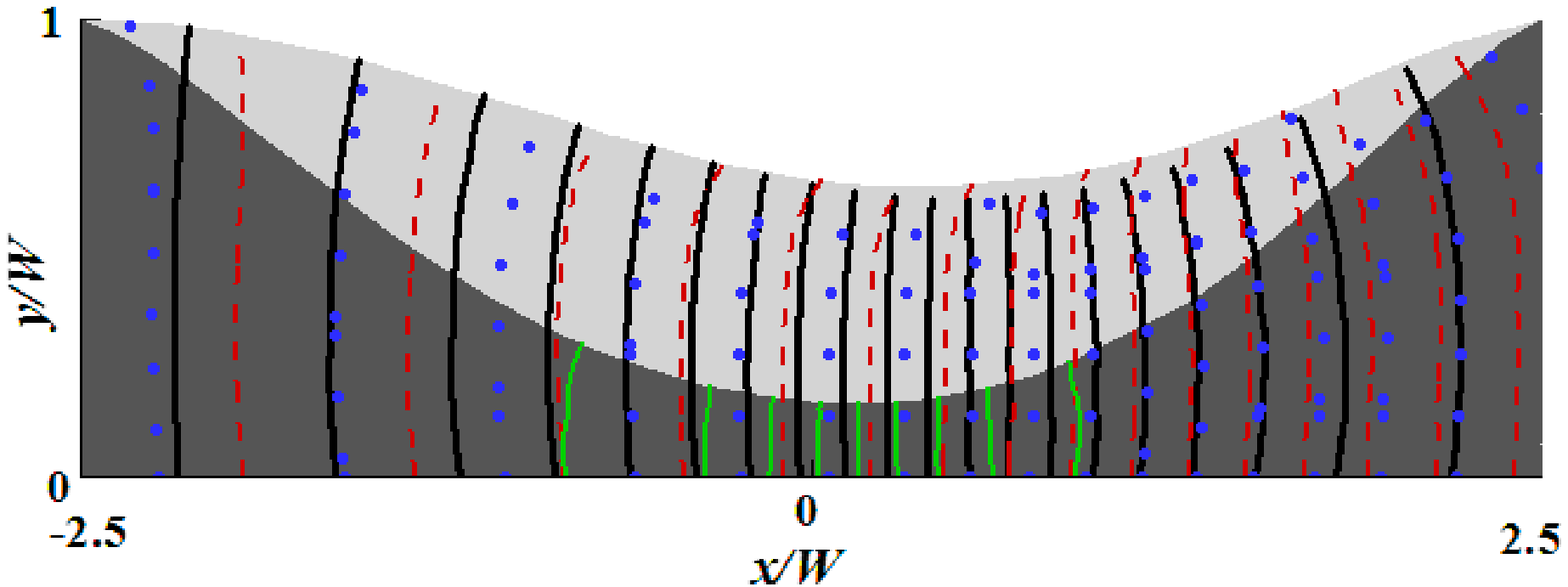}} &
\resizebox{8.0cm}{!} {\includegraphics*[width=10cm]{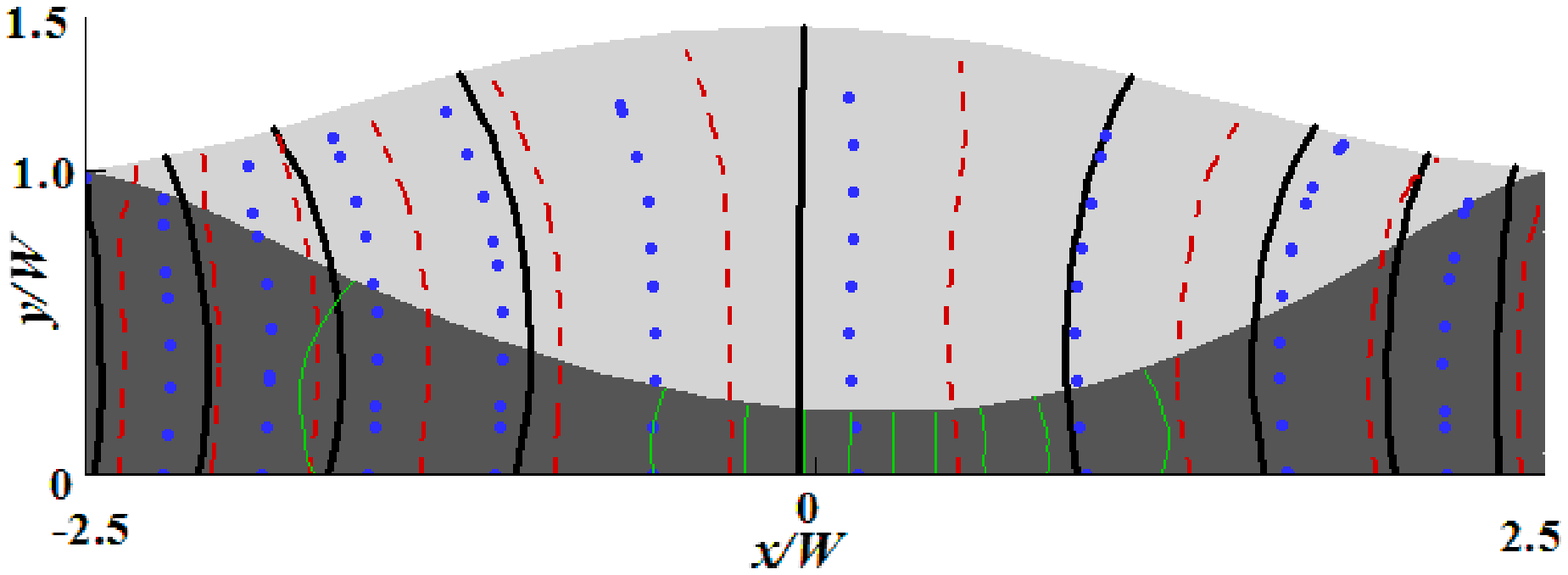}}\\
(b) & (d)  \\
\end{tabular}
\end{center}
%\begin{spacing}{1.5}
\caption{\small \label{figprescon} Contours of pressure in the flow domain, for 
Newtonian (black), Oldroyd-B (red), FENE-P (blue) and Owens (green) fluids at 
$P_e$ = $0.04$, $t = 0.4W$ for two different values of Weissenberg number $Wi 
=0.1$ ((a) and (c)) and $Wi =0.5$ ((b) and (d)). Note that $\Gamma = 1.98\times10^{-4}$ in (a)-(b) and $\Gamma$ = $4.95\times10^{-4}$ in (c)-(d).}
%\end{spacing}
\end{figure}

Figures~\ref{figprescon} display the pressure contours under the collapsible wall  for all the fluids at $P_e$ = 0.04 and $t = 0.4W$, for different values of $\Gamma$ and $Wi$. As was observed earlier in the case of velocity contours in figures~\ref{figvelcon}, the qualitative shape of the pressure contour depends strongly on whether the elastic solid lies within or outside the channel. Thus, shapes for all the fluids are similar to each other in figures~\ref{figprescon}~(a) and (b), while the shapes for the Newtonian, Oldroyd-B and FENE-P fluids differ qualitatively from that of Owens model fluid in figures~\ref{figprescon}~(c) and (d), since the elastic wall lies inside the channel in the latter case while lying outside for the former. 

The pressure contours for the Oldroyd-B and FENE-P fluids are quantitatively similar to those for the Newtonian fluid only for the case when $Wi = 0.1$ and $\Gamma = 1.98\times10^{-4}$ (figure~\ref{figprescon}~(a)). On the other hand, they depart from the Newtonian contours at the remaining values of $Wi$ and $\Gamma$ displayed in figures~\ref{figprescon}~(b) to (d). This is consistent with the behaviour of the pressure profiles along the fluid-solid interface observed in figures~\ref{figpresprof}~(a), (b), (d) and (e).  

\begin{figure}
\centering \subfigure[] {
  \label{figdeltaP:sub:a}
    \includegraphics[height=9cm]{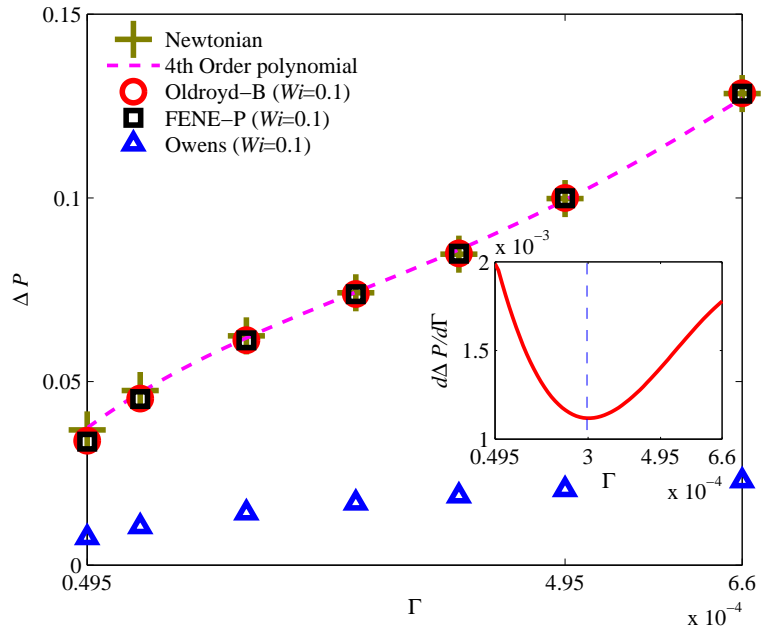}
}
 \subfigure[] {
    \label{figdeltaP:sub:b}
    \includegraphics[height=9cm]{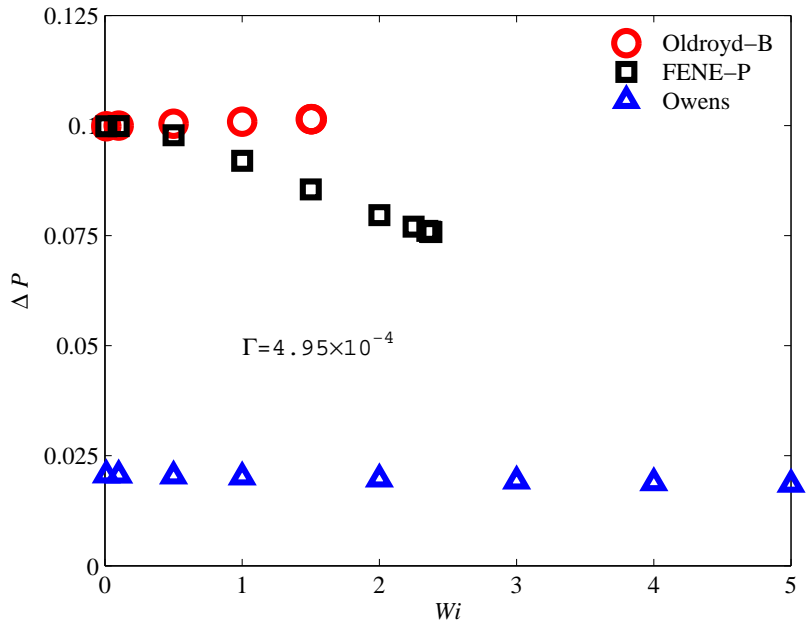}
}
%\begin{spacing}{1.5}
 \caption{\small \label{figdeltaP} Dependence of pressure drop $\Delta P$ in the 
channel for the Oldroyd-B, FENE-P and Owens models on (a) $\Gamma$ at a
fixed value of $Wi$=0.1 and (b) $Wi$ at $\Gamma = 4.95\times10^{-4}$. Note 
that for a Newtonian fluid, $\Delta P = 0.1$ in (b)micro-structural. The curves terminate at the limiting Weissenberg number for each model.}
%\end{spacing}
\end{figure}

A different perspective on fluid pressure in the channel is provided in figure~
\ref{figdeltaP}, where the pressure drop $\Delta P$ in the channel between the entrance and exit to the region beneath the elastic solid, is displayed. As seen earlier in figures~\ref{figintshape}~(a)-(b), with increasing $\Gamma$, the interface shape for the Oldroyd-B and FENE-P fluids moves from being concave
downwards to convex upwards. Figure~\ref{figdeltaP}~(a) shows that this
is accompanied by an increase in $\Delta P$. Interestingly, the rate of change of $\Delta P$ with $\Gamma$ has a point of inflection around $\Gamma \sim 3 \times 10^{-4}$, which is approximately the value at which the elastic solid becomes horizontal (see figures~\ref{figintshape}~(a)-(b), and inset to figure~\ref{figdeltaP}~(a)).

A striking manifestation of differences in the prediction of a macroscopic property, 
because of differences in fluid rheology, is displayed in  figure~\ref{figdeltaP}~(b), where the  dependence of pressure drop $\Delta P$ on Weissenberg number $Wi$ is plotted. The Owens model fluid has a nearly constant pressure drop because the fluid has undergone significant shear thinning, and has an almost constant viscosity under the deformable elastic solid for all values of $Wi $. For the Oldroyd-B model on the other hand, which is a constant viscosity fluid, there appears to be a very slight increase in $\Delta P$. Clearly, the decrease in pressure at the downstream end of the channel, is more than made up with the increase at the upstream end. For the FENE-P fluid, the increasing shear thinning with increasing $Wi$ is reflected in  figure~\ref{figdeltaP}~(b) with the observed decrease in $\Delta P$.

\begin{figure}
\begin{center}
\begin{tabular}{cc}
\resizebox{8.0cm}{!} {\includegraphics*[width=8.0cm]{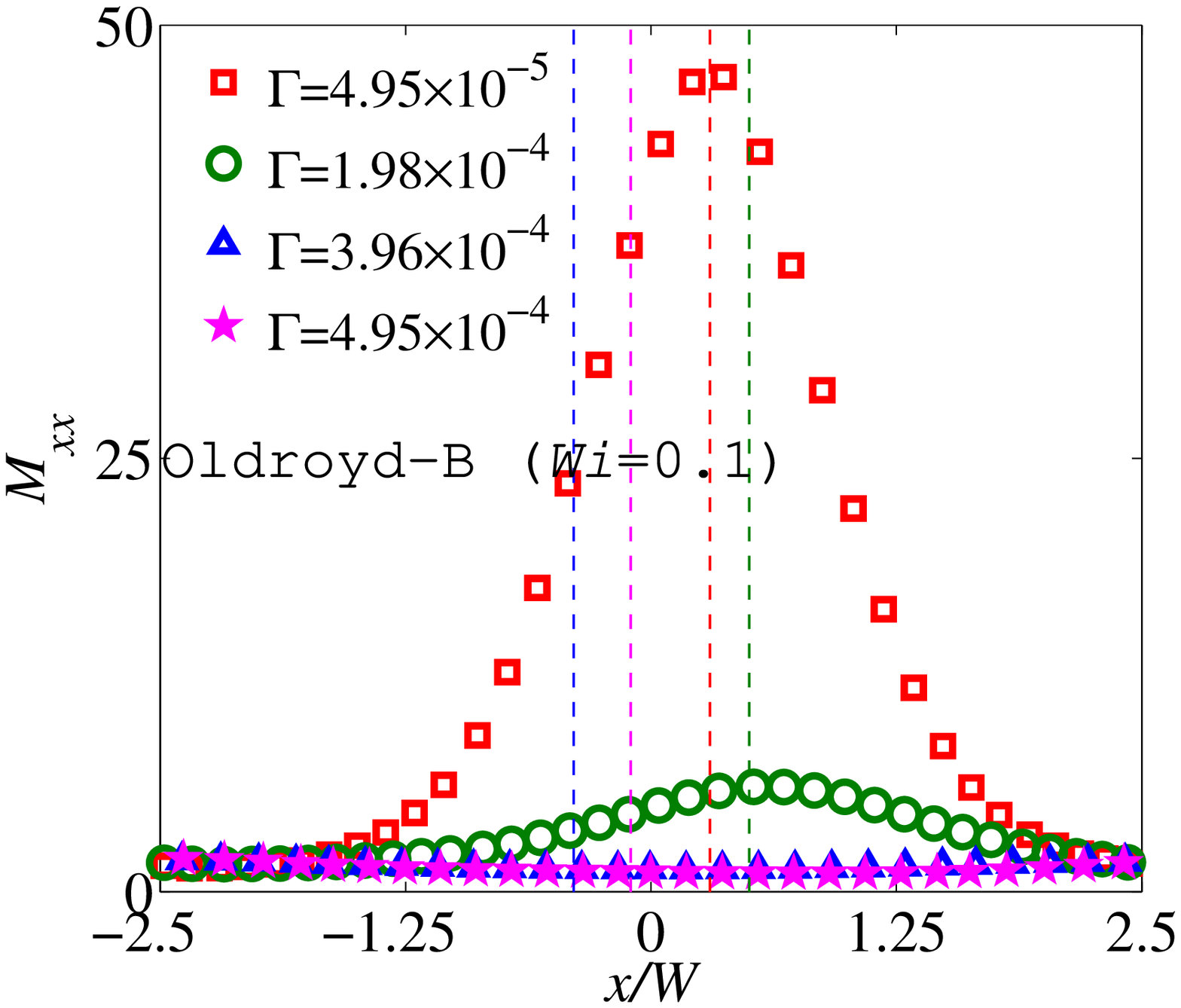}}
&
\resizebox{8.0cm}{!} {\includegraphics*[width=8.0cm]{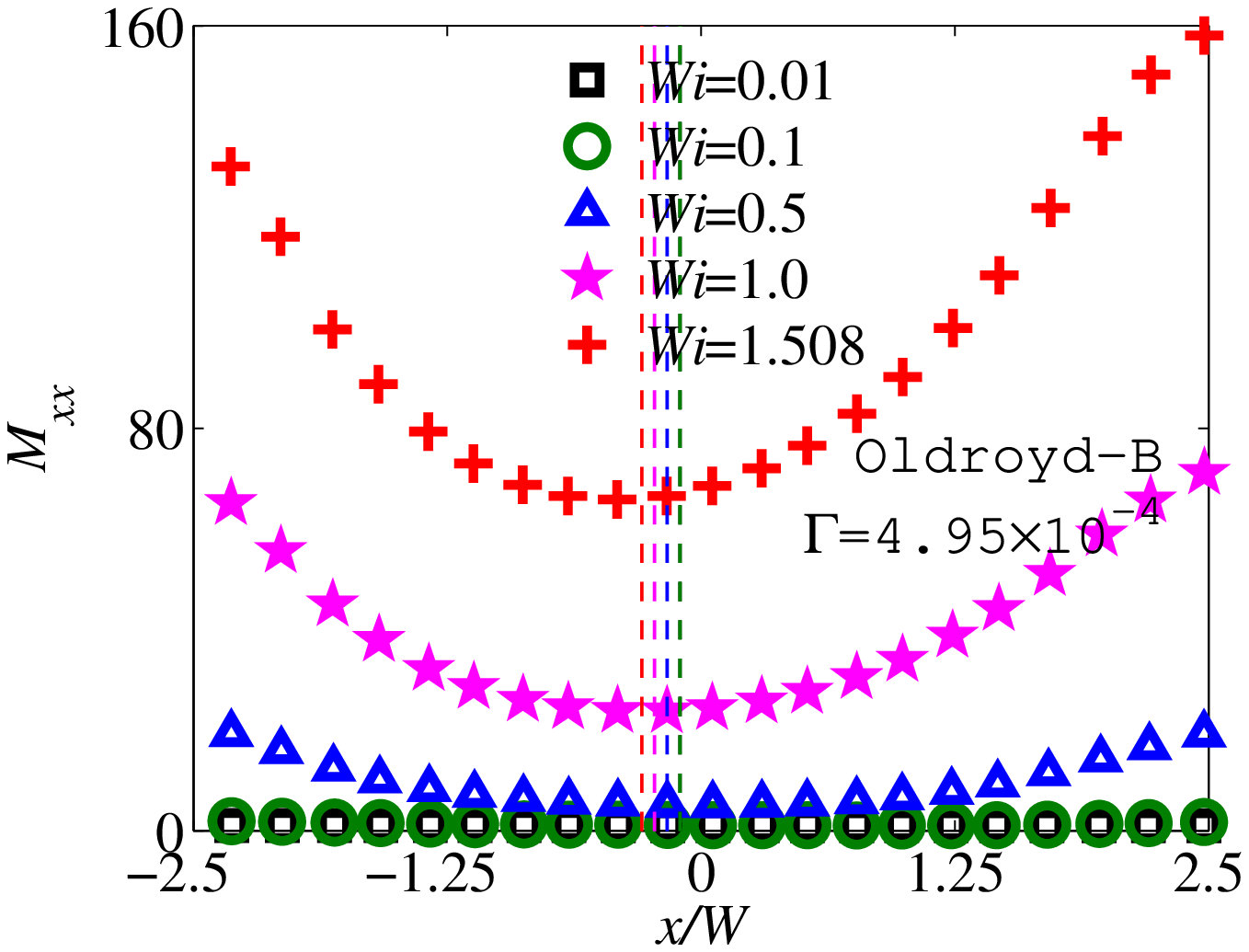}}\\
(a) & (d)  \\
\resizebox{8.0cm}{!} {\includegraphics*[width=8.0cm]{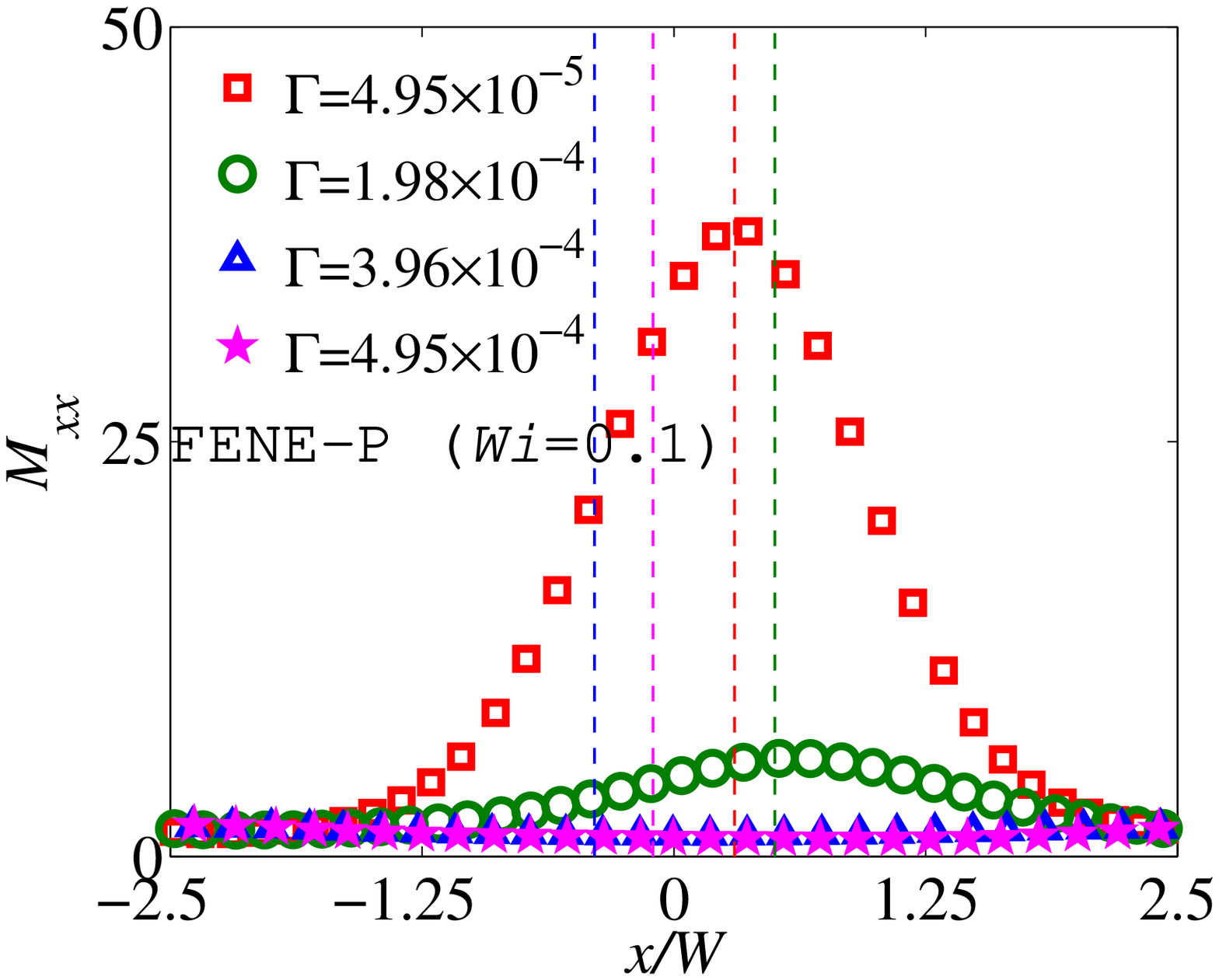}}
&
\resizebox{8.0cm}{!} {\includegraphics*[width=8.0cm]{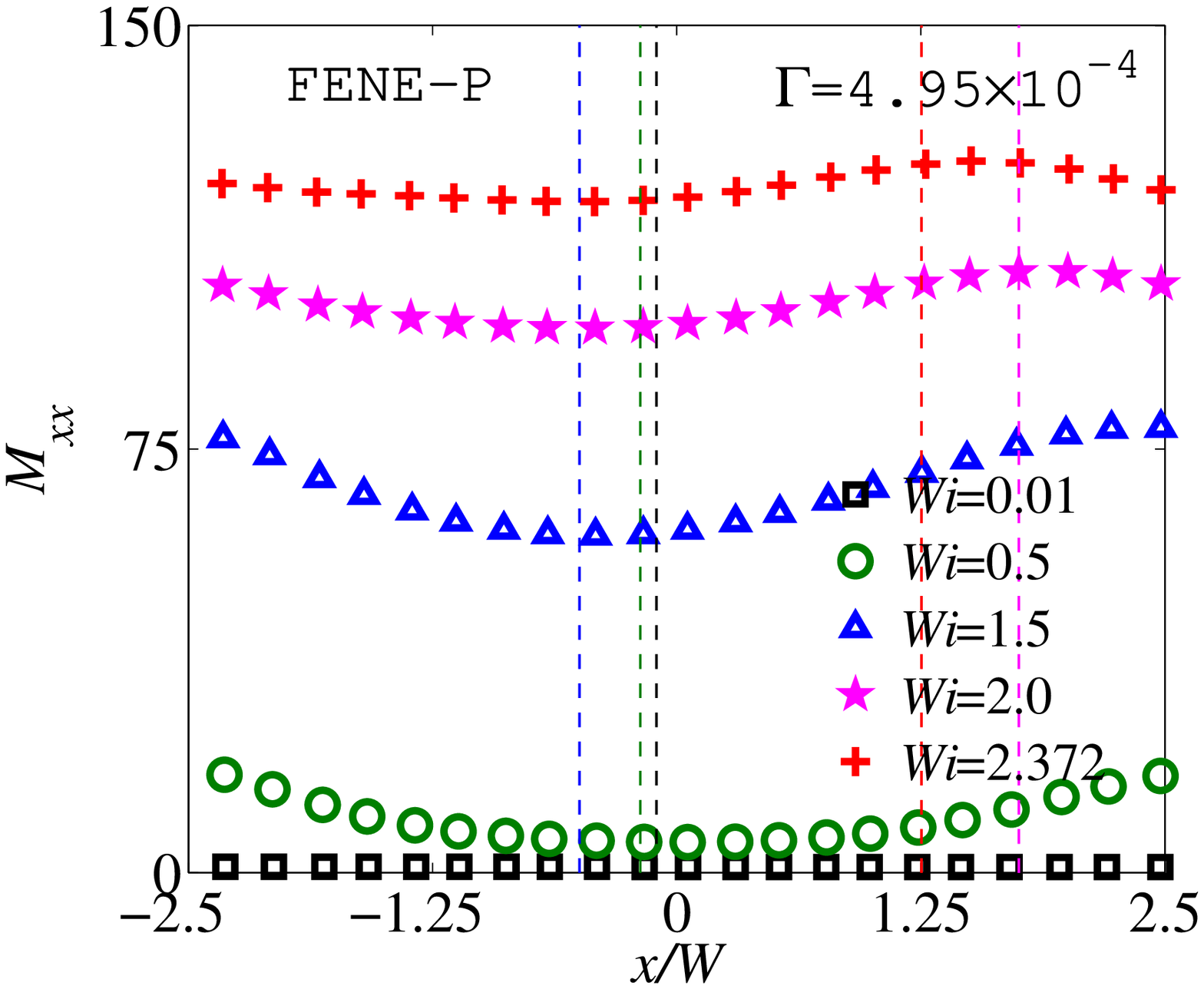}}\\
(b) & (e)  \\
\resizebox{8.0cm}{!} {\includegraphics*[width=8.0cm]{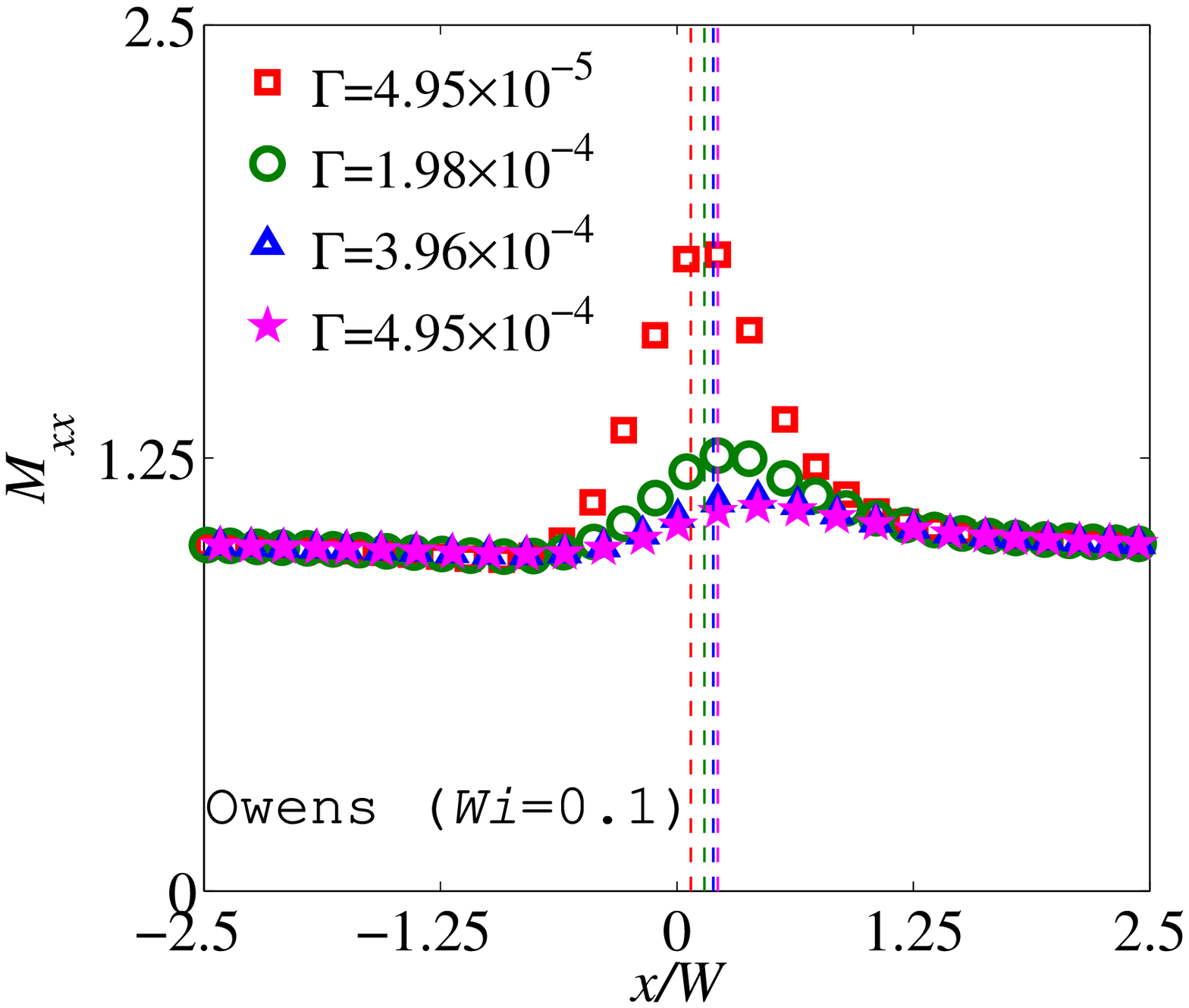}}
&
\resizebox{8.0cm}{!} {\includegraphics*[width=8.0cm]{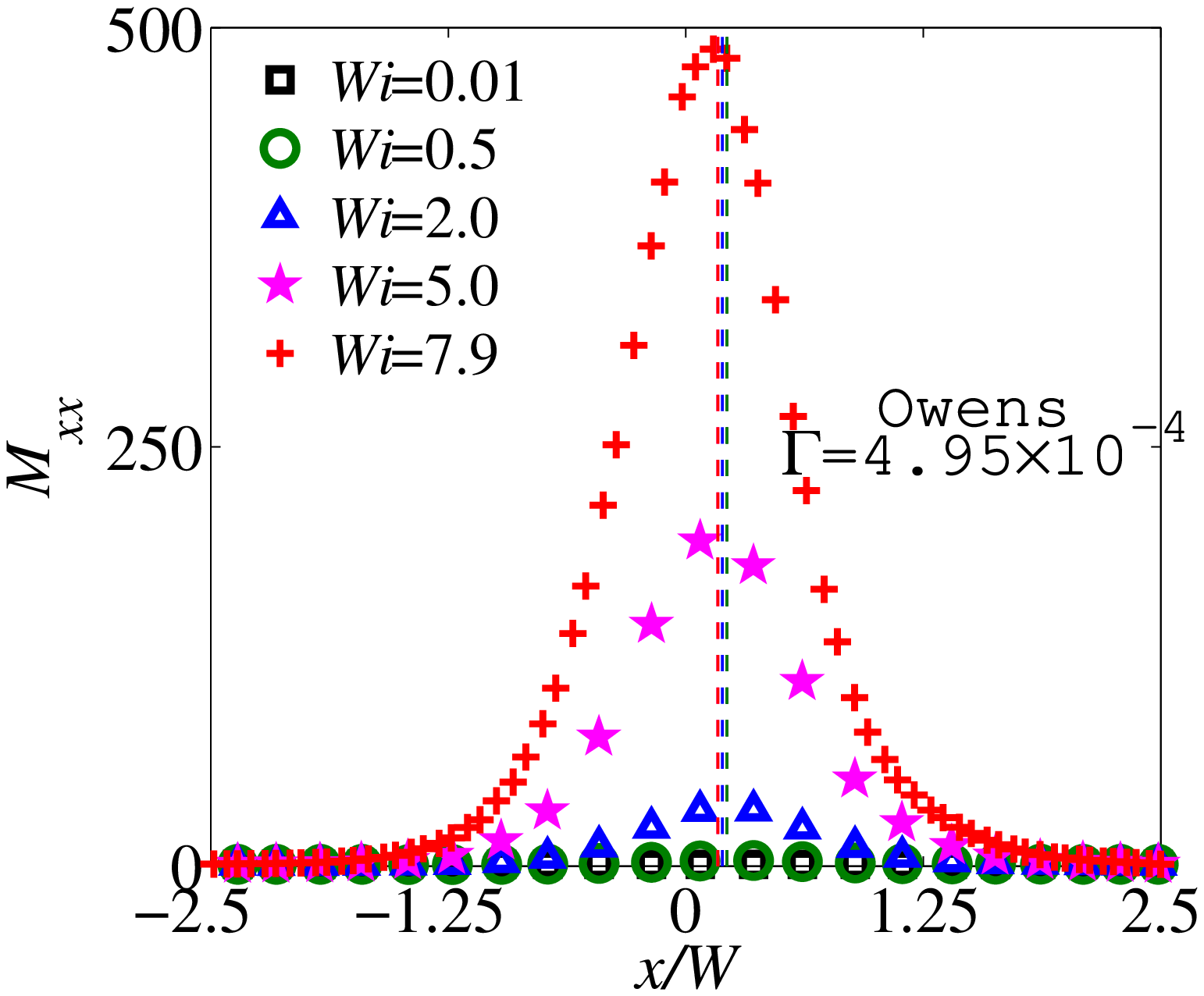}}\\
(c) & (f)  \\
\end{tabular}
\end{center}
%\begin{spacing}{1.5}
\caption{\small  \label{figMxxprof} Dependence of the axial component of
the conformation tensor $M_{xx}$ on $\Gamma$, for (a) Oldroyd-B, (b)
FENE-P,  and (c) Owens models, at $Wi = 0.1$, and dependence of
$M_{xx}$ on $Wi$, for (d) Oldroyd-B, (e) FENE-P,  and (f) Owens
models, at $\Gamma$ = $4.95\times10^{-4}$.}
%\end{spacing}
\end{figure}

Figure~\ref{figMxxprof} explores the dependence of the axial component of
the conformation tensor $M_{xx}$, along the flexible wall, on $\Gamma$
and $Wi$. Figures~\ref{figMxxprof}~(a)-(c) show that an
increase in $\Gamma$ leads to a decrease in the degree of stretching
experienced by the micro-structural elements, and that the value of
$M_{xx}$ in the Owens model is much less sensitive to the value of
$\Gamma$ compared to the Oldroyd-B and FENE-P fluids. For the latter two 
fluids, for values of  $\Gamma \lesssim  3  \times 10^{-4}$, the elastic solid is concave downwards. As a result, the $M_{xx}$ profile has a maximum at the location in the channel where the gap is narrowest. As the elastic solid moves out of the channel, there is a significant relaxation in the degree to which the micro-structural elements are stretched.

The correlation between interface shape and $M_{xx}$ profile is more strikingly
revealed in figures~\ref{figMxxprof}~(d)-(e), where the dependence of $M_{xx}$ 
on $Wi$ is explored at a constant value $\Gamma =  4.95  \times 10^{-4}$.
Since the interface always bulges outwards for the Oldroyd-B and FENE-P fluids at this value of $\Gamma$, the highest stretch occurs at the \emph{inlet} and \emph{outlet} to the deformable region, in contrast to
the situation for the Owens model, where the elastic solid is always concave
downwards, and consequently, the maximum stretch is always at the location of 
the narrowest gap. At high values of $Wi$, the shear thinning experienced by the FENE-P fluid appears to lead to a more uniform stretching along the length of the channel. In all cases, however, as might be anticipated, an increase in $Wi$ leads to an increase in stretching. 

\begin{figure}
\centering \subfigure[] {
    \label{figMxxcon:sub:a}
    \includegraphics[width=10cm,height=5cm]{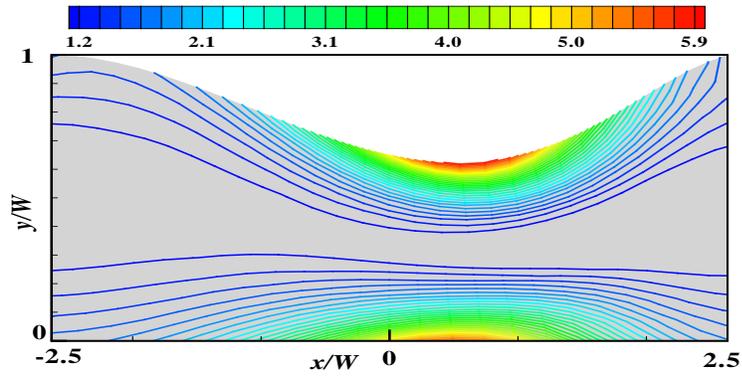}
}
 \subfigure[] {
    \label{figMxxcon:sub:b}
    \includegraphics[width=10cm,height=5cm]{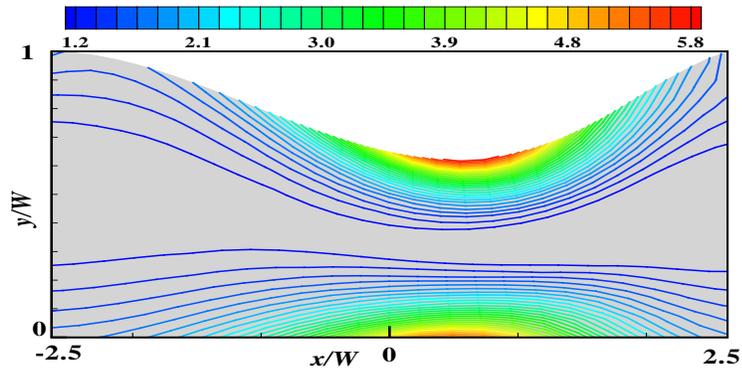}
}
 \subfigure[] {
    \label{figMxxcon:sub:c}
    \includegraphics[width=10cm,height=5cm]{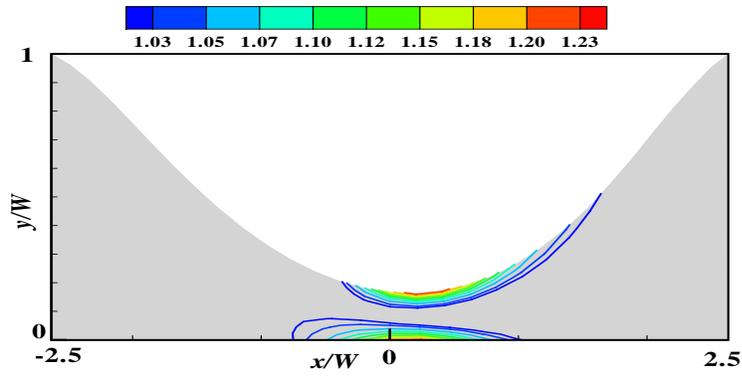}
}
%\begin{spacing}{1.5}
\caption{\small \label{figMxxcon} Contours of $M_{xx}$ in the flow domain for the 
Oldroyd-B, FENE-P and Owens models at $\Gamma$ = $1.98\times10^{-4}$, 
$P_e$ = $0.04$, $t = 0.4W$ and $Wi =0.1$.}
%\end{spacing}
\end{figure}

Figures~\ref{figMxxcon}~(a)-(c) show the contour plots of the mean streamwise 
molecular stretch $M_{xx}$ for the Oldroyd-B, FENE-P and Owens models,
at $\Gamma$ = $1.98\times10^{-4}$ and $Wi =0.1$. As can be seen from figures~\ref{figMxxprof}~(a)-(c), these parameter values correspond to the situation where the elastic wall lies within the channel for all the fluids, and the values of $M_{xx}$ for the Oldroyd-B and FENE-P models are close to each other along the entire length of the channel. Further, for all the models, the largest value of $M_{xx}$ occurs below the collapsible wall at the minimum gap location. All these observations are clearly reflected in figures~\ref{figMxxcon}~(a)-(c), both in the shape of the interface and in the values of $M_{xx}$ corresponding to the various contour lines. 

\begin{figure}
\begin{center}
\begin{tabular}{cc}
\resizebox{8.0cm}{!} {\includegraphics*[width=8.0cm]{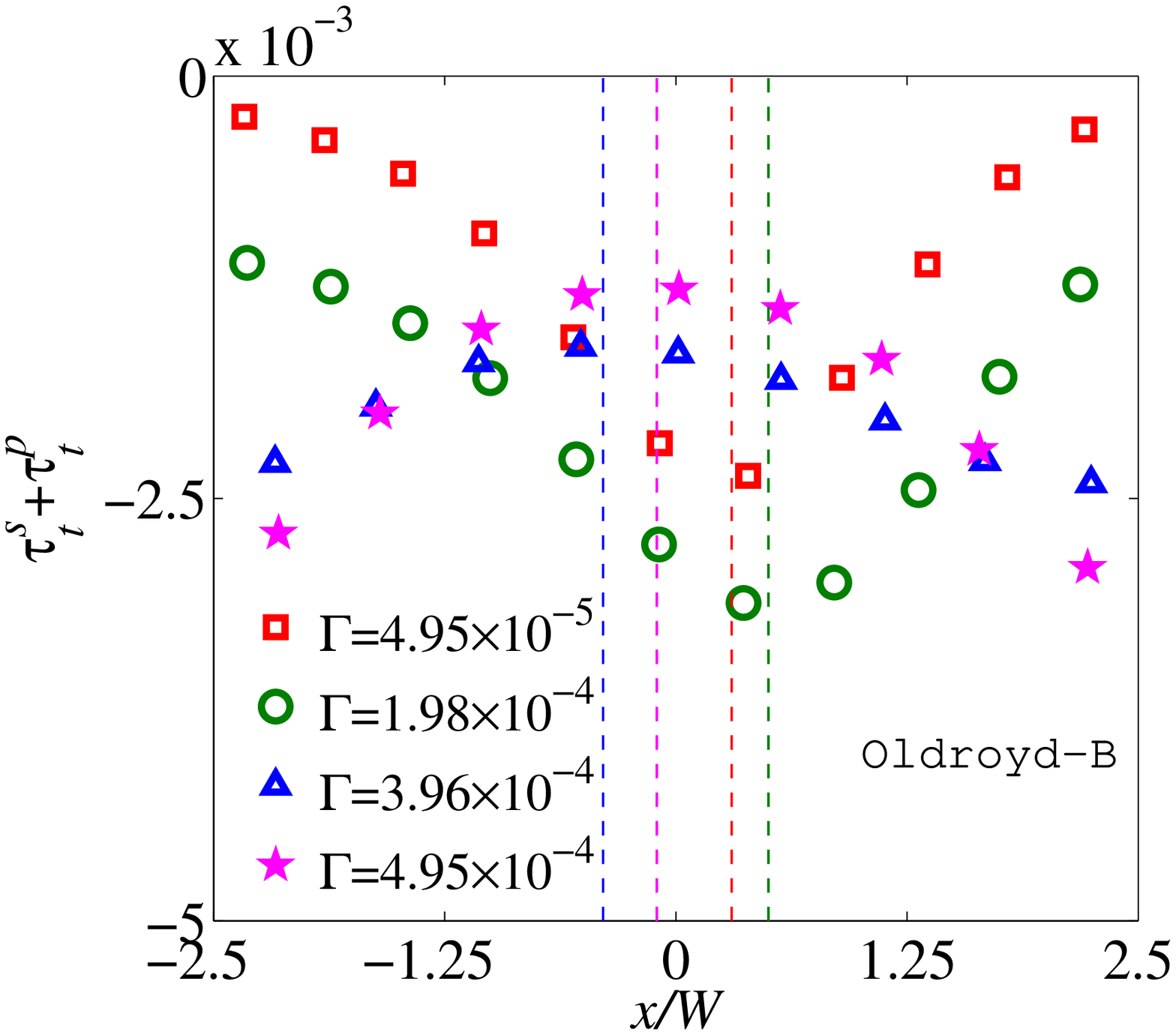}}
&
\resizebox{8.0cm}{!} {\includegraphics*[width=8.0cm]{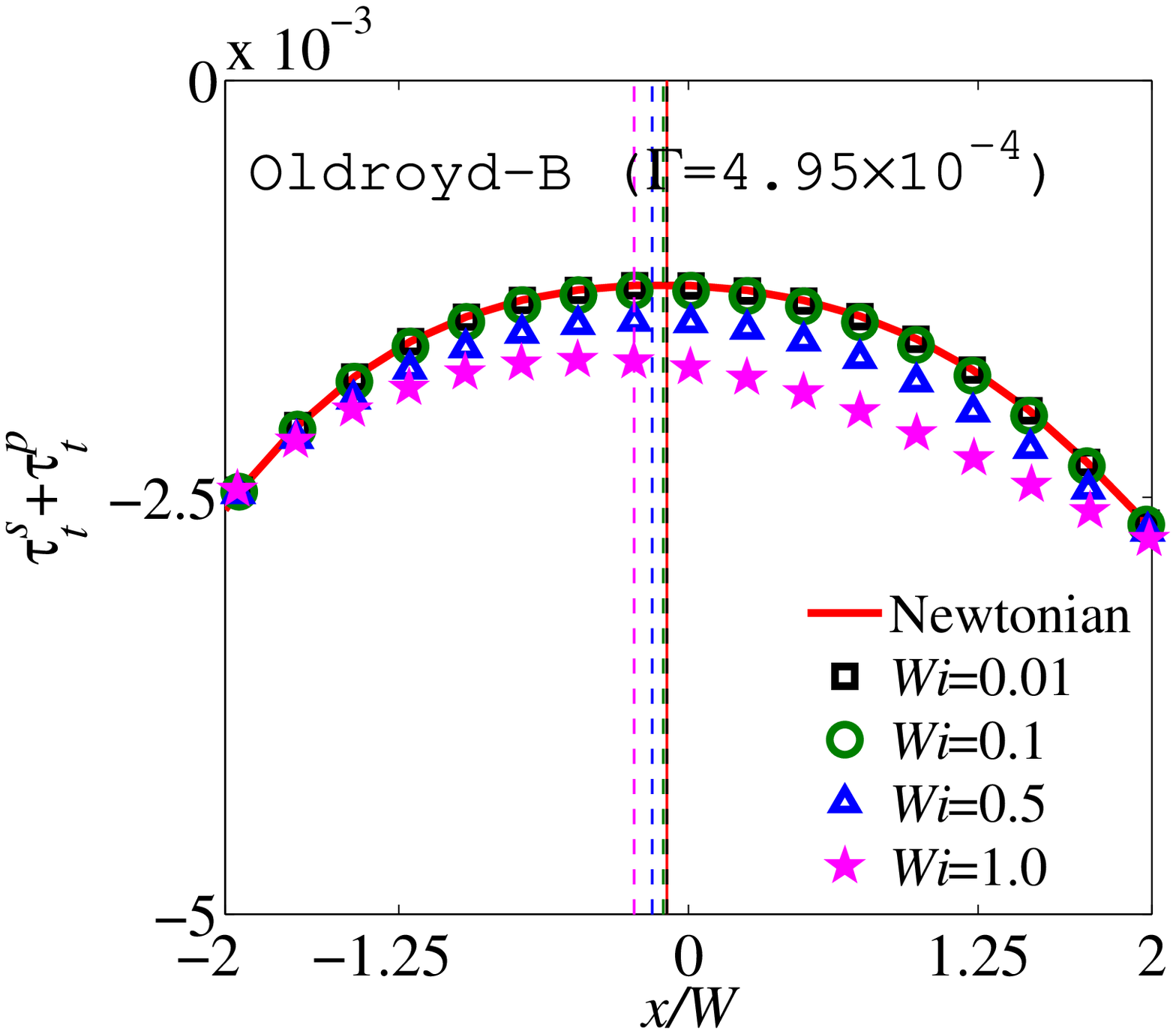}}\\
(a) & (d)  \\
\resizebox{8.0cm}{!} {\includegraphics*[width=8.0cm]{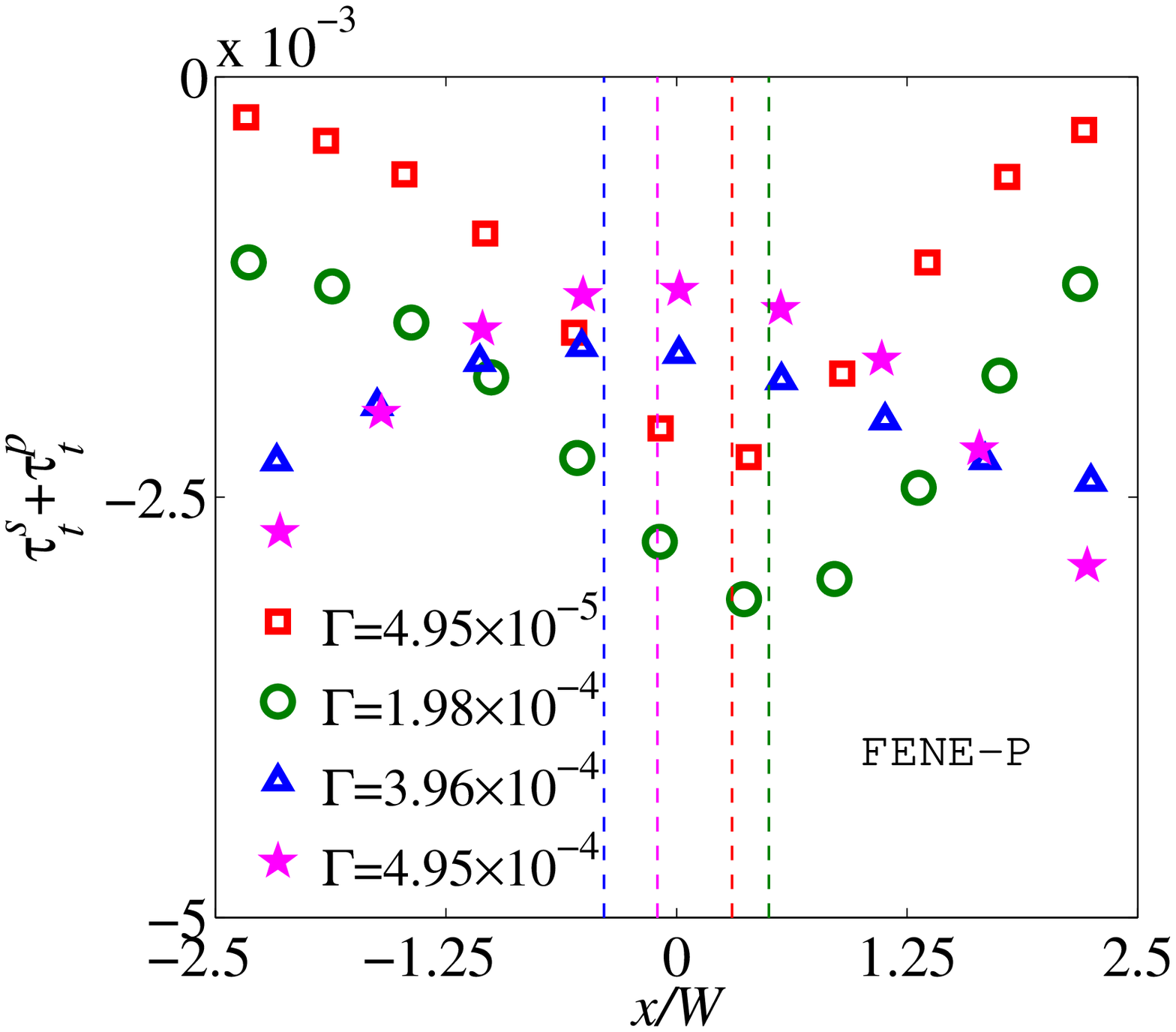}}
&
\resizebox{8.0cm}{!} {\includegraphics*[width=8.0cm]{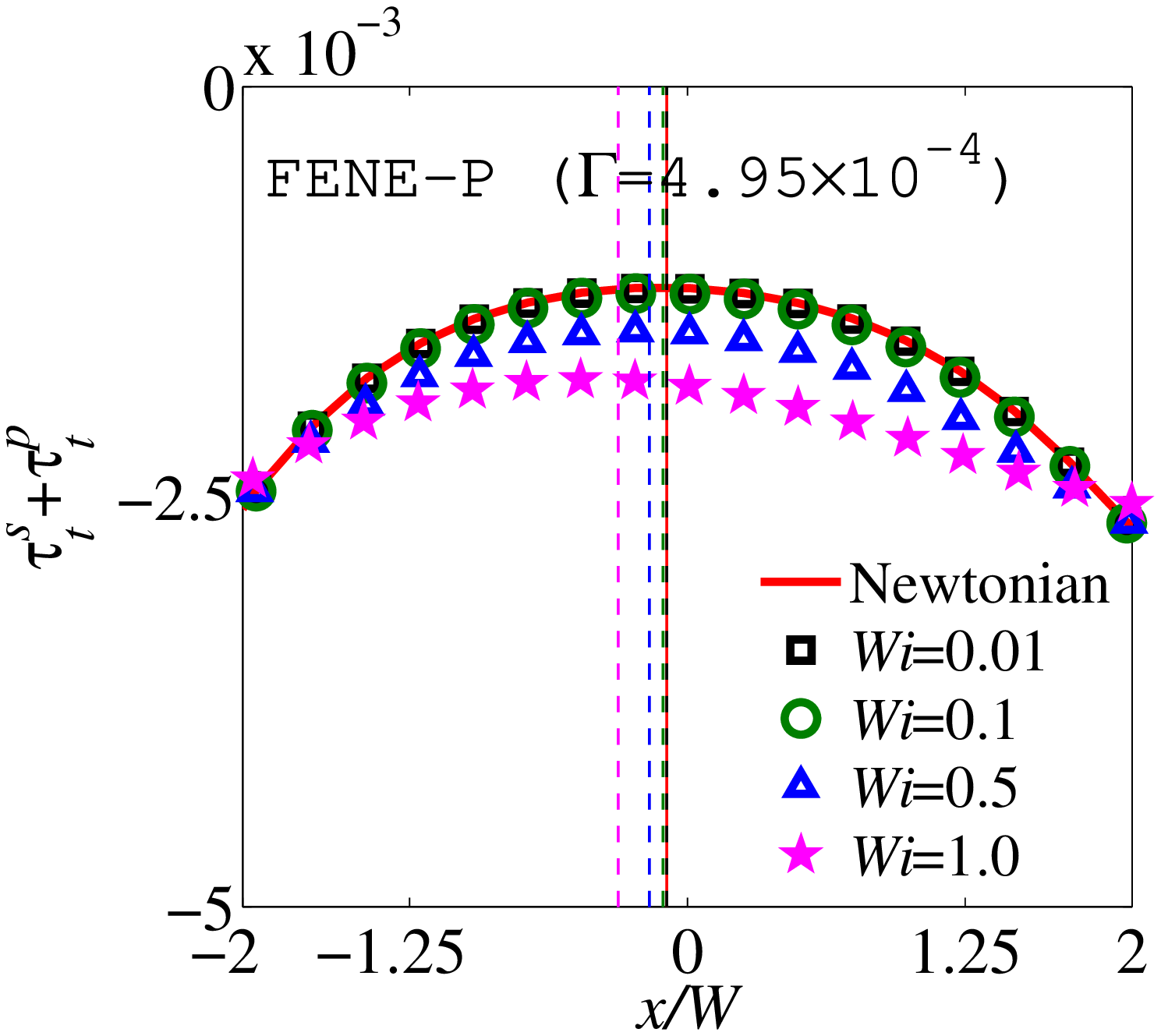}}\\
(b) & (e)  \\
\resizebox{8.0cm}{!} {\includegraphics*[width=8.0cm]{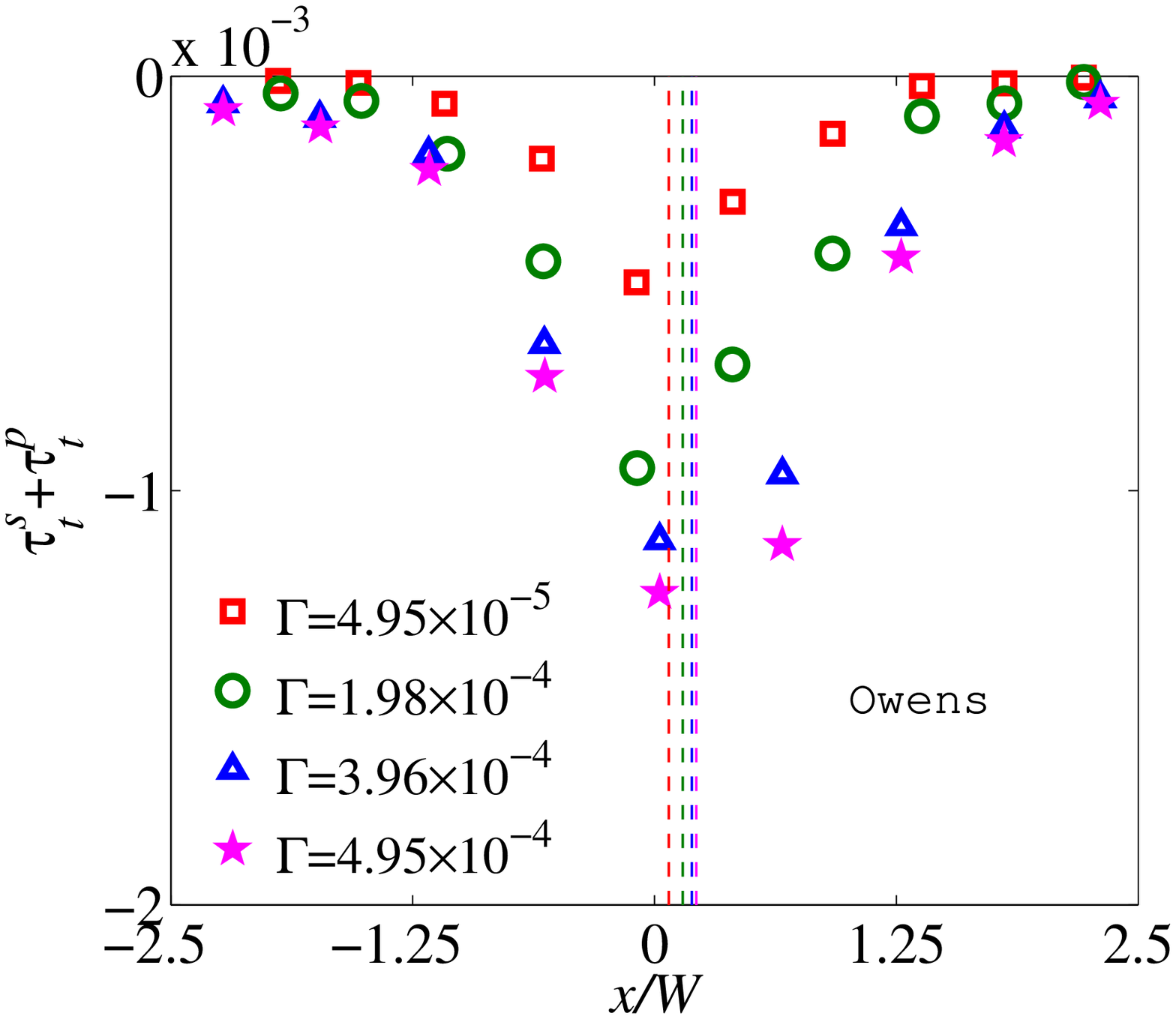}}
&
\resizebox{8.0cm}{!} {\includegraphics*[width=8.0cm]{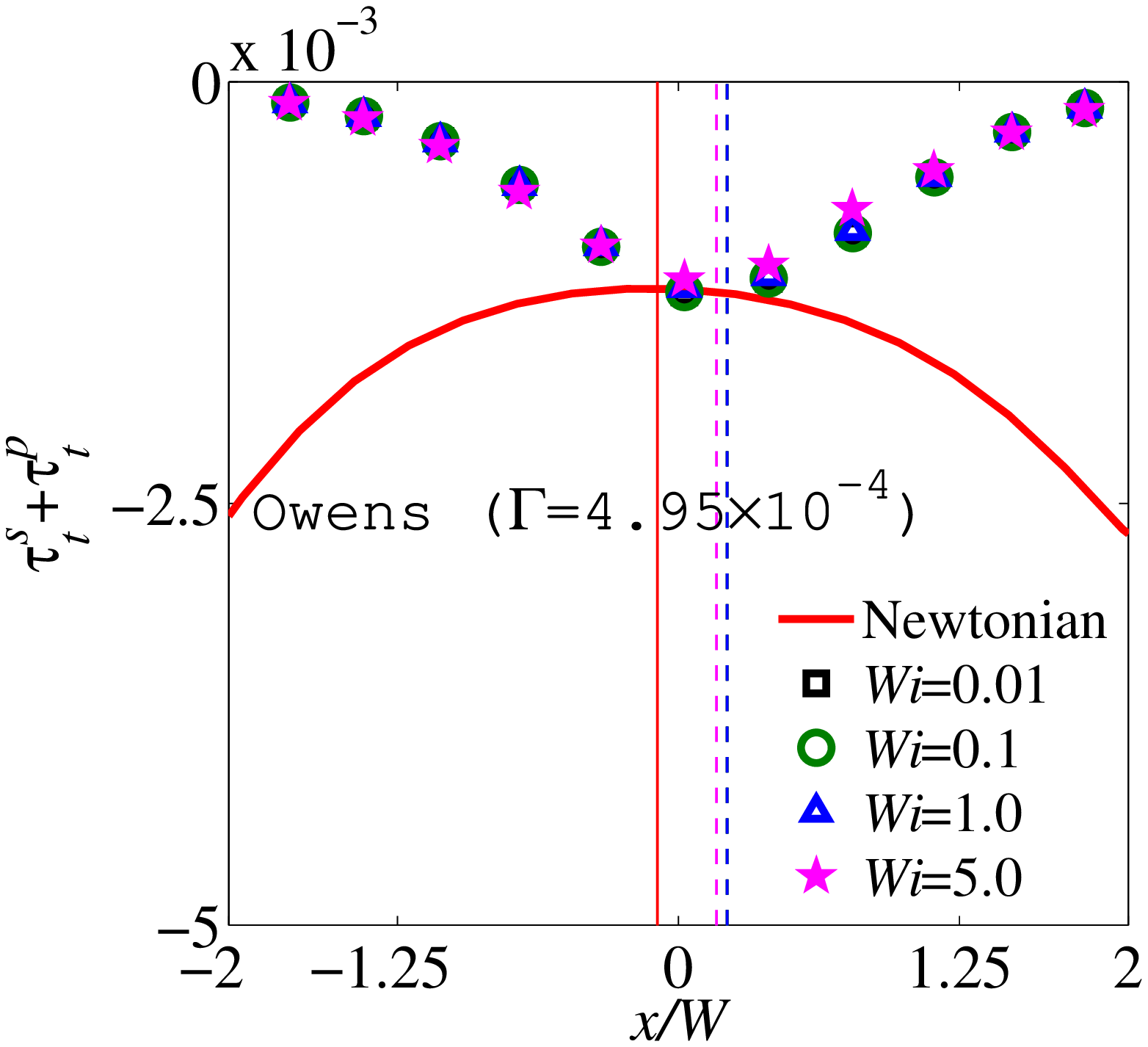}}\\
(c) & (f)  \\
\end{tabular}
\end{center}
%\begin{spacing}{1.5}
\caption{\small  \label{figtauprof} Dependence of the tangential
component of stress $\bm{\tau^{\text{s}}_t}+\bm{\tau^{\text{p}}_t}$
on $\Gamma$, for (a) Oldroyd-B, (b) FENE-P,  and (c) Owens models,
at $Wi = 0.1$, and dependence of
$\bm{\tau^{\text{s}}_t}+\bm{\tau^{\text{p}}_t}$ on $Wi$, for (d)
Oldroyd-B, (e) FENE-P,  and (f) Owens models, at $\Gamma$ =
$4.95\times10^{-4}$.}
%\end{spacing}
\end{figure}

Finally, the dependence of the total shear stress
on the elastic solid, $\bm{\tau^{\text{s}}_t}+\bm{\tau^{\text{p}}_t}$, on the 
parameters $\Gamma$ and $Wi$,
is examined in figure~\ref{figtauprof} for the three viscoelastic fluids.
Once again, there is close parallel between
the shape of the fluid-solid interface and the shear stress on the wall. Indeed, the
shear stress profiles are either concave downwards or convex upwards in 
complete
synchrony with the interface shape. In contrast to the zero-thickness membrane 
model,
where the shear stress on the membrane has no influence on membrane shape 
because of the
use of a boundary condition that only accounts for the influence of the normal 
stress, in the present model,
both the pressure and the shear stress are responsible for the membrane shape. 
As a result, a much
greater variety of interface shapes is observed for a finite thickness elastic solid.

\section{\label{sec:conclusion} Conclusions}
We have introduced a new geometry, whose central feature is the existence of 
fluid-structure interaction, into the lexicon of standard benchmark non-Newtonian 
flow computations. The role that the presence of a deformable membrane plays 
in the development of a complex flow field in the channel has been examined, 
and the relationship of the upper limit to the Weissenberg number to molecular 
conformations at various locations in the flow domain, has been delineated. The 
shape of the membrane as a function of a membrane elasticity parameter $
\Gamma$, and of the Weissenberg number $Wi$ has been studied, and the 
change in shape has been used as an indication of the extent of fluid-structure 
interaction. The nature of the coupling between macroscopic observables such 
as velocity, stress and conformation fields, and various rheological features of the 
three viscoelastic fluid models used in this study, has been explored in some 
detail.

There are many aspects of viscoelastic flows in two dimensional collapsible 
channels that remain to be studied. (i) The use of a constitutive model that 
accounts for thixotropy is an important feature, since the aggregation of blood 
cells in regions of low shear rate can lead to rheological properties that depend 
locally on micro-structural dynamics. Owens model in its most general form does 
account for thixotropy~\citep{owens06}, and as mentioned earlier, \citet{iolov11} have recently developed a finite element method for solving the Owens model in its complete generality. (ii) Even though there exists an upper 
limit to the Weissenberg number at which computations fail for each mesh, we 
have not encountered, in our admittedly limited simulations, a situation where this 
upper limit has not changed in spite of mesh refinement. It would be interesting to 
see if the use of a log-conformation tensor formalism leads to much higher upper 
limits to the Weissenberg number for all the models. (iii) The multiple modes of 
instabilities that arise for flow in collapsible channels, and the rich behaviour that 
occurs in unsteady flows, has been extensively investigated for Newtonian fluids. 
We hope that the present work provides a starting point for similar studies in the 
context of viscoelastic fluids.

\section{\label{sec:ack} Acknowledgments}
%\acknowledgments
We thank Matteo Pasquali and Marcio Carvalho for providing us with
their finite element code for simulating coating flows, which we
have modified and adapted to this work. This work was supported by
an award under the Merit Allocation Scheme on the NCI National
Facility at the Australian National University (ANU). The authors
also would like to thank the VPAC (Australia), and SUNGRID (Monash
University, Australia) for the allocation of computing time on their supercomputing 
facilities.

\bibliographystyle{elsarticle-num-names}

\bibliography{jnnfm}






% \end{thebibliography}

%\newpage

%\listoffigures

%\newpage

%\listoftables

%\newpage

\end{document}